\documentclass[aps,prb,twocolumn,
groupedaddress,superscriptaddress,showpacs]{revtex4}

\usepackage{amsmath}
\usepackage{graphicx}
\usepackage{bm}

\newcommand{\Hca}{\mathcal{H}}
\newcommand{\Kca}{\mathcal{K}}
\newcommand{\Mca}{\mathcal{M}}
\newcommand{\Pca}{\mathcal{P}}
\newcommand{\rA}{\text{A}}
\newcommand{\rB}{\text{B}}

\newcommand{\vk}{{\bf k}}
\newcommand{\vx}{{\bf x}}

\newcommand{\vq}{{\bf q}}
\newcommand{\vp}{{\bf p}}

\newcommand{\vf}{v_{\rm F}}

\newcommand{\la}{\langle}
\newcommand{\ra}{\rangle}

\newcommand{\tp}{t_{\perp}}
\newcommand{\om}{\omega}

\newcommand{\nimp}{n_{\rm i}}
\newcommand{\e}{\epsilon}

\newcommand{\nF}{n_{\rm F}}

\newcommand{\kp}{k_{\perp}}
\newcommand{\kpa}{k_{\parallel}}
\newcommand{\vkpa}{{\bf k}_{\parallel}}

\DeclareMathOperator{\sign}{sign}
\DeclareMathOperator{\real}{Re}
\DeclareMathOperator{\imag}{Im}

\newcommand{\SA}{\Sigma_{\text{A}}}
\newcommand{\SAo}{\Sigma_{\text{A}} (\omega)}
\newcommand{\SBo}{\Sigma_{\text{B}} (\omega)}
\newcommand{\SB}{\Sigma_{\text{B}}}

\newcommand{\GR}{{\rm G}}
\newcommand{\gr}{{\rm g}}
\newcommand{\grD}{{\rm g^D}}
\newcommand{\grND}{{\rm g^{ND}}}

\newcommand{\GDA}{{\rm G}_{\rm AA}^{\rm D}}
\newcommand{\GDB}{{\rm G}_{\rm BB}^{\rm D}}
\newcommand{\GDAa}{\overline{\rm G}_{\rm AA}^{\rm D}}
\newcommand{\GDBb}{\overline{\rm G}_{\rm BB}^{\rm D}}

\newcommand{\tE}{\widetilde{E}}
\newcommand{\tA}{\widetilde{A}}

\newcommand{\Eg}{E_{\text{g}}}

\newcommand{\VMF}{V_{\text{MF}}}

\newcommand{\PD}{\partial}

\newcommand{\Ri}{{\bf R}_i}

\newcommand{\kg}{k_{\text{g}}}

\newcommand{\Eb}{E_{\text{b}}}
\newcommand{\kapa}{\kappa_{\alpha}}

\begin{document}

\title{Electronic properties of bilayer and multilayer graphene}

\author{Johan Nilsson}
\affiliation{Department of Physics, Boston University, 590 
Commonwealth Avenue, Boston, MA 02215, USA}
\affiliation{Instituut-Lorentz, Universiteit Leiden,
P.O. Box 9506, 2300 RA Leiden, The Netherlands}

\author{A.~H. Castro Neto}
\affiliation{Department of Physics, Boston University, 590 
Commonwealth Avenue, Boston, MA 02215, USA}

\author{F. Guinea}

\affiliation{Instituto de  Ciencia de Materiales de Madrid, CSIC,
 Cantoblanco E28049 Madrid, Spain}

\author{N.~M.~R. Peres}

\affiliation{Center of Physics and Departamento de F{\'\i}sica,
Universidade do Minho, P-4710-057, Braga, Portugal}

\date{January 18, 2007}

\begin{abstract}
We study the effects of site dilution disorder on the electronic
properties in graphene multilayers, in particular the 
bilayer and the infinite stack.
The simplicity of the model allows for an easy implementation of the
coherent potential approximation and some analytical results.
Within the model we compute the self-energies, the density of states and
the spectral functions.
Moreover, we obtain the frequency and temperature dependence of the
conductivity as well as the DC conductivity. 
The c-axis response is unconventional in the sense that
impurities increase the response for low enough doping.
We also study the problem of impurities in the biased graphene bilayer.
\end{abstract}


\maketitle

\section{Introduction}

The isolation of single layer graphene by Novoselov {\it et al.} \cite{Novoselov2004} 
has generated enormous interest in the physics community. On the one hand, the 
electronic excitations of graphene can be described by the two-dimensional (2D)
Dirac equation, creating connections with certain theories in particle physics.\cite{pw}
Moreover, the ``relativistic'' nature of the quasiparticles,
albeit with a speed of propagation, $v_F$, 300 times smaller than the speed of light, 
leads to unusual spectroscopic, transport, and thermodynamic properties that are
at odds with the standard Landau-Fermi liquid theory of metals.\cite{ahcn_review} 
On the other hand, graphene opens the doors for an all-carbon based
micro-electronics.\cite{geim_review}

Due to the strong nature of the $\sigma$ bonds in graphene,
and strong mechanical stability of the graphene lattice, miniaturization can 
be obtained at sizes of order of a few nanometers, beyond what can obtained with
the current silicon technology (the smallest size being of the order of the benzene
molecule). Furthermore, the same stability allows for creation of entire devices
(transistors, wires, and contacts) carved out of the same graphene sheet, reducing
tremendously the energy loss, and hence heating, created by contacts between different
materials.\cite{Betal04} Early proposals for the control of the electronic properties
in graphene, such as the opening of gaps, were based on controlling its geometry,
either by reducing it to nanoribbons,\cite{Nakada1996} or producing graphene quantum
dots.\cite{Efetov06} Nevertheless, current lithographic techniques that can produce such 
nanostructures do not have enough accuracy to cut graphene to {\AA}ngstrom precision. 
As a result, graphene nanostructures unavoidably have rough edges which have strong
effects in the transport properties of nanoribbons.\cite{HOZK07} In addition, the
small size of these structures can lead to strong enhancement of the Coulomb 
interaction between electrons which, allied to the disorder at the edge of the nanostructures,
can lead to Coulomb blockade effects easily observable in transport and
spectroscopy.\cite{SGN07b}

Hence, the control of electronic gaps by finite geometry is still very unreliable at this 
point in time and one should look for control in bulk systems which are insensitive to
edge disorder. Fortunately, graphene is an extremely flexible material 
from the electronic point of view and electronic gaps can be controlled. This can be 
accomplished in a graphene bilayer with an electric field applied perpendicular to the
plane. It was shown theoretically \cite{Falko2006a,McCann2006a} and demonstrated experimentally
\cite{Castro_PRL_2007,OHLMV07} that a graphene bilayer is the only material with semiconducting properties
that can be controlled by electric field effect. The size of the gap between conduction
and valence bands is proportional to the voltage drop between the two graphene planes and
can be as large as $0.1-0.3$ eV, allowing for novel terahertz devices\cite{Castro_PRL_2007}
and carbon-based quantum dots\cite{Peeters_dots_2007} 
and transistors.\cite{Nilsson2006_BBB}

Nevertheless, just as single layer graphene,\cite{nuno2006_long} bilayer graphene is also
sensitive to the unavoidable disorder generated by the environment of the SiO$_2$ substrate:
adatoms, ionized impurities, etc. Disorder generates a scattering rate $\tau$ and hence
a characteristic energy scale $\hbar/\tau$ which is the order of the Fermi energy 
$E_F = \hbar v_F k_F$ ($k_F \propto \sqrt{n}$ is the Fermi momentum and $n$ is the planar
density of electrons) when the chemical potential is close to the Dirac point ($n \to 0$).
Thus, one expects disorder to have a strong effect in the physical properties of graphene.
Indeed, theoretical studies of the effect of disorder in unbiased \cite{Nilsson2006b} and
biased \cite{Nilsson2007} graphene bilayer (and multilayer) show that disorder leads to strong modifications
of its transport and spectroscopic properties. The understanding of the effects of disorder
in this new class of materials is fundamental for any future technological applications. 
In this context it is worth to mention the transport theories based on the 
Boltzmann equation,\cite{Katsnelson_pointscattering_2007,Adam_2007} 
a study of weak localization in bilayer graphene,\cite{Kechedzhi_2007} 
and also corresponding further experimental 
characterization.\cite{Morozov_2007,Gorbachev_2007}
DC transport in few-layer graphene systems have been studied in
Ref.~\onlinecite{Nakamura_08}, both without and in the presence of
a magnetic field.

In this paper, we study the effects of site dilution (or unitary scattering) on the electronic
properties of graphene multilayers within the well-known coherent potential approximation (CPA).
While the CPA does not take into account electron localization,\cite{note_localization,ziegler06} 
it does provide quantitative and qualitative information on the effect of disorder in
the electronic excitations. Furthermore, this approximation allows for analytical results
of electronic self-energies, allowing us to compute physical quantities such as 
spectral functions (measurable by angle resolved photoemission, ARPES 
\cite{Lanzara2006a,Lanzara2006b,Betal07,Ohta06,Lanzara_gap_2007}) and density
of states (measurable by scanning tunneling microscopy, 
STM\cite{SRRMKBHHF07,Metal07,LA07,Brar_STS_2007}),
besides standard transport properties such as the DC and AC conductivities.\cite{Nilsson2006b}
Furthermore, in the case of the semi-infinite stack of graphene planes we can compute
the c-axis response of the system which is rather unusual since it increases with disorder
at low electronic densities, in agreement with early transport measurements in graphite.\cite{graphitereview1}

The paper is organized as follows.
In Sec.~\ref{sec:bands} we discuss the band model of the
graphene bilayer within the tight-binding approximation. We also connect our
notation with the one established for graphite, namely the
Slonsczewki-Weiss-McClure (SWM) parameterization.
In Sec.~\ref{sec:simplebands} we introduce several simplified band models
and compare the electronic bands in different approximations.
The Green's functions that we use later on in the paper are given
in Sec.~\ref{sec:H_G_bilayer}.

We employ a simplified model for the disordered graphene bilayer in
 Sec.~\ref{sec:impurities_tmatrix} and work out the consequences on
the single particle properties encoded in the self-energies,
the density of states (DOS) and the spectral function.
Sec.~\ref{sec:G_multilayer} contains results for the graphene multilayer.
In Sec.~\ref{sec:electron_transport} we introduce the linear response formulas that
we use to calculate the electronic and optical response.
The results for the conductivities in the bilayer are presented in
Sec.~\ref{sec:conductivity_results_b}, while those for the multilayer can be found in
Sec.~\ref{sec:conductivity_results_g}.

The rest of the paper concerns the problem of impurities in the biased graphene
bilayer. The model of the system and some of its basic properties are discussed in
Sec.~\ref{sec:BGBimpurities}.
In  Sec.~\ref{sec:BGBi_bs_Dirac} we solve the problem of a Dirac delta impurity
exactly within the effective mass approximation.
A simple estimate of when the interactions among impurities becomes important
is presented in Sec.~\ref{sec:simple_estimate_nc}.
We treat more general impurity potentials with variational methods in
Sec.~\ref{sec:variational}, and the special case of a potential well with finite
range is studied in Sec.~\ref{sec:potential_well}.
In Sec.~\ref{sec:CPA_biased} we study the problem of a finite density of impurities
in the coherent potential approximation (CPA).
The effects of trigonal distortions on our results for the biased graphene bilayer
are discussed briefly in Sec.~\ref{sec:BGBi_trigonal}.
Finally, the conclusions of the paper are to be found in Sec.~\ref{sec:conclusions}.
We have also included four appendices with technical details of the calculations
of the minimal conductivity in bilayer graphene (App.~\ref{app:minimal}),
the DOS in multilayer graphene (App.~\ref{app:DOSgraphite}),
the conductivity kernels (App.~\ref{app:kernels}), and
the Green's function in the biased graphene bilayer (App.~\ref{app:Gdetail_biased}).

\section{Electronic bands of the graphene bilayer}
\label{sec:bands}

\begin{figure}[htb]
\centering
\includegraphics[scale=0.56]{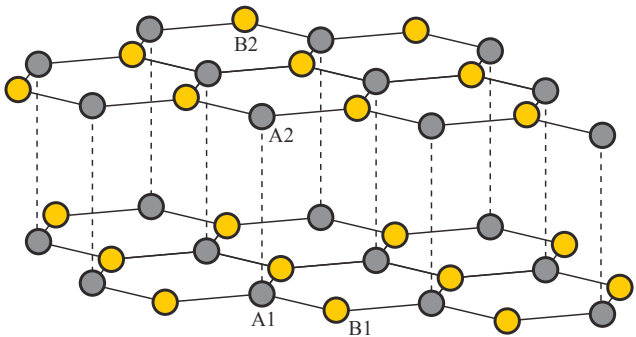}
\caption
  {[color online] Lattice structure of the graphene bilayer. 
  The $\rA$ ($\rB$) sublattices are indicated by the darker (lighter)
  spheres and the planes are labeled by 1 and 2.}
\label{fig_bilayer_unitcell}
\end{figure}

Many of the special properties of the graphene bilayer have their
origin in its lattice structure that leads to the peculiar
band structure that we discuss in detail in this section.
A simple way of arriving at the band structure of the graphene bilayer
is to use a tight-binding approximation. The positions of the different
atoms in the graphene bilayer are shown in
Fig.~\ref{fig_bilayer_unitcell} together with our labeling convention.

The advantage of this notation is that one can discuss collectively
about the $\rA$ ($\rB$) atoms that are equivalent in their physical
properties such as the weight of the wave functions and the
distribution of the density of states etc.
This notation was used in early work on graphite.\cite{McClure1957,SW58}
Many authors use instead a notation similar to
$\rA 1 \rightarrow \rA$, 
$\rB 1 \rightarrow \rB$, 
$\rA 2 \rightarrow \widetilde{\rB}$, and
$\rB 2 \rightarrow \widetilde{\rA}$.
In this notation the relative orientation within the planes of the
$\rA$ ($\widetilde{\rA}$) and $\rB$ ($\widetilde{\rB}$) atoms are the same;
but for the other physical properties the equivalent atoms are
instead $\rA$ ($\rB$) and $\widetilde{\rB}$ ($\widetilde{\rA}$).
Because the other physical properties are often more relevant for the
physics than the relative orientation of the atoms within the planes 
we choose to use the, in our view, most ``natural'' labeling convention.  

\subsection{Monolayer graphene}
\label{sec:graphene}
Let us briefly review the tight-binding model of 
monolayer graphene.\cite{Wallace47} The band structure can be described
in terms of a triangular lattice with two atoms per unit
cell. The real-space lattice vectors can be taken to be
${\bf a}_1 = \frac{a}{2}(3,\sqrt{3})$ and
${\bf a}_2 = \frac{a}{2}(3,-\sqrt{3})$. Here 
$a$ ($\approx 1.4 \, \text{\AA}$) denotes the
nearest neighbor carbon distance.
Three vectors that connect atoms that are nearest neighbors are
${\bm \delta}_1 = \frac{a}{2}(1,\sqrt{3})$, 
${\bm \delta}_2 = \frac{a}{2}(1,-\sqrt{3})$, and
${\bm \delta}_3 = a (-1,0)$;
we take these to connect the $\rA 1$ atoms to 
the $\rB 1$ atoms.
In terms of the operators that creates (annihilates) an electron on
the lattice site at position  $\Ri$ and lattice site $\alpha j$ 
[ $\alpha = (\rA,\,\rB)$ denotes the atom sublattice
and $j$ ($j=1$) denotes the plane ]:
$c_{\alpha j,\Ri}^{\dag}$ ($c_{\alpha j,\Ri}^{\ }$),
the tight-binding Hamiltonian reads:
\begin{equation}
  \label{eq:cintro_HTB_graphene}
  H_{\text{t.b.}} =
  t \sum_{\Ri}\sum_{j=1,2,3}
  \bigl( c_{\rA 1,\Ri}^{\dag}  c_{\rB 1,\Ri+\bm{\delta}_j}^{\   } +
  \text{h.c.} \bigr). 
\end{equation}
Here $t$ ($\approx 3 \, \text{eV}$) is the energy associated with
the hopping of electrons between neighboring $\pi$ orbitals.
We define the Fourier-transformed operators,
\begin{equation}
  \label{eq:cintro_FT_anihilaltion}
  c^{\ }_{\alpha j,\Ri} = \frac{1}{\sqrt{N}} \sum_{\vk}e^{i \vk \cdot \Ri} 
  c^{\ }_{\alpha j,\vk},
\end{equation}
where $N$ is the number of unit cells in the system.
Throughout this paper we use units such that 
$\hbar = k_{\text{B}} =1$ unless specified otherwise.

Because of the sublattice structure it is often convenient to describe
the system in terms of a spinor: 
$\Psi^{\dag}_{\vk}  = \bigl( c_{\rA 1, \vk }^{\dag} , \, 
c_{\rB 1, \vk }^{\dag} \bigr)$, 
in which case the Hamiltonian can be written as:
\begin{equation}
  \label{eq:cintro_graphene_TB2}
  H_{\text{t.b.}} = \sum_{\vk} \Psi^{\dag}_{\vk}
  \begin{pmatrix}
    0 & \zeta(\vk) \\
    \zeta^{*}(\vk) & 0
  \end{pmatrix}
  \Psi^{\,}_{\vk}.
\end{equation}
where
\begin{multline}
  \label{eq:cintro_zetadef}
  \zeta(\vk) = 
  t \sum_{i}e^{i \vk \cdot \bm{\delta}_i} 
  \\
  =
  t e^{i k_x a /2} \bigl[ 2  
  \cos( \frac{k_y a \sqrt{3}}{2}) + e^{-i 3 k_x a /2} \bigr].
\end{multline}
The reciprocal lattice vectors can be taken to be
${\bf b}_1 = \frac{2 \pi}{3 a}(1,\sqrt{3}) $ and
${\bf b}_2 = \frac{2 \pi}{3 a}(1,-\sqrt{3})$ as
is readily verified.
The center of the Brillouin zone (BZ) is denoted by
$\Gamma$, but for the low-energy properties one can
expand close to the K point of the BZ, which has coordinates
${\bf K}  = \frac{4 \pi}{3 \sqrt{3} a}(0,-1)$.
One then finds $\zeta({\bf K} + \vp) \equiv \sigma
= \vf p e^{i \alpha}$, 
where $\vf = 3 t a /2$ and $\alpha = - \arctan(p_x / p_y)$. Note that
$\alpha = 0$ along the $\text{K} - \Gamma$ line
of the BZ and that it increases anti-clockwise.
With these approximation one finds that the spectrum
of Eq.~(\ref{eq:cintro_graphene_TB2}) is that of
massless 2D Dirac fermions:
$E_{\pm} = \pm \vf p$.
\subsection{Bilayer graphene}

\begin{figure}[htb]
\centering
\includegraphics[scale=0.38]{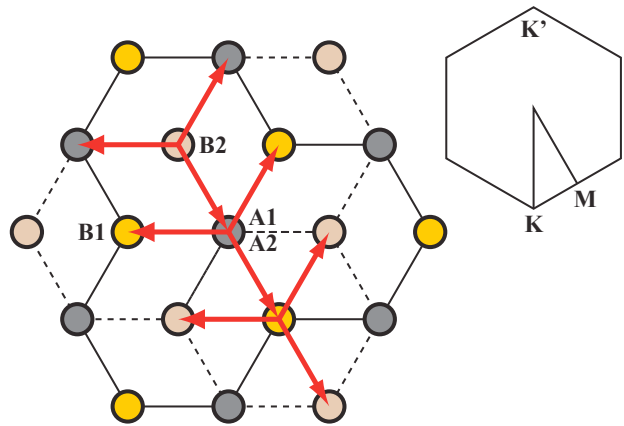}
\caption
  {[color online] The real space lattice structure of the graphene bilayer
  projected onto the $x$-$y$ plane showing the relative positions
  of the different sublattices. The upper right corner shows the BZ
  of the graphene bilayer including the labeling of the high symmetry
  points.} 
\label{fig_bilayer_projected_lattice}
\end{figure}

Since the system
is 2D only the relative position of the atoms
projected on to the $x$-$y$-plane enters into the model.
The projected position of the different atoms are shown in 
Fig.~\ref{fig_bilayer_projected_lattice} together with the
BZ.
Since the $\rA$ atoms are sitting right on top of each other in the
lattice, the hopping term between the $\rA 1$ and $\rA 2$ atoms are local in
real space and hence a constant that we denote by $\tp$ in momentum space.
Referring back to Section~\ref{sec:graphene} we note that the hopping
$\rB 1 \rightarrow \rA 1$ [$\rA 1 \rightarrow \rB 1$]
gives rise to the factor $\zeta(\vk)$ [$\zeta^{*}(\vk)$], with
$\zeta(\vk)$ defined in Eq.~(\ref{eq:cintro_zetadef}).
Since the geometrical role of the $\rA$ and $\rB$ 
atoms are interchanged between
plane 1 and plane 2 we immediately find that in Fourier space the hopping
$\rA 2 \rightarrow \rB 2$ [$\rB 2 \rightarrow \rA 2$]
gives rise to the factor $\zeta(\vk)$ [$\zeta^{*}(\vk)$].
Furthermore, the direction in the hopping 
$\rB 1 \rightarrow \rB 2$ (projected on to the $x$-$y$ plane)
is opposite to that of hopping
$\rB 1 \rightarrow \rA 1$.
Thus we associate a factor $v_3 \zeta^{*}(\vk)$ to the hopping
$\rB 1 \rightarrow \rB 2$, 
where the factor $v_3 = \gamma_3 / \gamma_0$ is needed because the
hopping energy is $\gamma_3$ instead of $\gamma_0 = t$.
Similarly, the direction of hopping
$\rB 1 \rightarrow \rA 2$ (projected on to the $x$-$y$ plane)
 is the same as
$\rB 1 \rightarrow \rA 1$ and therefore the term
$-v_4 \zeta(\vk)$ goes with the hopping
$\rB 1 \rightarrow \rA 2$.
The minus sign in front of $v_4$ follows from the conventional definition
of $\gamma_4$ in graphite, as are discussed below.
Continuing to fill in all the entries of the matrix the full
tight-binding Hamiltonian in the graphene bilayer becomes:
\begin{equation}
  \label{eq:cintro_HbilayerFull_TB}
  \Hca_{\text{t.b.}}(\vk) = 
  \begin{pmatrix}
    V/2 + \Delta & \zeta & \tp & -v_4  \zeta^{*} \\
    \zeta^{*}  & V/2 & -v_4  \zeta^{*} & v_3  \zeta \\
    \tp & -v_4 \zeta  & -V/2 + \Delta & \zeta^{*} \\
    -v_4  \zeta & v_3 \zeta^{*}  & \zeta & -V/2
  \end{pmatrix},
\end{equation}
where the spinor is
$\Psi^{\dag}_{\vk}  = \bigl( c_{\rA 1, \vk }^{\dag} , \, 
c_{\rB 1, \vk }^{\dag}        , \,
c_{\rA 2, \vk }^{\dag}        , \,
c_{\rB 2, \vk }^{\dag} \bigr)$.
Here we have also introduced the conventional (from graphite)
$\Delta$ that parametrizes the difference in energy between $\rA$ and
$\rB$ atoms. In addition we included the parameter $V$ which gives different values of
the potential energy in the two planes, such a term is generally
allowed by symmetry and is generated by an electric field that is
perpendicular to the two layers.
The system with $V \neq 0$ is called the biased graphene bilayer and
has a gap in the spectrum, in contrast the spectrum is gapless if $V=0$.
It is also possible to include further hoppings into the tight-binding
picture, this was done for graphite by Johnson and 
Dresselhaus.\cite{Johnson_Dresselhaus73} The inclusion of such terms is
necessary if one wants an accurate description of the bands throughout
the whole BZ.
If we expand the expression in Eq.~(\ref{eq:cintro_HbilayerFull_TB})
close to the K point in the BZ we obtain the matrix:
\begin{equation}
  \label{eq:cintro_HbilayerFull}
  \Hca_0(\vp) = 
  \begin{pmatrix}
    V/2 + \Delta & \sigma & \tp & - v_4 \, \sigma^*     \\
    \sigma^* & V/2 & -v_4  \sigma^* & v_3  \sigma \\
    \tp & -v_4  \sigma  & -V/2 + \Delta & \sigma^*  \\
    -v_4  \sigma  & v_3  \sigma^*  & \sigma & -V/2
  \end{pmatrix},
\end{equation}
where $\sigma$ was introduced after Eq.~(\ref{eq:cintro_zetadef}).

\begin{figure}[htb]
\includegraphics[scale=0.42]{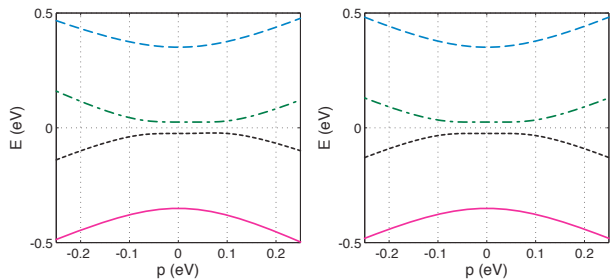}
\centering
\caption
  {[color online] Band dispersions near the K points in the bilayer along the direction $\alpha = 0$,
  with $V = 50 \, \text{meV}$ and $\vf =1$.
  Left: bands obtained from the full model in Eq.~(\ref{eq:cintro_HbilayerFull})
  with $\tp = 0.35 \, \text{eV}$, $v_3 = 0.1$ and $v_4 = 0.05$;
  Right: bands obtained from the simplified model in Eq.~(\ref{eq:bias_Hkin0bilayer}).
}
\label{fig_bilayer_bands1}
\end{figure}

The typical behavior of the bands obtained from 
Eq.~(\ref{eq:cintro_HbilayerFull})
is shown in Fig.~\ref{fig_bilayer_bands1}.
Two of the bands are moved away
from the Dirac point by an energy that is approximately
given by the interplane hopping term $\tp$ for $V \ll \tp$.
In the figure we have taken $V \neq 0$; but for $V=0$ 
there is no gap for the two bands closest to zero
energy (i.e. the Dirac point).

\subsection{The Slonsczewki-Weiss-McClure (SWM) model}
\label{sec:SWM}
First we make the observation that the graphene bilayer in the A-B 
stacking is just the unit cell of graphite that we depict in 
Fig.~\ref{fig_bilayer_unitcell}.
Therefore, if the two planes are equivalent much of the symmetry analysis 
of graphite is also valid for the graphene bilayer.
Thus we could alternatively use the SWM
for graphite with the proper identification of the parameters.
The SWM model for graphite,\cite{McClure1957,SW58}
is usually written as
\begin{equation}
  \label{eq:cintro_SWMcC1}
  \Hca_{\text{SWMC}} = 
  \begin{pmatrix}
    E_1 & 0 & H_{13} & H_{13}^* \\
    0 & E_2 & H_{23} & -H_{23}^* \\
    H_{13}^* & H_{23}^* & E_3 & H_{33} \\
    H_{13} & -H_{23} & H_{33}^* & E_3
  \end{pmatrix},
\end{equation}
where
\begin{subequations}
\begin{eqnarray}
  \label{eq:cintro_SWMcC2}
  E_1 & = & \Delta  + \gamma_1 \Gamma + \frac{1}{2}\gamma_5 \Gamma^2, \\
  E_2 & = & \Delta  - \gamma_1 \Gamma + \frac{1}{2}\gamma_5 \Gamma^2, \\
  E_3 & = & \frac{1}{2}\gamma_2 \Gamma^2, \\
  H_{13} & = & \frac{1}{\sqrt{2}} (-\gamma_0 + \gamma_4 \Gamma)e^{i
  \alpha} \zeta, \\
  H_{23} & = & \frac{1}{\sqrt{2}} (\gamma_0 + \gamma_4 \Gamma)
  e^{i \alpha} \zeta, \\
  H_{33} & = & \gamma_3 \Gamma e^{i \alpha} \zeta.
\end{eqnarray}
\end{subequations}
Here $\zeta = 3 a k /2$, and $\Gamma = 2 \cos(\kp d)$, with 
$d \approx 3.7 \, \text{\AA}$ being the interplane distance.
Typical values of the parameters from the graphite literature
are shown in Table~\ref{table_parameters}.
\begin{table}[htbp]
  \centering
  \begin{tabular}{|l|l|l|l|l|l|l|l|}
    \hline
    $\gamma_0$ & $\gamma_1$ & $\gamma_2$ & $\gamma_3$ & 
    $\gamma_4$ & $\gamma_5$ & $\gamma_6 = \Delta$ & $\e_{\text{F}}$  \\
    \hline
    3.16 & 0.39 & -0.02 & 0.315 & 0.044 & 0.038 & 0.008 & -.024 \\
    \hline
    3.12 & 0.377 & -0.020 & 0.29 & 0.120 & 0.0125 & 0.004 & -.0206 \\
    \hline
  \end{tabular}
  \caption[Values of the SWM parameters
    used to parametrize the band structure of graphite]
  {Values of the SWM parameters
    for the band structure of graphite.
    Upper row from Ref.~[\onlinecite{graphitereview1}] and lower row 
    from Ref.~[\onlinecite{Chung_review2002}].}
  \label{table_parameters}
\end{table}

It is straightforward to show that by identifying $\gamma_1 = \tp$ and taking
$\gamma_2 = \gamma_5 = 0$, $\Gamma = 1$ and $V=0$ the matrices in
Eq.~(\ref{eq:cintro_HbilayerFull}) and
Eq.~(\ref{eq:cintro_SWMcC1}) are equivalent
up to a unitary transformation. Hence they give rise to identical
eigenvalues and band structures. This completes the correspondence
between the tight-binding model and the SWM model
(see also Refs.~[\onlinecite{Johnson_Dresselhaus73,Partoens2006}] 
for a discussion on the
connection between the tight-binding parameters and those of SWM.)

\begin{figure}[htb]
\includegraphics[scale=0.4]{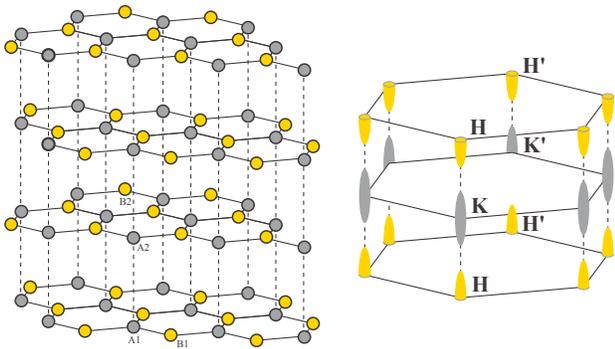}
\centering
\caption
  {[color online] Left: graphite lattice; Right: three-dimensional BZ with the
  symmetry points K and H indicated. The accepted parameters for
  graphite results in
  electron pockets near the K points and hole pockets
  near the H points as sketched in the figure.
}
\label{fig:graphite_lattice_BZ}
\end{figure}

The accepted parameters from the graphite literature
results in electrons near 
the K point [$\kp = 0$] and holes near 
the H point [$\kp = \pi/(2d)$] in the BZ
as sketched in Fig.~\ref{fig:graphite_lattice_BZ}.
These electron and hole pockets are mainly generated by
the coupling $\gamma_2$ that in the tight-binding model corresponds
to a hopping between the B-atoms of next-nearest planes. Note that
this process involves a hopping of a distance as large as 
$\sim 7 \, \text{\AA}$. 

Finally, it is interesting to note that at the H-point in the
BZ, $\Gamma = 0$, 
and therefore the two planes ``decouple'' at
this point. Furthermore, if one neglects $\Delta$ the spectrum is that
of massless Dirac fermions just like in the case of graphene.
Note that in graphite $\rA$ and $\rB$ atoms are different however, 
and that the term parametrized by $\Delta$, that breaks sublattice
symmetry in each plane, 
opens a gap in  the spectrum leading to massive Dirac
fermions at the H-point. Since the value of $\Delta$ in the
literature is quite small, the almost linear massless behavior should
be observed by experimental probes that are not sensitive to this
small energy scale.

The values of the parameters used in the graphite literature are
consistent with a large 
number of experiments. The most accurate ones are various
magneto-transport measurements discussed in Ref.~[\onlinecite{graphitereview1}]. 
More recently, angle-resolved photo-emission spectroscopy (ARPES)
 was used to directly visualize the dispersion of
massless Dirac quasi-particles near the H point and massive
quasi-particles near the K point in the 
BZ.\cite{Lanzara2006a,Lanzara2006b,Ohta06}

The band structure of graphite has been calculated and recalculated
many times over the years, a recent reference is 
Ref.~[\onlinecite{Charlier1991}].
It is also worth to mention that because of the
(relatively) large contribution of the non-local van der
Waals interaction to the interaction between the layers in graphite,
the usual local density approximation or
semilocal density approximation schemes are off by an order
of magnitude when the binding energy of the planes are calculated and
compared with experiments. For a discussion of this topic and
a possible remedy, see Ref.~[\onlinecite{Rydberg2003}].

\section{Simplified electronic band models}
\label{sec:simplebands}
In this section we introduce
three simplified models that we employ for
most of the calculations in this paper.
We also show how to obtain an effective two-band
model that is sometimes useful.
\subsection{Unbiased bilayer}
For the unbiased bilayer, a minimal model includes only the
nearest neighbor hopping energies within the planes and
the interplane hopping term between $\rA$ 
atoms; this leads to a Hamiltonian matrix of the form:
\begin{equation}
  \label{eq:nobias_Hkin0bilayer}
  \Hca_{B}(\vk) =
  \begin{pmatrix}
    0 & k e^{i \phi(\vk)} & \tp & 0 \\
    k e^{-i \phi(\vk)} & 0 & 0 & 0 \\
    \tp  & 0 & 0 & k e^{-i \phi(\vk)} \\
    0 & 0 & k e^{i \phi(\vk)} & 0
  \end{pmatrix},
\end{equation}
near the K point in the BZ.
Here we write $k_x + i k_y = k e^{i \phi(\vk)}$, where
$k=\sqrt{k_x^2 + k_y^2}$ and $\phi(\vk)$ is the appropriate
angle. Note that the absolute value of the angle can be
changed by a gauge transformation into a phase of the
wave functions on the B sublattices. This reflects the rotational
symmetry of the model. If one includes the ``trigonal
distortion'' term parametrized by
$\gamma_3$ the rotational symmetry is broken
and it is necessary to keep track of the absolute value of the
angle.
From now on in this paper, we most often 
use units such that $\vf = 1$ for simplicity.

This Hamiltonian has the advantage that it allows for relatively
simple calculations. Some of the fine details of the physics might not
be accurate but it works as a minimal model and capture most of
the important physics.
It is important to know the qualitative nature of the terms that are
neglected in this approximation, this is discussed later in this
section.
It is also an interesting toy model as it allows for (approximately)
``chiral'' particles with mass (i.e., a parabolic spectrum) at low
energies.\cite{Falko2006a}

\subsection{Biased graphene bilayer}
\begin{figure}[htb]
\includegraphics[scale=0.42]{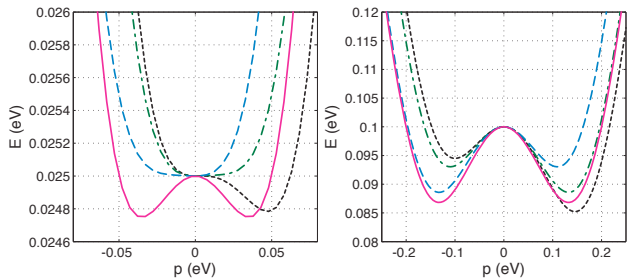}
\centering
\caption
  {[color online] Band dispersions near the band edge (note the energy scale)
   in the biased graphene bilayer.
  Left: $V = 50 \, \text{meV}$, Right: $V = 200 \, \text{meV}$.
  The solid line is the simplified model in Eq.~(\ref{eq:bias_Hkin0bilayer})
  that is cylindrically symmetric.
  The other lines are along different directions in the BZ for the full model
  in Eq.~(\ref{eq:cintro_HbilayerFull}):
  $\alpha = 0$ (dotted),  $\alpha = \pi/9$ (dash-dotted), and  $\alpha = 2 \pi/9$ (dashed).
  The parameters are the same as in Fig.~\ref{fig_bilayer_bands1} }
\label{fig_bilayer_bands2}
\end{figure}

\begin{figure}[htb]
\includegraphics[scale=0.42]{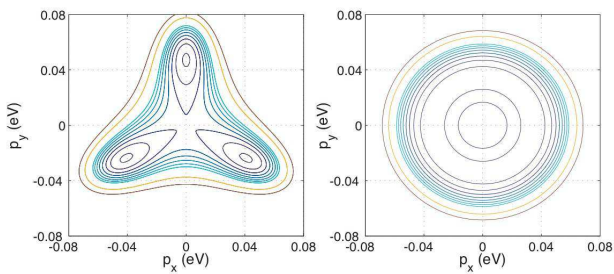}
\centering
\caption
  {[color online] Contour plots of the band dispersions near the band edge in the biased
  graphene bilayer for $V = 50 \, \text{meV}$.
  Left: full model in Eq.~(\ref{eq:cintro_HbilayerFull}), 
  Right: simplified model in Eq.~(\ref{eq:bias_Hkin0bilayer}).
  The parameters are the same as in Fig.~\ref{fig_bilayer_bands1} }
\label{fig_bilayer_bands_contour}
\end{figure}

For the biased bilayer, a minimal model employs 
Eq.~(\ref{eq:nobias_Hkin0bilayer}) augmented with
the bias potential $V$:
\begin{equation}
  \label{eq:bias_Hkin0bilayer}
  \Hca_{BB}(\vk) =
  \begin{pmatrix}
    V/2 & k e^{i \phi(\vk)} & \tp & 0 \\
    k e^{-i \phi(\vk)} & V/2 & 0 & 0 \\
    \tp  & 0 & -V/2 & k e^{-i \phi(\vk)} \\
    0 & 0 & k e^{i \phi(\vk)} & -V/2
  \end{pmatrix}.
\end{equation}
This model was introduced in Refs.~
[\onlinecite{Paco06a,McCann2006a}].
It correctly captures the formation of an electronic gap of size
$\sim V$ at the K point and the overall features of the bands
as can be seen in Fig.~\ref{fig_bilayer_bands1}.
Nevertheless, the fine details of the bands close to the band edge
are not captured correctly in this simple model; this fact is illustrated in
Figs.~\ref{fig_bilayer_bands2}-\ref{fig_bilayer_bands_contour}.
In particular the simple model is cylindrically symmetric; whereas the
"trigonal distortion" breaks this symmetry. Thus the inclusion of $v_3$ 
results in a "trihorn" structure for small values of $V$ and a weaker modulation of the band
edge for larger values of $V$ as illustrated in Fig.~\ref{fig_bilayer_bands2}.

\subsection{Multilayer graphene}
\label{sec:multilayer_mod}
In the graphene multilayer, a minimal model
for the bands is again given by Eq.~(\ref{eq:nobias_Hkin0bilayer})
with the understanding that the
momentum label also includes the perpendicular direction:
$\vk \rightarrow (\vkpa,\kp)$.
The only change is that we must make the substitution 
$\tp \rightarrow 2 \tp \cos(\kp d)$ everywhere.
Note that this is exactly the $\Gamma$-factor appearing in the SWM model
discussed in Section~\ref{sec:SWM}.
In the following we often use units such that the interplane distance
$d$ is set to $1$,
then -- since the unit cell holds two layers -- the allowed values of
$\kp$ lies in the interval $[-\pi/2,\pi/2]$.
We note that this band model was used already in the seminal
paper by Wallace as a simple model for graphite.\cite{Wallace47}
More recent works on the band structure of few-layer graphene systems
include Refs.~\onlinecite{Latil2006,Partoens2006,Paco06a,Koshino_Ando_2007a,Nakamura_08}.

\subsection{Approximate effective two-band models}
\label{sec:effective_models}
There are two main reasons for constructing approximate two-band
models. 
Firstly, on physical grounds the high-energy bands (far away
from the Dirac point) should not be very important for the low-energy
properties of the system.
Secondly, it is sometimes easier to work with $2\times2$ instead
of $4\times4$ matrices.
Nonetheless, it is not always a simplification to use the 2-band model
when one is studying inhomogeneous systems
as it generally leads to two coupled second order differential equations
whereas the 4-band model involves four coupled linear differential
equations. The matching of the wave functions in the 2-band case
then involves both the continuity of the wave function and its derivative
whereas in the 4-band model only continuity of the wave function
is necessary. We note that the two-band description of the problem
is only valid as long the electronic density is low enough, that is,
when the Fermi energy is much smaller than $t_{\perp}$. At intermediate
to high densities a 4-band model is required in order to obtain the
correct physical properties.\cite{KNCN07}

In this section, we derive the low-energy effective model by doing
degenerate second order perturbation theory. The quality of the
expansion is good as long as $\vf p \ll \tp \approx 0.35 \, \text{eV}$.
We first present the general expression for the second-order 
$2 \times 2$ effective Hamiltonian, thereafter various simplified
forms are introduced. 
Analyses similar to the one here were previously described
in Refs.~[\onlinecite{Falko2006a,Nilsson2006a}].

\begin{figure}[htb]
\includegraphics[scale=0.42]{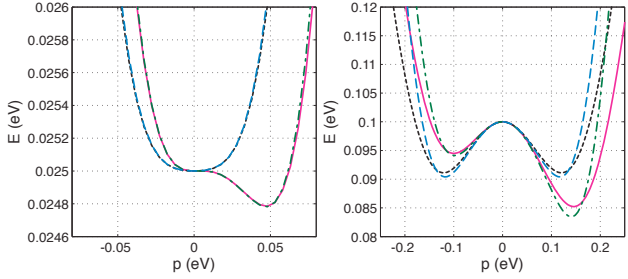}
\centering
\caption
  {[color online] Band dispersions near the band edge in the biased graphene bilayer along
  two directions in the BZ.
  Left: $V = 50 \, \text{meV}$, Right: $V = 200 \, \text{meV}$.
  The solid (dash-dotted) line is the effective model in 
  Eq.~(\ref{eq:BGBbands_ElowBB}) along $\alpha = 0$ ($\alpha = \pi/6$).
  The dotted (dashed) line is the full model in 
  Eq.~(\ref{eq:cintro_HbilayerFull}) along $\alpha = 0$ ($\alpha = \pi/6$).
  The parameters are the same as in Fig.~\ref{fig_bilayer_bands1}. 
  For $V = 50 \, \text{meV}$ the different curves are almost not discernible.
}
\label{fig_bilayer_bands3}
\end{figure}

First we introduce the projector matrices
 $\Pca_0 = \text{Diag}[0, \, 1, \, 0, \, 1]$
($\Pca_1 = \text{Diag}[1, \, 0, \, 1, \, 0]$) that projects
onto (out of) the low-energy subspace of the B atoms.
Then we split the Hamiltonian in Eq.~(\ref{eq:cintro_HbilayerFull})
according to:
$\Hca_0 = \Kca_0 + \Kca_1 + \Kca_2$,
with
\begin{equation}
  \label{eq:BGBbands_Kca0BB}
  \Kca_0 =
    \begin{pmatrix}
    \Delta+V/2 & 0 & \tp & 0 \\
    0 & V/2 & 0 & 0 \\
    \tp & 0 & \Delta-V/2 & 0 \\
    0 & 0 & 0 & -V/2
  \end{pmatrix},
\end{equation}
\begin{equation}
  \label{eq:BGBbands_Kca1BB}
  \Kca_1 = \vf
    \begin{pmatrix}
    0 & p e^{i \phi} & 0 & -v_4  p e^{-i \phi} \\
    p e^{-i \phi} & 0 & -v_4  p e^{-i \phi} & 0 \\
    0 & -v_4 p e^{i \phi} & 0 & p e^{-i \phi} \\
    -v_4  p e^{i \phi} & 0 & p e^{i \phi} & 0
  \end{pmatrix},
\end{equation}
\begin{equation}
  \label{eq:BGBbands_Kca2BB}
  \Kca_2 = \vf
    \begin{pmatrix}
    0 & 0 & 0 & 0 \\
    0 & 0 & 0 & v_3  p e^{i \phi} \\
    0 & 0 & 0 & 0 \\
    0 & v_3  p e^{-i \phi} & 0 & 0
  \end{pmatrix}.
\end{equation}
Introducing the vectors $\bigl\{ |l\ra \bigr\}$ that to zeroth order only
have components in the low energy subspace 
(i.e., $\Pca_1 |l^{(0)} \ra = 0$) and
following the standard procedure (see e.g., Ref.~[\onlinecite{Sakurai_QM}]) for
degenerate perturbation theory we arrive at:
\begin{multline}
  \label{eq:BGBbands_Secondorder_BB}
  \la l | \Pca_0 \, \Hca_0 \, \Pca_0 | l' \ra \approx 
  \la l | \Pca_0 \bigl[ \Kca_0 + \lambda^2 \Kca_2 \bigr] \Pca_0 | l' \ra
\\
 + \lambda^2 \la l | \Pca_0 \Kca_1 \Pca_1 
                \frac{1}{\hat{E}-\Kca_0} 
                     \Pca_1 \Kca_1  \Pca_0 | l' \ra,
\end{multline}
where we explicitly assume that $\Kca_2$ is of the same order as
$\Kca_1^2 / \Kca_0$. This expression is correct to second order in
$\lambda$.
Note that we are doing second order perturbation theory for all of the
components of the $2 \times 2$-matrix in the low-energy subspace.
Working to first order in $\lambda^2$ (and then setting $\lambda =1$)
one obtains for this matrix (taking $\vf =1$ for brevity):
\begin{multline}
  \label{eq:BGBbands_KcaLowBB}
  \Kca_{\text{low}} =
  \begin{pmatrix}
    V/2 & v_3  p e^{i \phi} \\
    v_3  p e^{-i \phi} & -V/2
  \end{pmatrix}
\\
 + 
    \begin{pmatrix}
    \frac{-V+2 \tp v_4 + \Delta (1+v_4^2)}{\tp^2 - \Delta (\Delta-V)} &
    - \frac{ \tp(1+v_4^2) +2 v_4 \Delta }
           {\tp^2 + V^2/4 - \Delta^2}  e^{- 2 i \phi}
  \\
    - \frac{ \tp(1+v_4^2) +2 v_4 \Delta }
           {\tp^2 + V^2/4 - \Delta^2} e^{2 i \phi} &
    \frac{V+2 \tp v_4 + \Delta (1+v_4^2)}{\tp^2 - \Delta (\Delta+V)}
  \end{pmatrix} p^2 .
\end{multline}
Taking $\Delta = 0$ leads to an even simpler expression,
in particular the effective spectrum becomes:
\begin{widetext}
  \begin{equation}
    \label{eq:BGBbands_ElowBB}
    E_{\text{low},\pm} \approx
    \frac{2 v_4  p^2}{\tp} 
    \pm 
    \sqrt{
      \Bigl[1 - \frac{2 p^2}{\tp^2} \Bigr]^2 
      \frac{V^2}{4}
     +
      ( v_3 p)^2
     + 
      \Bigl\{
      \frac{p^2 \bigl[ \tp(1+v_4^2)\bigr]}
      {\tp^2 +V^2/4 }  \Bigr\}^2
    - \frac{2 v_3 p^3 \bigl[ \tp(1+v_4^2)\bigr]}
      {\tp^2 +V^2/4 }  \cos(3 \phi)
    }.
  \end{equation}
\end{widetext}
That this approximation to the bands is excellent near the band edge 
for small values of the bias $V$ is illustrated in Fig.~\ref{fig_bilayer_bands3}.
For larger values of the bias the agreement is less accurate because the
assumption of smallness of certain terms in the perturbation expansion
is no longer valid.

\section{Green's function in the graphene bilayer}
\label{sec:H_G_bilayer}
As discussed in Section~\ref{sec:simplebands} we use
the minimal model Hamiltonian in Eq.~(\ref{eq:nobias_Hkin0bilayer}).
We note that
the phases $\phi \equiv \phi(\vk)$ can be gauged away by an
application of a unitary transformation defined by the matrix
\begin{equation}
  \label{eq:unitary_M1}
  \Mca_1(\vk) =
  \begin{pmatrix}
    1 & 0 & 0 & 0 \\
    0 & e^{-i \phi(\vk)} & 0 & 0\\
    0 & 0 & 1 & 0 \\
    0 & 0 & 0 & e^{i \phi(\vk)}
  \end{pmatrix}.
\end{equation}
It is also easy to compute the energy eigenvalues that are given by
$\tp/2 \pm \sqrt{\tp^2/4 + k^2}$ and $-\tp/2 \pm \sqrt{\tp^2/4 + k^2}$.
Before we solve for the Green's functions it is convenient to allow
for a local frequency dependent self-energy in the problem. 
In the general case the self-energies on all of the inequivalent sites
in the problem are allowed to be different, 
and we explicitly introduce the matrix
\begin{equation}
  \label{eq:BGBi_self_energy_matrix}
  \Hca_{\Sigma}(\om) =
  \begin{pmatrix}
    \Sigma_{\rA 1} (\om) & 0 & 0 & 0 \\
    0 & \Sigma_{\rB 1}     (\om) & 0 & 0\\
    0 & 0 & \Sigma_{\rA 2} (\om) & 0 \\
    0 & 0 & 0 & \Sigma_{\rB 2}     (\om) 
  \end{pmatrix},
\end{equation}
to describe this. 
The Green's function matrix is then given by the equation
\begin{equation}
  \label{eq:dis_nobias_green1}
  \GR^{-1}(\om,\vk) = \om - \Hca_0 (\vk) - \Hca_{\Sigma}(\om).
\end{equation}

In the case of the unbiased bilayer the A (B) sites in both
of the layers are equivalent and we only need two
self-energies: $\SAo$ and $\SBo$ which are local but
we allow for a frequency ($\om$) dependence.
In this case 
the matrix inversion is simple since it factorizes into two 
$2\times 2$ matrices. An explicit form is given by
\begin{equation}
  \label{eq:dis_nobias_green2}
  \GR(\om,\vk)  = \Mca_1(\vk)
  \begin{pmatrix}
    \grD(\om , k) & \grND(\om , k) \\
    \grND(\om, k) & \grD(\om,k )
  \end{pmatrix}
  \Mca_1^{\dag}(\vk),
\end{equation}
where $k=|\vk|$.
Here $\text{D}$ ($\text{ND}$) stands for diagonal
(non-diagonal) in the layer index.
The components of the g-matrices are given by
\begin{subequations}
\label{eq:dis_nobias_green3}
\begin{eqnarray}
  {\rm g_{AA}^{D,ND}} &=&  
  \frac{\om - \SB}{2 D_{-}} 
  \pm \frac{\om - \SB}{2 D_{+}} ,
\\
  {\rm g_{BB}^{D,ND}} &=& 
  \frac{\om - \tp - \SA}{2 D_{-}} 
  \pm
  \frac{\om + \tp - \SA}{2 D_{+}} ,
\\
  {\rm g_{AB}^{D,ND}} &=&
  \frac{k}{2 D_{-}} \pm \frac{k}{2 D_{+}} ,
\end{eqnarray}
\end{subequations}
where
\begin{equation}
  \label{eq:dis_nobias_green4}
    D_{\pm}(\om,p) 
    = \bigl[\om \pm \tp - \SAo \bigl] \bigl[\om -\SBo \bigr] - k^2.
\end{equation}
Note that we often suppress the momentum and frequency
dependence in the following when no confusion arises.
We will come back to the biased case in 
Section~\ref{sec:CPA_biased}.

\section{Impurity scattering: t-matrix and
  coherent potential approximation}
\label{sec:impurities_tmatrix}
We are interested in the influence of disorder in the bilayer.
To model the impurities we use the standard t-matrix approach and the
Coherent Potential Approximation (CPA).
The effect of repeated scattering from a single impurity can be
encoded in a self-energy which can be computed 
from:\cite{JonesMarch2}
\begin{multline}
  \label{eq:dis_nobias_sigma1}
  \Sigma_j = V_j + V_j N \overline{\GR} V_j 
  + V_j N \overline{\GR} V_j N \overline{\GR} V_j + \ldots 
\\
= V_j \bigl[1 - N \overline{\GR} V_j]^{-1}.
\end{multline}
Here $V_j$ is a matrix that encodes both the strength and the lattice
site of the impurity in question.
For example, an impurity on an A1 lattice site of strength $U$ at the 
origin is encoded in Fourier space by the matrix
 \begin{equation}
  \label{eq:dis_nobias_potential1}
  V_{1} = \frac{U}{N}
  \begin{pmatrix}
    1 & 0 & 0 & 0 \\
    0 & 0 & 0 & 0 \\
    0 & 0 & 0 & 0 \\
    0 & 0 & 0 & 0
  \end{pmatrix},
\end{equation}
implying that the potential is located only on a single site.
We have also introduced the quantity:
\begin{multline}
  \label{eq:BGBi_Gbar1}
  \overline{\GR}_{\alpha j }(\om) 
= \frac{1}{N}\sum_{\vk} \GR_{\alpha j  \alpha j}(\om,\vk)
\\
\approx 
  \frac{1}{\Lambda^2}\int_{0}^{\Lambda^2} d(k^2) \int \frac{d \phi}{2 \pi}
   \, \GR_{\alpha j  \alpha j}(\om,\vk),
\end{multline}
which is the local propagator at the impurity site; and in the second
step the $\vk$-sum is to be taken over the whole BZ.
The last line is an approximate expression that
is obtained by expanding the
propagator close to the K points and taking the continuum
limit with the introducing of the cutoff $\Lambda$.
We estimate the cutoff by a Debye
approximation that conserves the number of states in the BZ. Then 
$\Lambda \approx 7 \text{eV}$ and in units of the cutoff we have
$\tp \approx 0.05$ (taking $\tp \approx 0.35 \, \text{eV}$).
Due to the special form of the propagator and the impurity potential
the self-energy we get from this is diagonal.
The result for site dilution (or vacancies) is obtained
by taking the limit $U \rightarrow \infty$, so that the electrons are
not allowed to enter the site in question. 
We also introduce a finite density $\nimp$ of impurities in the
system. To leading order in the impurity concentration the equations
for the self-energies then become:
\begin{subequations}
\label{eq:dis_nobias_Gbars}
\begin{eqnarray}
  \label{eq:dis_nobias_GbarAA1}
  \frac{n_i}{\SA}
  &=& - \overline{\GR}_{\rA},
\\
  \label{eq:dis_nobias_GbarBB1}
  \frac{n_i}{\SB}
  &=& - \overline{\GR}_{\rB}.
\end{eqnarray}
\end{subequations}
The explicit form of the propagators in 
Eq.~(\ref{eq:dis_nobias_green3}) makes it
easy to compute the $\overline{\text{G}}$'s.
The t-matrix result for the self-energies is obtained by 
using the bare propagators on the right hand side of
Eq.~(\ref{eq:dis_nobias_Gbars}). 
In the CPA one assumes that the electrons are moving in an effective
medium with recovered translational invariance
which in this case is characterized by the local
self-energies. To determine what the medium 
is, one must solve the equations self-consistently with
the full propagators on the right hand side of
Eq.~(\ref{eq:dis_nobias_Gbars}).
Because of the simple form of the propagators this is a simple 
numerical task in the model we are using.
%
To simplify the equations further we assume that $\Lambda \gg
\om,\tp,\SA,\SB$. This is a physical assumption since when the
self-energies becomes of the order of the cutoff the effective 
theory breaks down. The self-consistent equations then reads:
\begin{subequations}
\label{eq:dis_nobias_CPAbilayer}
\begin{multline}
  \label{eq:dis_nobias_CPAbilayer_A}
  \bigl[ \frac{\SA}{n_i} \bigr]^{-1}
  = - \overline{\GR}_{\rA} =
  \\ 
  \frac{\om - \SB}{2 \Lambda^2}
  \sum_{\alpha =\pm}
  \log \Bigl[
  \frac{\Lambda^2}
  {- (\om + \alpha \tp -\SA) (\om -\SB)}
  \Bigr],
\end{multline}
\begin{multline}
  \label{eq:dis_nobias_CPAbilayer_B}
  \bigl[ \frac{\SB}{n_i} \bigr]^{-1}
  = - \overline{\GR}_{\rB}
  \\
  = 
  \frac{\om - \SA}{2 \Lambda^2}
  \sum_{\alpha =\pm}
  \log \Bigl[
  \frac{\Lambda^2 }
  {- (\om -\alpha \tp -\SA) (\om -\SB)}
  \Bigr]
  \\
  +
  \frac{\tp}{2 \Lambda^2}
  \log \Bigl[
  \frac{- (\om -\tp -\SA) (\om -\SB)}
  {- (\om +\tp -\SA) (\om -\SB)}
  \Bigr].
\end{multline}
\end{subequations}
This includes inter-valley scattering in the intermediate states.
It is easy to obtain the non-disordered density of states from 
these equations by taking $\SA = \SB =0$ and 
$\om \rightarrow \om +i\delta$ 
(here $\delta$ is a positive infinitesimal)
resulting in:
\begin{subequations}
\label{eq:dis_nobias_DOS0bilayer}
\begin{equation}
  \label{eq:dis_nobias_DOS0bilayerA1}
  \rho_{\rA}^{0}(\om) 
  = -\frac{1}{\pi} \imag \overline{\GR}_{\text{A}}
  = \frac{|\om|}{2 \Lambda^2}
  \bigl[1+\Theta(|\om|- \tp ) \bigr],
\end{equation}
\begin{multline}
  \label{eq:dis_nobias_DOS0bilayerB1}
  \rho_{\rB}^{0}(\om) 
  = -\frac{1}{\pi} \imag \overline{\GR}_{\text{B}}
\\
  = \frac{|\om|}{2 \Lambda^2}
  \bigl[1+\Theta(|\om|- \tp ) \bigr]
  +
  \frac{\tp}{2 \Lambda^2} 
  \bigl[1-\Theta(|\om|- \tp) \bigr].
\end{multline}
\end{subequations}
Observe in particular that the density of states on the A sublattice
goes to zero in the limit of zero frequency, this fact is responsible
for much of the unconventional physics in the graphene bilayer.
In contrast the density
of states on the B sublattice is finite at $\om = 0$.
We discuss how this result is changed with disorder and the
solution of Eq.~(\ref{eq:dis_nobias_CPAbilayer}) in the following.

\subsection{Zero frequency limit}
\label{sec:zerofreqlimit}
One interesting feature of the CPA equations in 
Eq.~(\ref{eq:dis_nobias_CPAbilayer}) is that it is easy to see that 
they do not allow for a finite $\SA$ in the limit of
$\om \rightarrow 0$. Since by setting $\om = 0$ the last term in
Eq.~(\ref{eq:dis_nobias_CPAbilayer_B}) must vanish,
and this is not possible for finite values of $\SA$. 
Then one also must have that $\SB \rightarrow 0$ there. 
This implies that the density of states on
sublattice A is still zero even within the CPA in the limit $\om
\rightarrow 0$. 
More explicitly, by defining $\SA \SB = - \xi \Lambda^2$ one can show (assuming
$\SA \gg \tp$ and $\SB \gg \om$) that $\SA$ and $\SB$ are given 
asymptotically by:
\begin{subequations}
  \label{eq:dis_nobias_S_smallOm}
\begin{eqnarray}
  \SA &=& \Bigl( \frac{\tp^2 \xi^2 \Lambda^2}{n_i \om}\Bigr)^{1/3} 
  e^{-i \pi/3}, \\
  \SB &=& \Bigl( \frac{n_i \Lambda^4 \xi \om}{\tp^2} \Bigr)^{1/3}
  e^{-i 2 \pi/3},
\end{eqnarray}
\end{subequations}
and $\xi$ satisfies
\begin{equation}
  \label{eq:dis_nobias_14}
  n_i = \xi \log(1 / \xi).
\end{equation}
Notice that $\sqrt{\xi} \Lambda \sim \sqrt{n_i} \Lambda$ is exactly
the energy scale that is
generated by disorder of the same kind in the single layer case.\cite{nuno2006_long}
We have checked that
the expressions in Eq.~(\ref{eq:dis_nobias_S_smallOm}) seem to agree with the
numerical calculations in the small frequency limit, and
the frequency range in which they holds grows with increasing $n_i$.

\subsection{Uncompensated impurity densities}
\label{sec:uncompensated}
The divergence of the self-energy $\SA$ on the A sublattice in the
CPA in the above is due to the fact that there
is a perfect compensation between the number of impurities on
the two sublattices: $n_{{\rm i},\rA} = n_{{\rm i},\rB}$.
For the more general case where $n_{{\rm i},\rA} \neq n_{{\rm i},\rB}$ the
divergence is not present so that $\SA$ may become finite at $\om=0$.
To make comparison with other work on the graphene bilayer it is
fruitful to consider another extreme limit where only the B sites
are affected by the disorder.
Explicitly this means that we take $n_{{\rm i},\rA} = 0$
and $n_{{\rm i},\rB} \neq 0$.
The generalization of the CPA equations in 
Eq.~(\ref{eq:dis_nobias_CPAbilayer}) for this case then immediately imply that
$\SA(\om) \equiv 0$. In the limit of $\om \rightarrow 0$, $\SB$ is finite, 
purely imaginary, and given by:
\begin{equation}
  \label{eq:dis_nobias_SBonlyB}
  \SB(\om = 0) = -i \frac{2 \Lambda^2}{\pi \tp} n_{{\rm i},\rB} 
  \equiv -i \Gamma.
\end{equation}

\subsection{Born scattering}
\label{sec:born_self_energy}

Another often studied limit is the one of weak impurities, in particular
Koshino and Ando have studied electron transport in the graphene bilayer
in this approximation.\cite{Ando06a} This is the
Born limit and it can be studied using perturbation theory in
the strength of the impurities $U$. The leading non-trivial
contribution to the self-energies is given by the contribution
to second order:
\begin{subequations}
  \label{eq:dis_nobias_Born_eq1}
\begin{eqnarray}
  \SA &=& \nimp U \overline{\GR}_{\text{A}} U , \\
  \SB &=& \nimp U \overline{\GR}_{\text{B}} U.
\end{eqnarray}
\end{subequations}
If one substitutes the bare propagators on the right
hand side one finds $\SA = 0$ and 
$\SB = -i \pi \nimp \tp (U/\Lambda)^2 /2$ at the Dirac point.
Thus, to leading
order Born scattering is formally equivalent to the
previous case with vacancies on only sublattice B
exactly at $\om = 0$. The frequency range for which the $\om=0$
result is valid is different however.

\subsection{Self-energy comparisons and the density of states}

\begin{figure}[htb]
\centering
 \includegraphics[scale=.40]{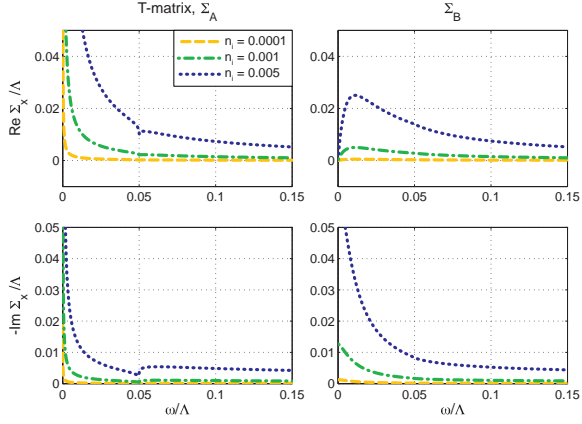}
 \caption
   {[color online] Self-energies within the t-matrix
   approximation in the bilayer as a function of
   the frequency $\omega$. Left: sublattice A; Right: sublattice B.}
 \label{fig_Sigma_Tmat_bilayer}
 \end{figure}

\begin{figure}[htb]
\centering
 \includegraphics[scale=.40]{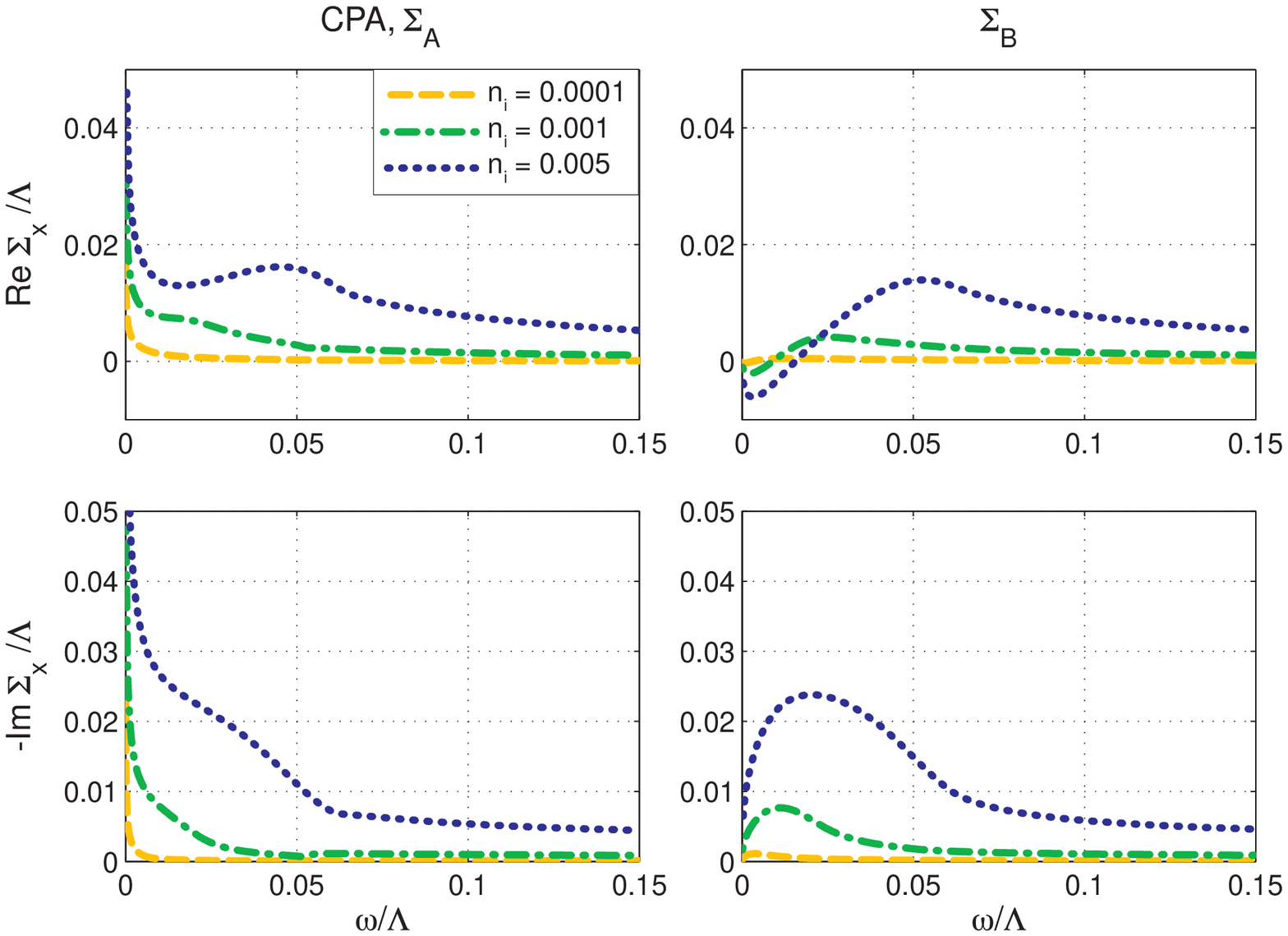}
 \caption
 {[color online] Self-energies within the CPA
   in the bilayer as a function of
   the frequency $\omega$. Left: sublattice A; Right: sublattice B.}
 \label{fig_Sigma_CPA_bilayer}
 \end{figure}

\begin{figure}[htb]
\centering
 \includegraphics[scale=.40]{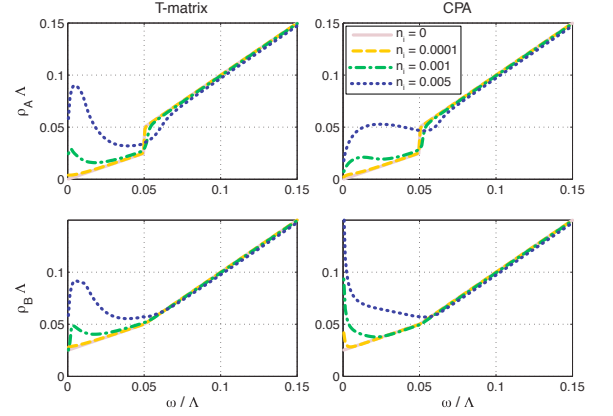}
 \caption
 {[color online] Local density of states $\rho$ on the different sublattices
   in the bilayer. Left: t-matrix; Right: CPA. Top: sublattice A; Bottom: sublattice B.}
 \label{fig_DOS_bilayer}
 \end{figure}

We compare the self-energies obtained from the t-matrix and
the CPA. Within the t-matrix the self-energies are given by
\begin{subequations}
\label{eq:dis_nobias_TmatSigma}
\begin{eqnarray}
  \label{eq:dis_nobias_TmatSigmaA}
  \SA 
  &=& \frac{n_i}{F_{\rA}^0 + i \pi \rho_{\rA}^{0}(\om) },
\\
  \label{eq:dis_nobias_TmatSigmaB}
  \SB
  &=& \frac{n_i}{F_{\rB}^0 + i \pi \rho_{\rB}^{0}(\om) },
\end{eqnarray}
\end{subequations}
where the $\rho^{0}$'s are given in
Eq.~(\ref{eq:dis_nobias_DOS0bilayer}) and
\begin{subequations}
\begin{eqnarray}
  \label{eq:dis_nobias_TmatFA}
  F_{\rA}^0 \equiv -\real \bigl[\, \overline{\GR}^0_{\text{AA}} \, \bigr]
  &=& \frac{\om}{2 \Lambda^2} \log 
  \Bigl( \frac{\Lambda^4}{\om^2 |\om^2 - \tp^2|} \Bigr),
\\
  \label{eq:dis_nobias_TmatFB}
  F_{\rB}^0 \equiv -\real \bigl[\, \overline{\GR}^0_{\text{BB}} \, \bigr]
  &=& F_{\rA}^0 
  + \frac{\tp}{2 \Lambda^2} \log
  \Bigl| \frac{\om - \tp}{\om + \tp} \Bigr|.
\end{eqnarray}
\end{subequations}
The results for the self-energies in the two different approximations
are shown in Figs.~\ref{fig_Sigma_Tmat_bilayer} 
and~\ref{fig_Sigma_CPA_bilayer}.
Note that at least on the scale of the figures the $\Sigma_A$ diverges
as $\om \rightarrow 0$ in the CPA, as discussed above. 
The solution to the self-consistent equations also does not converge
very well when they are pushed close to the limit of $\om \rightarrow 0$.
The total DOS on the A-sublattice and B-sublattice is pictured in
Fig.~\ref{fig_DOS_bilayer}. Note in particular that the case of 
$n_i =.0001$ closely resemble the non-interacting case except for the new
low-energy feature.
A possible interpretation of the enhancement of the DOS 
on the B sublattice
near $\om =0$ is in terms of the ``half-localized'' states 
(meaning they do not decay fast enough to be normalizable
at infinity)
that have been studied for monolayer graphene.\cite{Vitor2006}
Because these states have weight on only one sublattice
(the opposite one of the vacancy) the construction in
Ref.~[\onlinecite{Vitor2006}]
is valid also in the graphene bilayer when there is a
vacancy on one of the A sublattices.
For a discussion of the related problem of edge states in bilayer graphene
see Ref.~[\onlinecite{Eduardo_edge_2007}].

\subsection{Spectral function}

\begin{figure}[htb]
\centering
 \includegraphics[scale=.40]{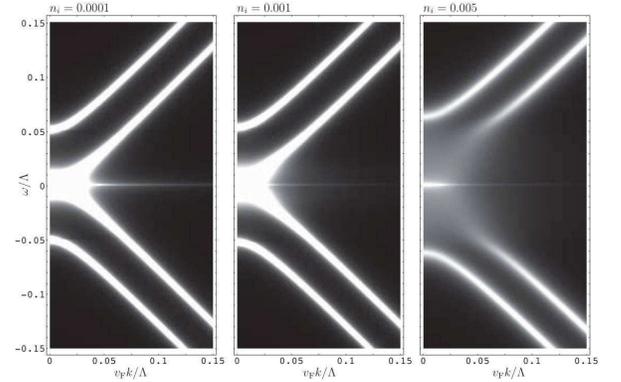}
 \caption[Intensity plot of the spectral
   function in the $k \times \omega$ plane (normalized by the cutoff) 
in the bilayer for different impurity concentrations in the
   CPA approximation.] 
{Intensity plot of the spectral
   function in the $k \times \omega$ plane (normalized by the cutoff) 
in the bilayer for different impurity concentrations in the
   CPA approximation. From left to right: $n_i = 10^{-4}$, $10^{-3}$, $5 \times 10^{-3}$.}
 \label{fig_Arpes_bilayer}
 \end{figure}

The electron spectral function $A(\vk,\om)$,
which is observable in ARPES experiments,
is defined by
$A(\vk,\om) \equiv -\text{Trace}[\imag \, \GR(\vk,\om) ] / \pi$,
so that in our case:
\begin{equation}
  \label{eq:dis_nobias_spectral_bilayer}
  A(\vk,\om) = -\frac{2}{\pi} 
  \Bigl\{ \imag \bigl[ \GDA(\vk,\om)  \bigl] 
  +  \imag \bigl[  \GDB(\vk,\om) \bigl] \Bigr\}.
\end{equation}
The spectral function in the $k \times \omega$ plane,
calculated within the CPA, 
is pictured in Fig.~\ref{fig_Arpes_bilayer}.
As can be seen in the figures the low-energy branch becomes
significantly blurred, especially for the higher impurity 
concentrations.
Note also that the gap to the high-energy branch
becomes slightly larger as the disorder value increases due to the
fact that $\SA$ is not negligible there.

\begin{figure}[htb]
\centering
 \includegraphics[scale=.40]{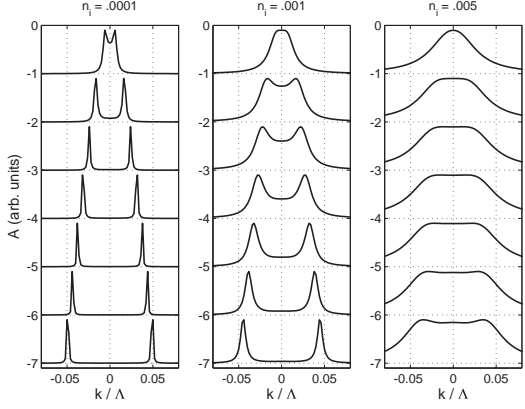}
 \caption
 {MDC's  in the graphene bilayer.
 The three panels are for different impurity concentrations and are calculated
 in the CPA.
 From the top the energy cuts are at the energies:
 .0005, 
 .0055, 
 .0105, 
 .0155, 
 .0205, 
 .0255, and
 .0305
 in units of  the cutoff $\Lambda$.
The curves are uniformly displaced for clarity.
 }
 \label{fig_MDC_bilayer}
 \end{figure}

\begin{figure}[htb]
\centering
 \includegraphics[scale=.40]{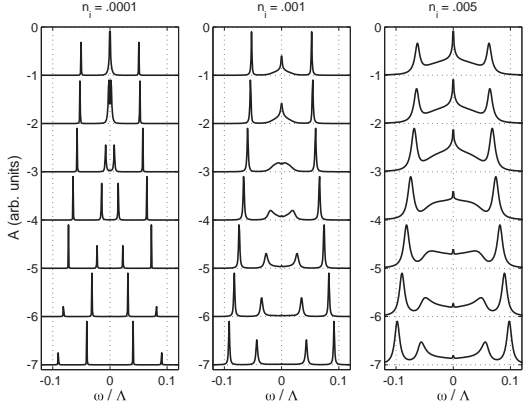}
 \caption
 {EDC's  in the graphene bilayer.
 The three panels are for different impurity concentrations and are calculated
 in the CPA.
 From the top the cuts are at fixed values of the in-plane momentum $k$:
 .0001, 
 .01, 
 .02, 
 .03, 
 .04, 
 .05, and
 .06
 in units of  the cutoff $\Lambda$.
The curves are uniformly displaced for clarity.
 }
 \label{fig_EDC_bilayer}
 \end{figure}

Examples of the momentum distribution curves (MDC's) 
and the energy distribution curves (EDC's)
in the disordered
graphene bilayer are shown in Figs.~\ref{fig_MDC_bilayer} and \ref{fig_EDC_bilayer}.
The evolution of the peaks from delta functions to broader peaks with
increasing disorder is clear in the figure.

\section{Green's function and one-particle properties in multilayer graphene}
\label{sec:G_multilayer}
We will use the extension of the bilayer model to the multilayer
that we introduced in Section~\ref{sec:multilayer_mod}.
As discussed there
we can immediately use the Hamiltonian in
Eq.~(\ref{eq:nobias_Hkin0bilayer}) with the understanding that the
momentum label also includes the perpendicular direction:
$\vk \rightarrow (\vkpa,\kp)$ and by substituting 
$\tp \rightarrow 2 \tp \cos(\kp d)$ everywhere.
In particular
the Green's function including the self-energies are again given by
the expressions in Eq.~(\ref{eq:dis_nobias_green3}) 
and Eq.~(\ref{eq:dis_nobias_green4})
with the substitution $\tp \rightarrow 2 \tp \cos(\kp d)$.

\subsection{Self-energies and the density of states}

\begin{figure}[htb]
\centering
 \includegraphics[scale=.40]{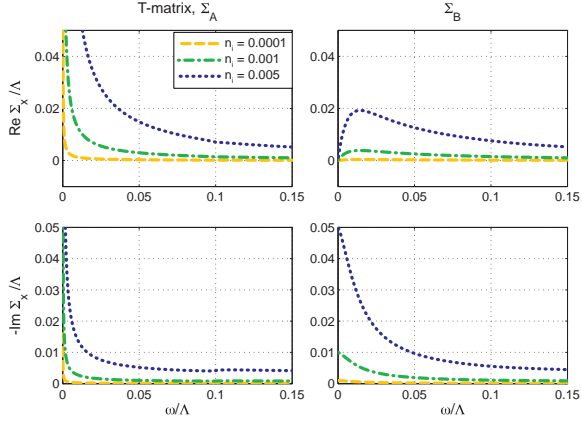}
 \caption
 {[color online] Self-energies within the t-matrix
   in the multilayer as a function of
   the frequency $\omega$. Left: sublattice A; Right: sublattice B.}
 \label{fig_Sigma_Tmat_graphite}
 \end{figure}

\begin{figure}[htb]
\centering
 \includegraphics[scale=.40]{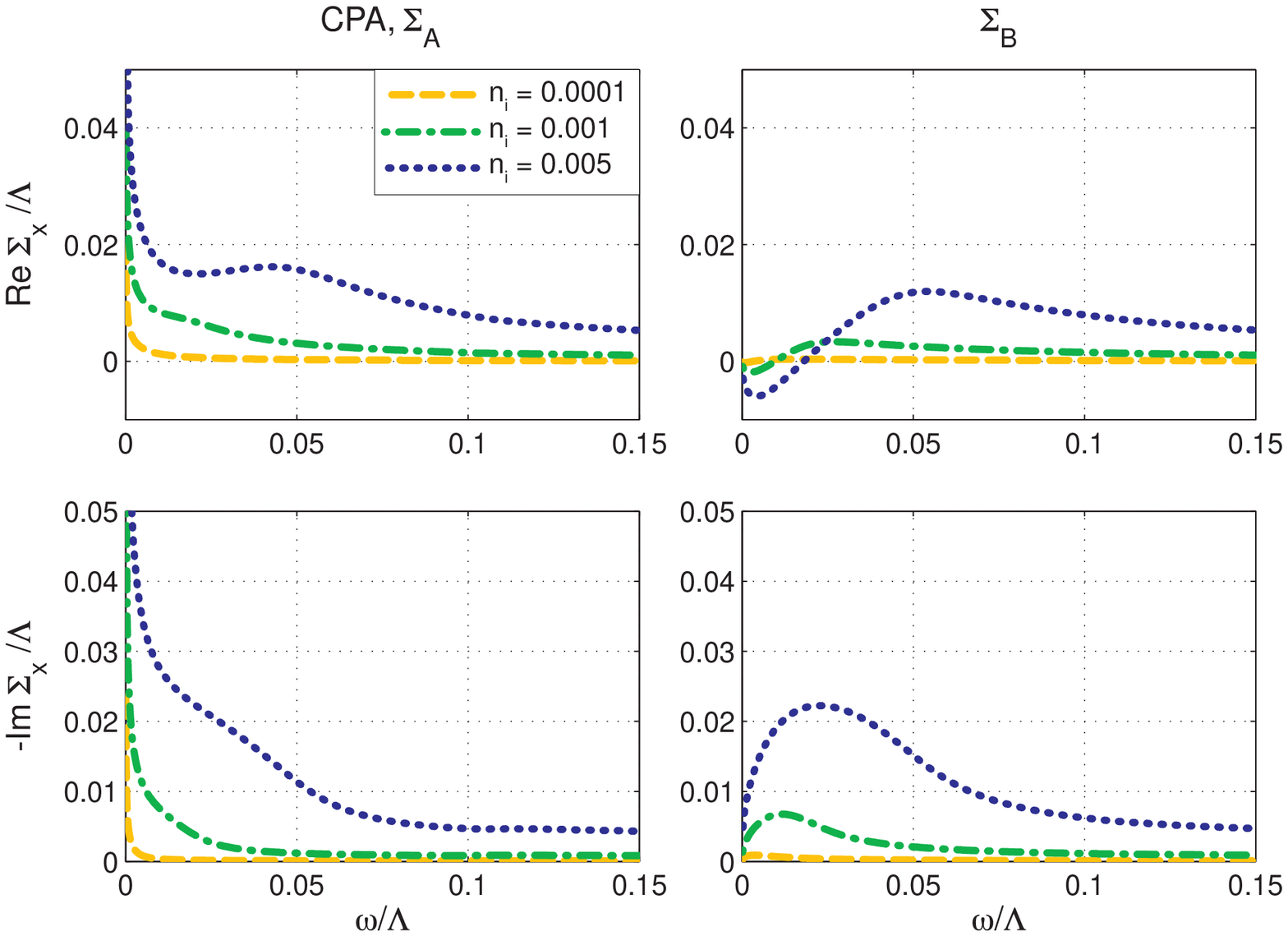}
 \caption
 {[color online] Self-energies within the CPA
   in the multilayer as a function of
   the frequency $\omega$.  Left: sublattice A; Right: sublattice B.}
 \label{fig_Sigma_CPA_graphite}
 \end{figure}

\begin{figure}[htb]
\centering 
\includegraphics[scale=0.40]{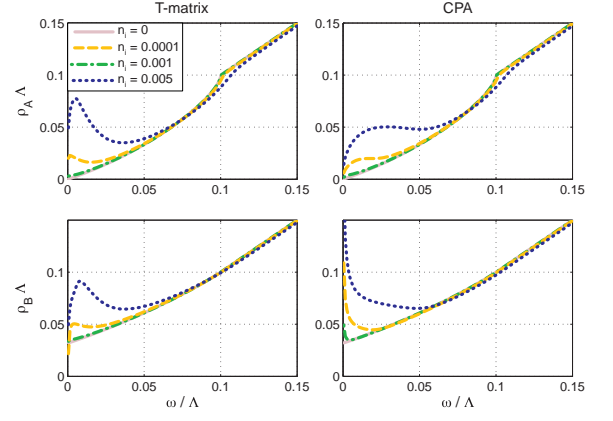}
 \caption
 {[color online] Local density of states $\rho$ on the different sublattices
   as a function of
   the frequency $\omega$ in the multilayer. Top: sublattice A; Bottom: sublattice B.}
 \label{fig_DOS_graphite}
 \end{figure}

To get the CPA equations we must evaluate the local propagator 
$\overline{\GR}$ that is given by the straightforward generalization of
Eq.~(\ref{eq:BGBi_Gbar1}) to include an extra sum over $\kp$:
\begin{equation}
  \label{eq:dis_nobias_Gbar1multi}
  \overline{\GR}(\om) = \frac{1}{N}\sum_{\vkpa} 
  \frac{1}{N_c}\sum_{\kp} \GR(\om).
\end{equation}
Here $N_c$ is the number of unit cells in the perpendicular direction.
This extra sum can be transformed into an integral using the relation 
$\frac{1}{N_c} \sum_{k_{\perp}} \rightarrow
\int_{-\pi/2}^{\pi/2} \frac{d \kp}{\pi}$.
It is possible to perform the integrals analytically as we explain
in App.~\ref{app:DOSgraphite}.
Using the integrals defined there ($I_1$ and $I_2$)
we obtain for  $\om, \SA, \SB, \tp \ll \Lambda$
\begin{subequations}
\label{eq:dis_nobias_CPAgraphite}
\begin{eqnarray}
  \label{eq:dis_nobias_CPAgraphite1}
  \frac{n_i}{\SA} &=& - \GDAa = 
  -\frac{\om - \SB}{\Lambda^2} I_1 ,
  \\
  \frac{n_i}{\SB} &=& - \GDBb = 
  -\frac{\om - \SA}{\Lambda^2} I_1
  -\frac{1}{\Lambda^2} I_2 .
\end{eqnarray}
\end{subequations}
From these equations one can easily obtain the non-interacting density
of states by considering the clean retarded case and
take $\Sigma_A = \Sigma_B = 0 $
and $\om \rightarrow \om +i \delta$, from which we get:
\begin{subequations}
\label{eq:dis_nobias_DOSgraphite0}
\begin{multline}
  \label{eq:dis_nobias_DOSgraphiteA0}
  \rho_{\rA}^{0}
=
    \frac{|\om|}{\pi \Lambda^2} \Bigl[\frac{\pi}{2}
    +\tan^{-1} (\frac{|\om|}{\sqrt{4\tp^2 - \om^2}}) \Bigr]
     \Theta (2 \tp - |\om|)
 \\
+
    \frac{|\om|}{\Lambda^2}
     \Theta( |\om| - 2 \tp ),
\end{multline}
\begin{equation}
   \label{eq:dis_nobias_DOSgraphiteB0}
  \rho_{\rB}^{0} = \rho_{\rA}^{0} + 
  \frac{\sqrt{4\tp^2 - \om^2}}{\pi \Lambda^2}
  \Theta (2 \tp - |\om|).
\end{equation}
\end{subequations}
The equivalent expression were previously obtained in 
Ref.~[\onlinecite{Paco06a}] using a different method.
The self-energy within the t-matrix is again given by the expression in
Eq.~(\ref{eq:dis_nobias_TmatSigma}), 
with the $\rho$'s given by the non-interacting density of states in
Eq.~(\ref{eq:dis_nobias_DOSgraphite0}).
The F's are obtained from the real parts of the non-interacting local
propagators: 
\begin{subequations}
\label{eq:dis_nobias_Fgraphite0}
\begin{multline}
  \label{eq:dis_nobias_FgraphiteA0}
  F_{\rA}^{0} =
    \frac{\om}{\Lambda^2}
    \log \Bigl( \frac{\Lambda^2}{\tp |\om|} \Bigr)
    \Theta( 2 \tp - |\om| )
\\
    +
    \frac{\om}{\Lambda^2} \log \Bigl(
    \frac{2 \Lambda^2}{|\om|(|\om|+\sqrt{\om^2 - 4\tp^2})} \Bigr)    
    \Theta ( |\om| - 2 \tp )
    ,
\end{multline}
\begin{multline}
  \label{eq:dis_nobias_FgraphiteB0}
    F_{\rB}^{0} =
    F_{\rA}^{0} - \frac{\om}{\Lambda^2}
    +  \sign(\om) \frac{\sqrt{\om^2 - 4 \tp^2}}{\Lambda^2} 
    \Theta ( |\om| - 2 \tp ) .
\end{multline}
\end{subequations}
The self-energies obtained 
within the t-matrix are shown in Fig.~\ref{fig_Sigma_Tmat_graphite}
while those obtained from the
CPA are shown in Fig.~\ref{fig_Sigma_CPA_graphite}.
A comparison between the density of states in the different approximations
is shown in Fig.~\ref{fig_DOS_graphite}.
In general the curves are similar to the ones in the bilayer but
somewhat smoother.

The behavior of the self-energies at the Dirac point in the multilayer
are similar to the case of the bilayer treated in 
Section~\ref{sec:zerofreqlimit},~\ref{sec:uncompensated}
and~\ref{sec:born_self_energy}. We have more to say about this when
we discuss the perpendicular transport in the multilayer in
Section~\ref{sec:perp_transport}.

\subsection{Spectral function}

\begin{figure}[htb]
\centering
 \includegraphics[scale=.40]{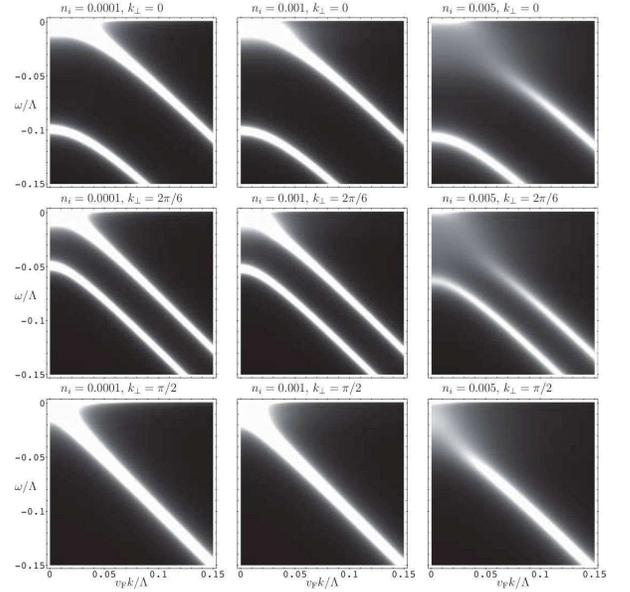}
 \caption[Intensity plot of the spectral
   function in the multilayer for different impurity concentrations and
   different values of the perpendicular momentum]
 {
   Intensity plot of the electron spectral
   function in the $k \times \omega$ plane for the multilayer for different disorder values and
   different values of $k_{\perp}$ in the CPA. From left to right: $n_i = 10^{-4}$, $10^{-3}$, $5 \times 10^{-3}$.
From top to bottom: $k_{\perp} = 0$, $\pi/(3 d)$, $\pi/(2 d)$.}
 \label{fig_Arpes_multilayer}
 \end{figure}

The spectral function for the graphene multilayer is given by a 
generalization of Eq.~(\ref{eq:dis_nobias_spectral_bilayer}):
\begin{multline}
  \label{eq:dis_nobias_spectral_graphite}
  A(\vkpa,\kp,\om) = -\frac{2}{\pi} 
  \Bigl\{ \imag \bigl[ \GDA (\kpa,\kp,\om)  \bigl] 
\\
  +  \imag \bigl[  \GDB (\kpa,\kp,\om) \bigl] \Bigr\}.
\end{multline}
Given the form of the Green's function and the CPA self-energies it is
straightforward to obtain this quantity. The spectral function is
depicted in Fig.~\ref{fig_Arpes_multilayer} for three values of the
perpendicular momentum, since the model we use is electron-hole
symmetric we only show the results for negative frequencies.
We would like to stress that for a large part of the BZ the spectra
are reminiscent of the bilayer spectra. At the edges of the BZ however,
where $2 \tp \cos(\kp d) = 0$, the spectrum is that of massless Dirac
fermions. 

\begin{figure}[htb]
\centering
 \includegraphics[scale=.40]{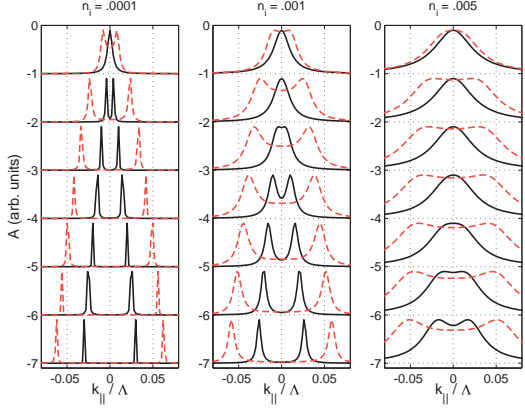}
 \caption
 {[color online] MDC's in the graphene multilayer
 for two values of the perpendicular momentum:
 $k_{\perp} = 0$ (i.e. at the K point, parabolic dispersion, dashed line) 
 and
 $k_{\perp} = \pi /2$ (i.e. at the H point, linear dispersion, solid line).
 The three panels are for different values
 of the density of impurities in the CPA.
  From the top the energy cuts are at the energies:
 .0005, 
 .0055, 
 .0105, 
 .0155, 
 .0205, 
 .0255, 
and  .0305
 in units of  the cutoff $\Lambda$.
 The curves are uniformly displaced for clarity.
}
 \label{fig_MDC_multilayer}
 \end{figure}

\begin{figure}[htb]
\centering
 \includegraphics[scale=.40]{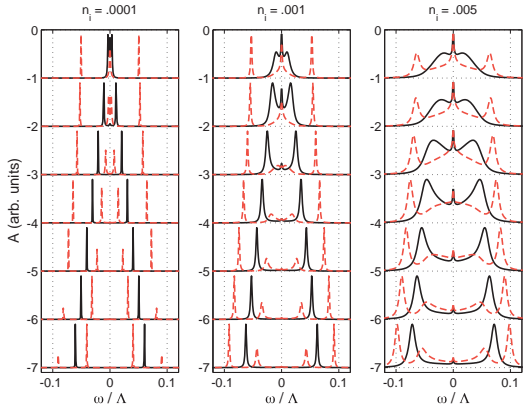}
 \caption
 {[color online] EDC's in the graphene multilayer
 for two values of the perpendicular momentum:
 $k_{\perp} = \pi / 3$ (i.e. between the K and H point, parabolic dispersion, dashed line) 
 and
 $k_{\perp} = \pi /2$ (i.e. at the H point, linear dispersion, solid line).
 The three panels are for different values
 of the density of impurities in the CPA.
 From the top the cuts are at fixed values of the in-plane momentum $k_{||}$:
 .0001, 
 .01, 
 .02, 
 .03, 
 .04, 
 .05, and
 .06
 in units of  the cutoff $\Lambda$.
 The curves are uniformly displaced for clarity.
}
 \label{fig_EDC_multilayer}
 \end{figure}

Examples of the momentum distribution curves (MDC's) 
and the energy distribution curves (EDC's)
in the disordered graphene
multilayer are shown in Figs.~\ref{fig_MDC_multilayer}
and~\ref{fig_EDC_multilayer}
for two different values of $k_{\perp}$.
The evolution of the peaks from delta functions to broader peaks with
increasing disorder is clearly seen.
One can also note that the influence of the impurities is more severe close
to the H point of the BZ since the overlap of the peaks is larger there.
The reason for this is that the particles there are dispersing linearly
leading to peaks that are closer together than for particles with 
a parabolic dispersion.

\section{Electron transport in bilayer and multilayer graphene}
\label{sec:electron_transport}
Having worked out the self-energies in the previous sections we now
turn to linear response (Kubo formula) to study electron transport.
We saw in Sections~\ref{sec:impurities_tmatrix}
and~\ref{sec:G_multilayer} that the low-energy
states are mainly residing on the B sublattice. Nevertheless,
electron transport coming from nearest neighbor hopping must go over
the A sublattice. This is particularly important for the case of perpendicular
transport, since in the simple model that we are using, hopping
comes exclusively from states with weight on the A sublattice. 

This feature implies that although the total density of states at
half-filling is finite, because the density of states on the A
sublattice goes to zero as the Dirac point is approached,
the in-plane and out-of-plane transport properties are unconventional.
The main purpose of this section and the two following sections
is to show how this comes about in
detail through concrete calculations.
We calculate conductivities (or optical response) parallel and perpendicular
to the planes in both bilayer graphene and multilayer graphene.
The resulting conductivities are very anisotropic and we find a universal 
nonzero minimum value for the in-plane DC conductivity as a function of the
chemical potential.

\subsection{Conductivity formulas}
\label{sec:conductivity_formulas}
To calculate the conductivity we use the Kubo formula.\cite{mahan}
We only consider the homogeneous ($\vq = 0$) response, but we
investigate both the temperature dependence and the frequency 
dependence of the various conductivities.

The conductivity is computed from the Kubo formula:
\begin{equation}
  \label{eq:dis_nobias_kubo1}
  \sigma(\om) = \frac{1}{S \om } \int_{0}^{\infty} dt 
  e^{i \om t} \la [J^{\dag}(t),J(0)] \ra
  = \frac{i}{\om + i\delta} \Pi(\om).
\end{equation}
Here $S$ is the area of the system,
$J$ is the current operator of interest, and $\Pi$ the appropriate
current-current correlation function.
A contribution to $\Pi$ from a term of the form $\la [A(t),B(0)]\ra$
where $A=\sum_{\vk} \alpha(\vk)a_{1 \vk}^{\dag} a_{2 \vk}$ and
$B=\sum_{\vk} \beta(\vk)b_{1 \vk}^{\dag} b_{2 \vk}$ can be shown by
the usual methods to give a contribution to the real part of the
conductivity of the form: 
\begin{multline}
  \label{eq:dis_nobias_conductivity1}
  \real \bigl[ \sigma(\om) \bigr] =
  \frac{1}{S}\sum_{\vk}
  \int
  \frac{d\e}{\pi} \Bigl[- \frac{\nF(\e+\om)-\nF(\e)}{\om} \Bigr]
\\ \times
  \imag \bigl[ \GR_{b2,a1}(\e,\vk) \bigr]
  \imag \bigl[ \GR_{a2,b1}(\e+\om,\vk) \bigr]
  \alpha(\vk) \beta(\vk).
\end{multline}
Here the imaginary parts only involve the frequency
part and not the angular (spatial) parts of the propagators.
In terms of the expression in Eq.~(\ref{eq:dis_nobias_green2})
this imply that the imaginary parts involves $\grD$ and $\grND$ but
not the spatial information encoded in
$\Mca_1^{\ }(\vk)$ and $\Mca_1^{\dag}(\vk)$.
With the inclusion of the two spin projections and two valleys we
get (putting back $h = 2 \pi \hbar$ and extracting the electric charge
from the current operators):
\begin{equation}
  \label{eq:dis_nobias_conductivity2}
  \real \bigl[ \sigma(\om) \bigr] = \frac{2 e^2}{\pi h} \int
  d\e \Bigl[- \frac{\nF(\e+\om)-\nF(\e)}{\om} \Bigr]
  \Xi(\e, \e + \om).
\end{equation}
Here $\nF$ is the Fermi distribution function. We have also
introduced the kernel $\Xi$ that for the case of
the operators above becomes:
\begin{multline}
  \label{eq:dis_nobias_kernel1}
  \Xi(\e, \e + \om) = 
  \int_{0}^{\Lambda} d(k^2) 
  \int \frac{d \phi(\vk)}{2 \pi}
\\
  \times
  \imag \bigl[ \GR_{b2,a1}(\e,\vk) \bigr]
  \imag \bigl[ \GR_{a2,b1}(\e+\om,\vk) \bigr]
  \alpha(\vk) \beta(\vk).  
\end{multline}
Thus the contribution to the in-plane DC conductivity at zero temperature is
\begin{equation}
  \label{eq:dis_nobias_conductivityDC1}
  \sigma_{\text{DC},\parallel} = \frac{2 e^2}{\pi h} \Xi(0,0)
  \equiv \sigma_{0 \parallel} \Xi(0,0).
\end{equation}
Finally we also note that -- since we are using the approximation of
purely local impurities -- there are no vertex corrections appearing
in the model.

\subsection{Bilayer graphene}
The current operators can be obtained from the Peierls'
substitution,\cite{kotliar2003,nuno2006_long} 
and an expansion close to the K (or K') point in the BZ.
Alternatively the current operators can be obtained
directly from the Hamiltonian in 
Eq.~(\ref{eq:nobias_Hkin0bilayer}) using
${\bf J} = e n {\bf v}$ with the velocity being given by
${\bf v} = \partial \Hca_0(\vk)/ \partial \vk$.
In any case, the current operators needed for the calculation
of the conductivities are given by:
\begin{subequations}
  \label{eq:dis_nobias_current_operator_x}
\begin{equation}
  J_{x1} =  \vf e \sum_{\vk} \bigl[
    c_{\rA 1,\vk}^{\dag} c_{\rB 1,\vk}^{\, }
 +  c_{\rB 1,\vk}^{\dag} c_{\rA 1,\vk}^{\, }
 \bigr] ,
\end{equation}
\begin{equation}
  J_{x2} =  \vf e \sum_{\vk} \bigl[
    c_{\rA 2,\vk}^{\dag} c_{\rB 2,\vk}^{\, }
 +  c_{\rB 2,\vk}^{\dag} c_{\rA 2,\vk}^{\, }
 \bigr] ,
\end{equation}
\begin{equation}
  \label{eq:dis_nobias_Jperp_bilayer}
  J_{\perp} = i e d \tp \sum_{\vk} \bigl[
  c_{\rA 1, \vk}^{\dag} c_{\rA 2, \vk}^{\, } -
  c_{\rA 2, \vk}^{\dag} c_{\rA 1, \vk}^{\, }
  \bigr].
\end{equation}
\end{subequations}
From the contributions of the form $\la [J_{x1},J_{x1}] \ra$ to the current
correlator we get a
contribution to the kernel which is:
\begin{multline}
  \label{eq:dis_nobias_kernel2}
  \Xi_{x1,x1} = 
  \int_{0}^{\Lambda ^2} d(\vf^2 k^2)  
  \Bigl\{
  \imag \bigl[ \gr_{\text{AA}}^{\text{D}}(\e,\vk) \bigr]
  \imag \bigl[ \gr_{\text{BB}}^{\text{D}}(\e+\om,\vk) \bigr]
  \\
  +
  \imag \bigl[ \gr_{\text{BB}}^{\text{D}}(\e,\vk)      \bigr]
  \imag \bigl[ \gr_{\text{AA}}^{\text{D}}(\e+\om,\vk)  \bigr]
  \Bigr\}.
\end{multline}
Similarly from the cross term $\la [J_{x1},J_{x2}] \ra$ we get:
\begin{equation}
  \label{eq:dis_nobias_kernel3}
  \Xi_{x1,x2} = 
  2 \int_{0}^{\Lambda ^2} d(\vf^2 k^2)  
  \imag \bigl[ \gr_{\text{AB}}^{\text{ND}}(\e,\vk) \bigr]
  \imag \bigl[ \gr_{\text{AB}}^{\text{ND}}(\e+\om,\vk) \bigr],
\end{equation}
while for the interplane optical response the contribution from 
$\la [J_{\perp},J_{\perp}] \ra$ is:
\begin{multline}
  \label{eq:dis_nobias_kernel4}
  \Xi_{\perp} = 
  2 \Bigl( \frac{2 d \tp}{3 a t} \Bigr)^2
  \int_{0}^{\Lambda ^2} d(\vf^2 k^2)
  \\ \times
  \Bigl\{
  \imag \bigl[ \gr_{\text{AA}}^{\text{D}}(\e,\vk) \bigr]
  \imag \bigl[ \gr_{\text{AA}}^{\text{D}}(\e+\om,\vk) \bigr]
  \\
  -
  \imag \bigl[ \gr_{\text{AA}}^{\text{ND}}(\e,\vk)      \bigr]
  \imag \bigl[ \gr_{\text{AA}}^{\text{ND}}(\e+\om,\vk)  \bigr]
  \Bigr\}.
\end{multline}
Due to the phases in the Green's functions the other terms such as
those involving 
$\GR_{\text{AB}}^{\text{D}} \GR_{\text{AB}}^{\text{D}}$ vanish upon
performing the angle average.
To get the total $\sigma_{\parallel}$ per plane in the bilayer one
should add the contributions from $\Xi_{x1,x1}$ and $\Xi_{x1,x2}$.

\subsection{Multilayer graphene} 
The expressions for the current operators in the multilayer
are obtained in a similar way.
$J_{\parallel}$ is given by the same expression as in
Eq.~(\ref{eq:dis_nobias_current_operator_x}) 
except that the momentum variable is now three-dimensional.
The current operator in the perpendicular direction is
\begin{equation}
  \label{eq:dis_nobias_current_perp_graphite}
  J_{\perp} = -2 e \tp d \sum_{\vkpa,\, \kp} \sin(k_{\perp})
  \bigl[ c_{\rA 1,\vk }^{\dag} c_{\rA 2,\vk} 
+ c_{\rA 2,\vk }^{\dag} c_{\rA 1,\vk} \bigr].
\end{equation}
To get the conductivities in the multilayer, 
we should divide by the volume $V = S \times 2 d N_c$ instead of the
area $S$. Here $N_c$ is the number
of unit cells in the perpendicular direction. 
We then turn the sums into integrals using 
$\frac{1}{N_c} \sum_{k_{\perp}} \rightarrow 
\int_{-\pi/2}^{\pi/2} \frac{dk}{\pi}$.
Thus to get $\sigma_{\parallel}$ we use the expressions in
Eq.~(\ref{eq:dis_nobias_conductivity2}) and 
Eq.~(\ref{eq:dis_nobias_kernel2}-\ref{eq:dis_nobias_kernel3}) 
and add the perpendicular integral according to: 
\begin{equation}
  \label{eq:dis_nobias_conductivity_par_graphite}
  \Xi_{\parallel,\text{multi}} 
  = \frac{1}{d}\int_{-\pi/2}^{\pi/2}\frac{dk_{\perp}}{\pi} 
  \bigl[ \Xi_{x1,x1} + \Xi_{x1,x2} \bigr].
\end{equation}
Similarly for the perpendicular conductivity we use 
Eq.~(\ref{eq:dis_nobias_conductivity2}) with:
\begin{multline}
  \label{eq:dis_nobias_conductivity_perp_graphite}
  \Xi_{\perp,\text{multi}}
  = \frac{1}{d}
  \Bigl( \frac{4 d \tp}{3 a t} \Bigr)^2
  \int_{-\pi/2}^{\pi/2}\frac{dk_{\perp}}{\pi} 
  \int_{0}^{\Lambda ^2} d(\vf^2 k^2)
  \\
  \times
  \sin^2(k_{\perp})
  \Bigl\{
  \imag \bigl[ \gr_{\text{AA}}^{\text{D}}(\e,\vk) \bigr]
  \imag \bigl[ \gr_{\text{AA}}^{\text{D}}(\e+\om,\vk) \bigr]
  \\
  +
  \imag \bigl[ \gr_{\text{AA}}^{\text{ND}}(\e,\vk)      \bigr]
  \imag \bigl[ \gr_{\text{AA}}^{\text{ND}}(\e+\om,\vk)  \bigr]
  \Bigr\}.
\end{multline}
The numerical value of the dimensionless prefactor in $\Xi_{\perp}$ is approximately $0.15$
(using $d\approx 2.5 \,a$). When we present the results in the following sections it is
convenient to use a unit that
combines the prefactor in this kernel with the factor from 
Eq.~(\ref{eq:dis_nobias_conductivityDC1})
according to 
$
 \sigma_{\perp 0} = [2 e^2 /(\pi h) ] (1 / d) [ 4 d \tp  /( 3 a t) ]^2 
$.

\section{Results for the conductivities in the graphene bilayer}
\label{sec:conductivity_results_b}

Using the formulas for the kernels (i.e., the $\Xi$'s) from the
previous section and the explicit forms of the propagators from
Section~\ref{sec:H_G_bilayer}  [in particular 
Eqs.~(\ref{eq:dis_nobias_green2})-(\ref{eq:dis_nobias_green4})]
we have calculated the kernels for arbitrary values of the self-energies.
Details of this calculation are provided in App.~\ref{app:kernels}.
In this section we present results for the conductivities
[via Eq.~(\ref{eq:dis_nobias_conductivity2})
and  Eq.~(\ref{eq:dis_nobias_conductivityDC1})] 
using the kernels obtained with the t-matrix and CPA self-energies 
discussed in Sec.~\ref{sec:impurities_tmatrix}.

\subsection{Chemical potential dependence}
\begin{figure}[htb]
\centering
 \includegraphics[scale=.40]{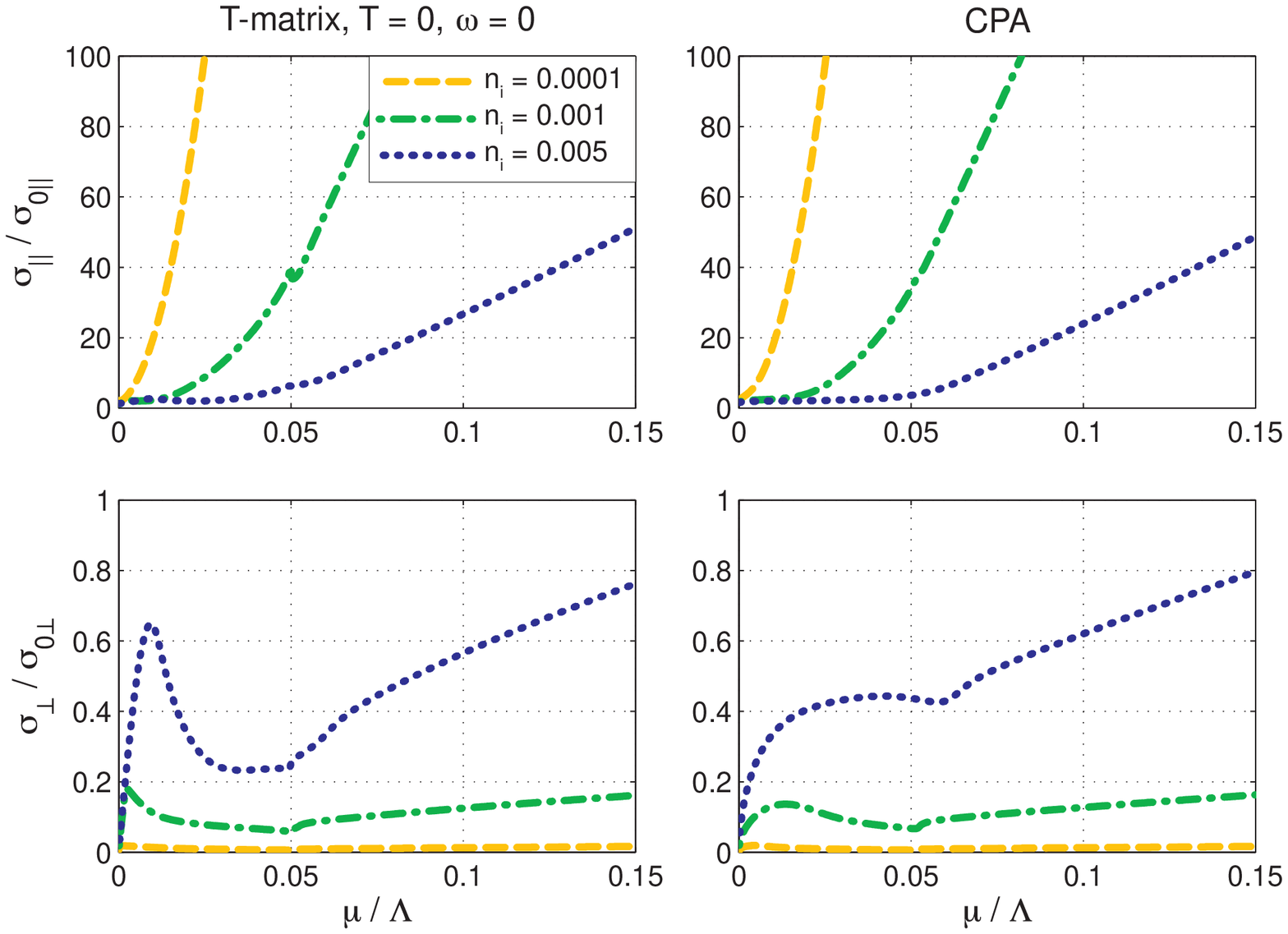}
 \caption
 {[color online] DC conductivities in the bilayer as a function of the chemical
   potential (in units of the cutoff) at zero temperature. Left: t-matrix; Right: CPA.
Top: in plane; Bottom: c-axis.}
 \label{fig_DCcond_bilayer1}
 \end{figure}

\begin{figure}[htb]
\centering
 \includegraphics[scale=.40]{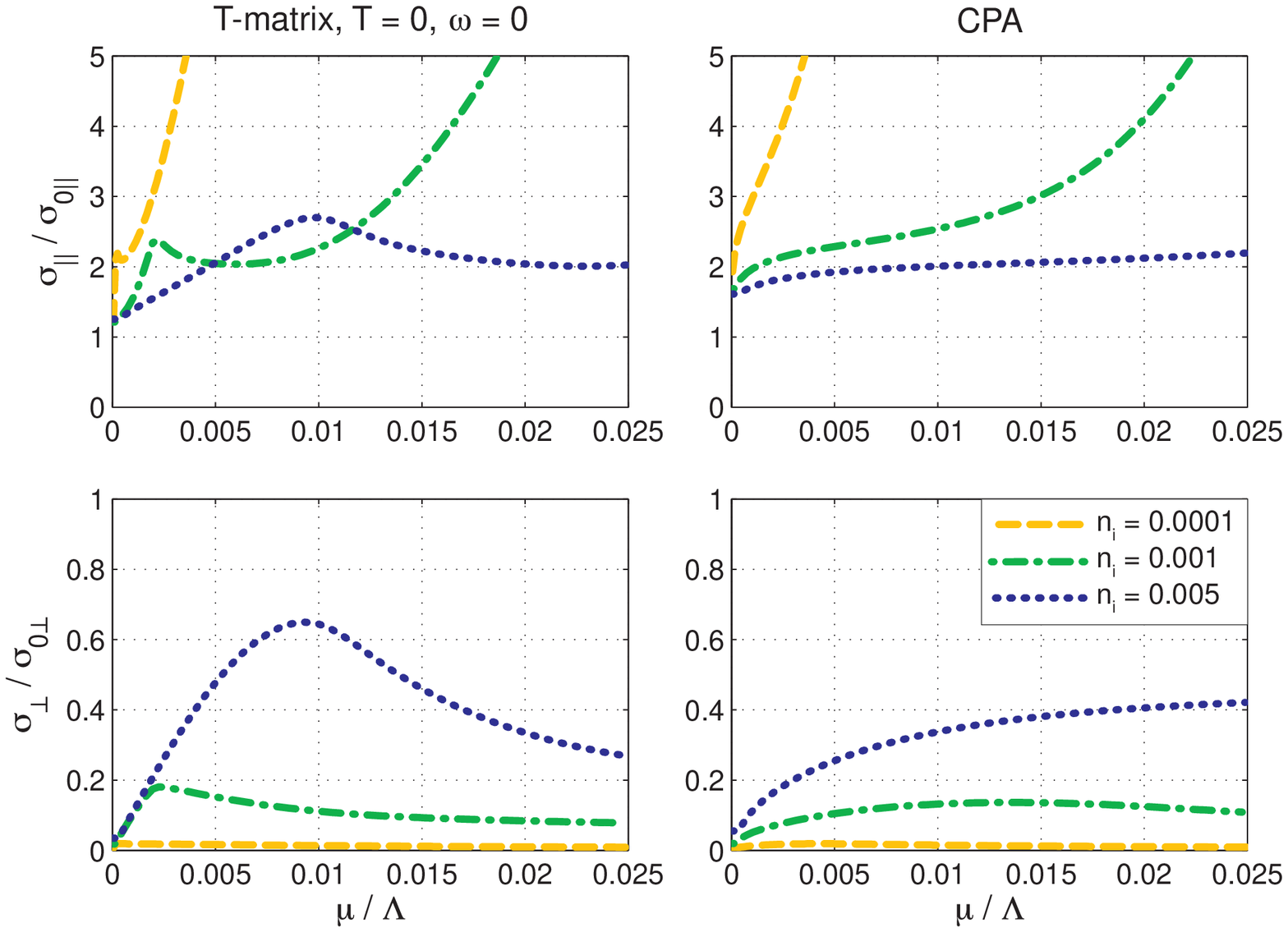}
 \caption{[color online] 
   DC conductivities in the bilayer as a function of the chemical
   potential (in units of the cutoff) at zero temperature. Left: t-matrix; Right: CPA.
Top: in plane; Bottom: c-axis.}
 \label{fig_DCcond_bilayer3}
 \end{figure}

The results for the DC conductivities at $T=0$ in the t-matrix and CPA
approximations for different values of the chemical 
potential are shown in 
Figs.~\ref{fig_DCcond_bilayer1}
and~\ref{fig_DCcond_bilayer3}.
The only difference between these figures are in the scales of the axes.
It looks as if the minimum value for $\sigma_{\parallel}$ per plane
in the bilayer is approximately 
$2 \sigma_{\parallel 0} = 4 e^2 /(\pi h)$, which is identical to the
result one obtains using the same methods in single layer 
graphene.\cite{nuno2006_long} We discuss the minimum conductivity
later in this section.
%
%

%
The low-energy feature in the t-matrix curves comes about at the
energy scale at which the two planes start to decouple.
The scale at which this takes place ($\SA \approx \tp$) is easily
found numerically with the results shown in 
Table~\ref{tab:decoupling_scale}.
The local maximum in the conductivities 
is readily identified with this energy scale.
\begin{table}[ht]
  \centering
  \begin{tabular}{|c|c|c|c|c|}
    \hline
    $ \nimp$ & 0.01 & 0.005 & 0.001 & 0.0001 \\
    \hline
    $\om/\Lambda$ & $1.4 \times 10^{-2}$ & $6.5 \times 10^{-3}$ &
    $1.1 \times 10^{-3}$ & $8.4 \times 10^{-5}$  \\ 
    \hline
  \end{tabular}
  \caption[Energy scale at which the planes in the bilayer start to
    decouple within the t-matrix approximation] 
  {Energy scale $\om$ at
    which the planes start to decouple within the t-matrix
  approximation [this happens when $\SA(\om) \approx \tp$].}
  \label{tab:decoupling_scale}
\end{table}
In the CPA this scale is suppressed and the curves show no peak.
Quite generally it is seen that the CPA curves are smoother than
the ones for the t-matrix. 

Another interesting feature of $\sigma_{\perp}$ is that it is
increased by disorder. This is due to the fact that disorder
enhance the DOS on sublattice A where all the contribution to
$\sigma_{\perp}$ is coming from.
At the lowest values of the chemical potential the perpendicular 
conductivity still goes to zero however.

\subsection{Minimal DC conductivity}
By studying  the curves more closely, it looks as if the CPA curves
actually gives a value that is smaller than 2 in the limit
 $\om \rightarrow 0$.
In fact, we find that the minimum in the in-plane DC conductivity is
again (as in the single layer case of massless Dirac 
fermions~\cite{nuno2006_long}) universal in the sense of being
independent of the particular impurity concentration. 
In the bilayer the minimum value per plane obtained from the CPA is
\begin{equation}
  \label{eq:dis_nobias_sigmamin_cpa}
  \sigma_{\parallel \text{min,CPA}} = \frac{3}{\pi} \frac{e^2}{h}.
\end{equation}
%
This value is obtained by using the form of the self-energies
in the low frequency limit that are given in
Eq.~(\ref{eq:dis_nobias_S_smallOm}).
Explicitly one finds for the propagators via Eq.~(\ref{eq:dis_nobias_green3}):
\begin{subequations}
  \label{eq:dis_nobias_asymptotic_g1}
\begin{eqnarray}
  \gr_{\text{AA}}^{\text{D}}(\om \rightarrow 0,\vk) &\sim& 
  \frac{\SB}{\xi \Lambda^2+k^2} , \\
  \gr_{\text{BB}}^{\text{D}}(\om \rightarrow 0,\vk) &\sim& 
  \frac{\SA}{\xi \Lambda^2+k^2} , \\
  \gr_{\text{AB}}^{\text{ND}}(\om \rightarrow 0,\vk) &\sim& 0.
\end{eqnarray}
\end{subequations}
Using these asymptotic forms in Eq.~(\ref{eq:dis_nobias_kernel2}) 
and Eq.~(\ref{eq:dis_nobias_kernel3}) the contribution from the
latter equation drops out. 
The value in Eq.~(\ref{eq:dis_nobias_sigmamin_cpa}) is obtained
from the first term after employing the relation: 
$\imag[\SA] \imag[\SB] \sim 
(\sqrt{3}/2)^2 |\SA \SB| \sim 3 \xi \Lambda^2 /4$.

We note that our value for the minimal conductivity is different
from the values obtained in works by other groups. In particular
both Koshino and Ando (using a 2-band model in conjunction with
a second-order self-consistent Boltzmann approximation) and
Snyman and Beenakker (using the conductance formula for coherent
transport) both finds the value $4 e^2 /(\pi h)$ per 
plane.\cite{Ando06a,Snyman2007} (The minimal conductivity problem
in bilayer graphene has also been discussed in 
Ref.~[\onlinecite{Katsnelson06c,Cserti2007}]). 
We can reproduce their result in our formalism by considering
the case that the impurities only affect the B sublattice,
as discussed in Sec.~\ref{sec:uncompensated}
(or the case of Born scattering discussed
in Sec.~\ref{sec:born_self_energy}).
In particular, taking $\SA = 0$ and
$\SB(\om=0) = -i \Gamma$ [from Eq.~(\ref{eq:dis_nobias_SBonlyB})]
one finds
\begin{subequations}
\begin{eqnarray}
  \label{eq:dis_nobias_asymptotic_g2}
  \gr_{\text{AA}}^{\text{D}}(\om = 0,\vk) &=& 
  \frac{-i \Gamma k^2}{(\tp \Gamma)^2 +k^4} , \\
  \gr_{\text{BB}}^{\text{D}}(\om = 0,\vk) &=& 
  \frac{-i \Gamma \tp^2}{(\tp \Gamma)^2 +k^4} , \\
  \gr_{\text{AB}}^{\text{ND}}(\om = 0,\vk) &=& 
  \frac{i \Gamma \tp k}{(\tp \Gamma)^2 +k^4}.
\end{eqnarray}
\end{subequations}
Using these expressions in Eq.~(\ref{eq:dis_nobias_kernel2}) 
and Eq.~(\ref{eq:dis_nobias_kernel3}), $\Xi_{x1,x1}$ and
$\Xi_{x1,x2}$ are found to contribute equally to the conductivity with
the total value being $4e^2 /(\pi h)$.
This result shows that the minimal conductivity is not really
``universal'' in the graphene bilayer since it actually
depends on how the impurities are distributed among the
inequivalent sites of the problem.
Furthermore, the ballistic results corresponds to the case
where the disorder is only affecting the B sublattice.
Nevertheless, the general conclusion is that there is a non-zero
minimum in the in-plane conductivity of the order of $e^2/h$.
Further evidence for this conclusion is hinted by the 
related issue of how other hopping terms, such as
$\gamma_3$ and $\gamma_4$, affect the value of the minimal
conductivity.
The case of trigonal warping (i.e. $\gamma_3 \neq 0$) have
been discussed in detail by Cserti and 
collaborators,\cite{Cserti2007b} and
they find that the minimal value of the conductivity per plane is
$(12/ \pi) (e^2/h)$ in this case 
(See App.~\ref{app:minimal} for an alternative derivation
of this result).
The introduction of $\gamma_4$ (or $\Delta$, or 
a next-nearest neighbor hopping
within the planes) -- which breaks the symmetry
in energy between the central Dirac point and the three
elliptical cones away from $\vk = 0$ will likely further
influence the minimal value.

\subsection{Frequency dependence}

\begin{figure}[htb]
\centering
 \includegraphics[scale=.40]{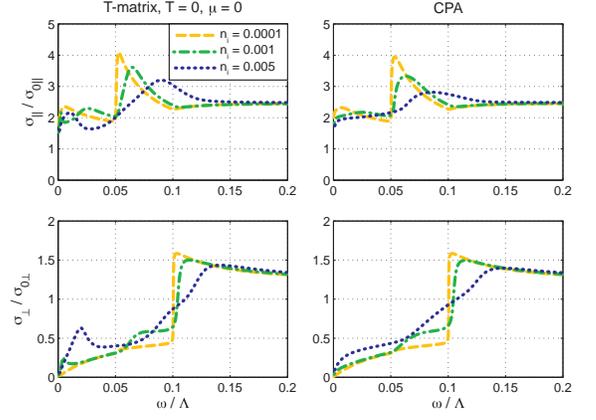}
 \caption
 {[color online] 
   Conductivity as a function of frequency (in units of the cut-off) for the bilayer
   at the Dirac point $\mu =0$ for $T=0$. Left: t-matrix; Right: CPA. Top: in plane; Bottom: c-axis.}
 \label{fig_freq_cond_bilayer4}
 \end{figure}

\begin{figure}[htb]
\centering
 \includegraphics[scale=.40]{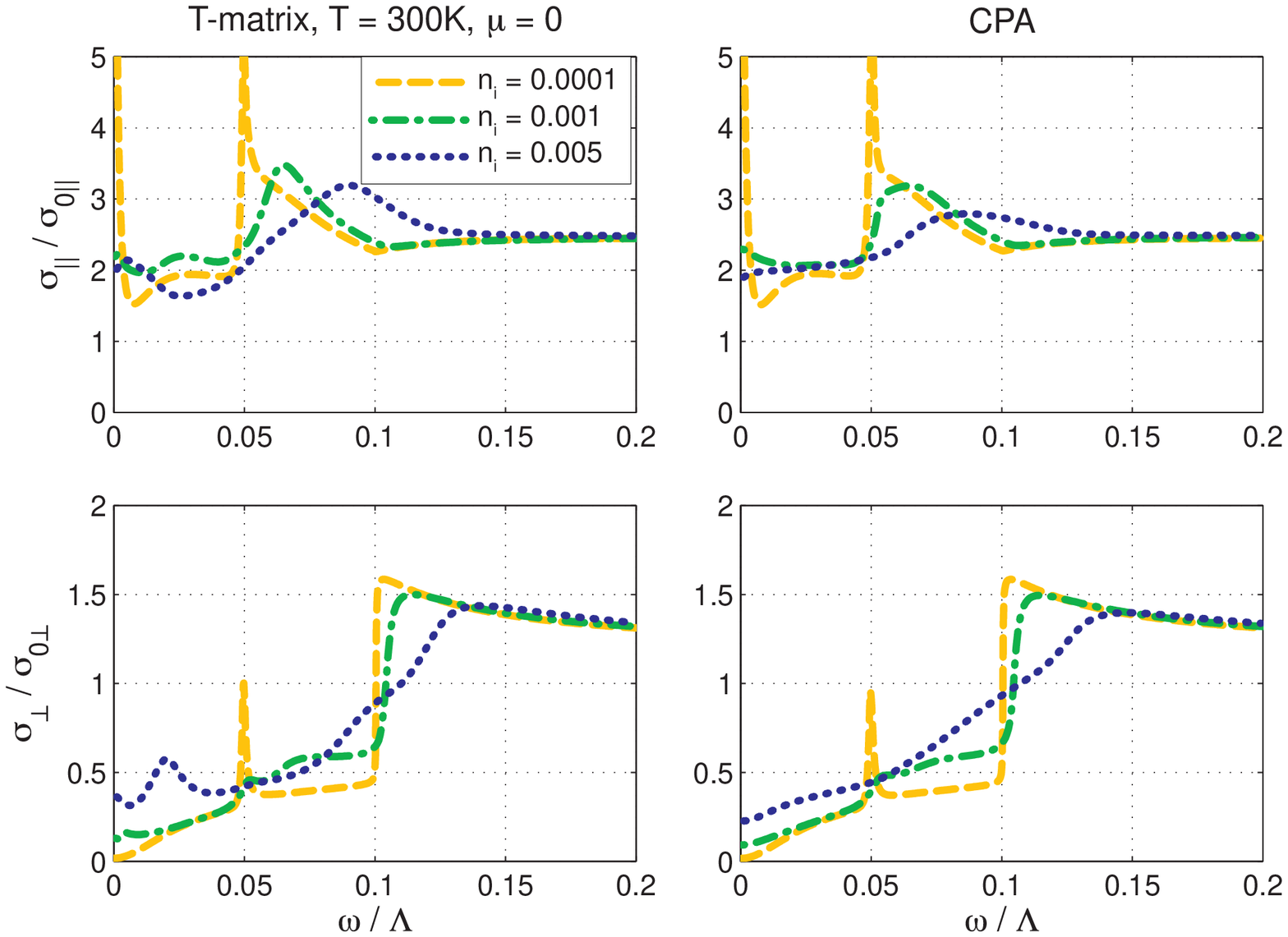}
 \caption{[color online] 
 Conductivity as a function of frequency (in units of the cut-off) for the bilayer
 at finite temperature
  $T = 300 K$ near the Dirac point $\mu = 0$. Left: t-matrix; Right: CPA. Top: in plane; Bottom: c-axis.}
 \label{fig_freq_cond_bilayer5}
 \end{figure}

\begin{figure}[htb]
\centering
 \includegraphics[scale=.40]{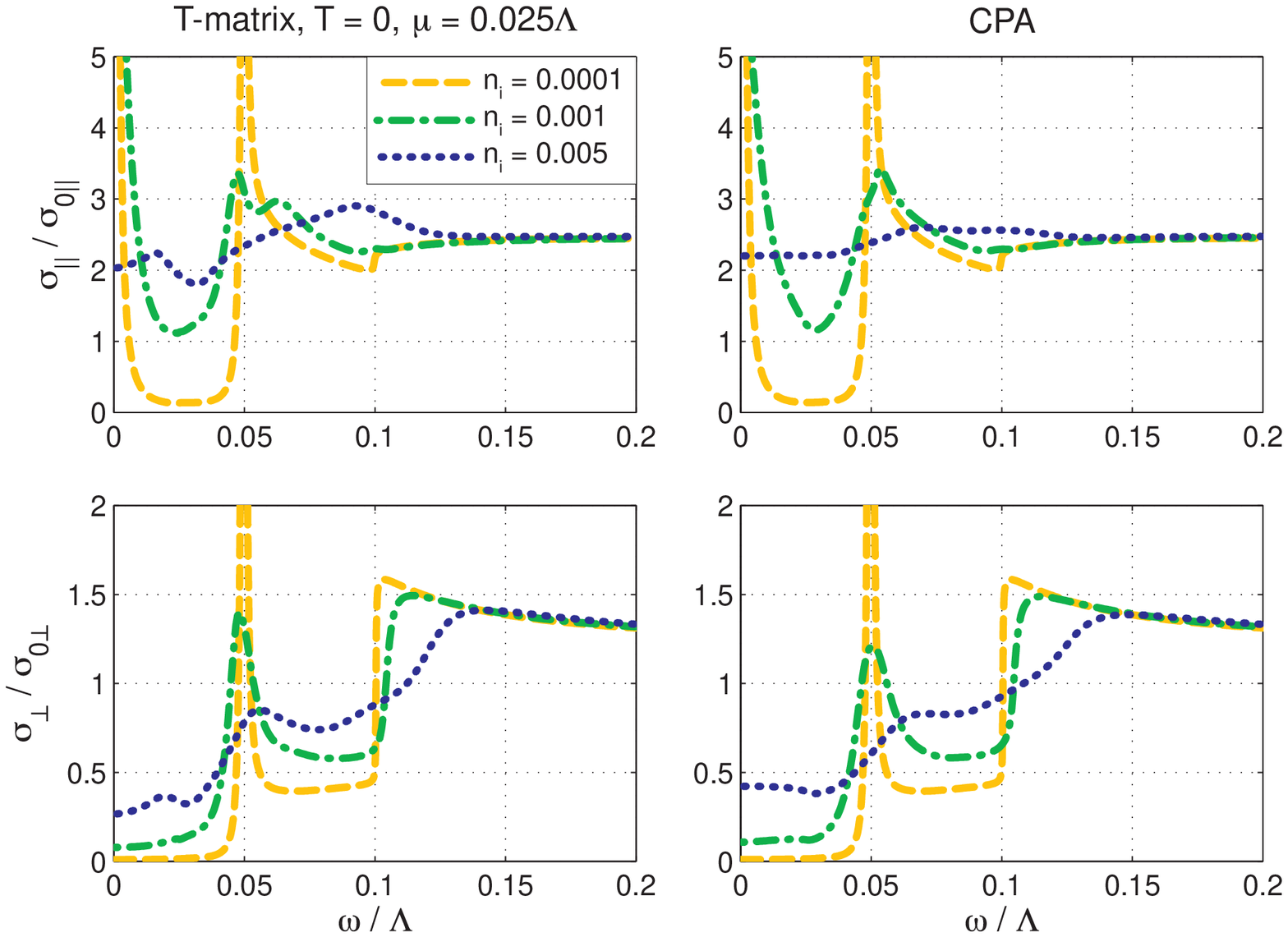}
 \caption
 {[color online]
 Conductivity as a function of frequency (in units of the cut-off) for the bilayer
  at finite chemical potential $\mu = 0.025 \Lambda$ at $T=0$. Left: t-matrix; Right: CPA. Top: in plane; Bottom: c-axis.}
 \label{fig_freq_cond_bilayer6}
 \end{figure}

\begin{figure}[htb]
\centering
 \includegraphics[scale=.40]{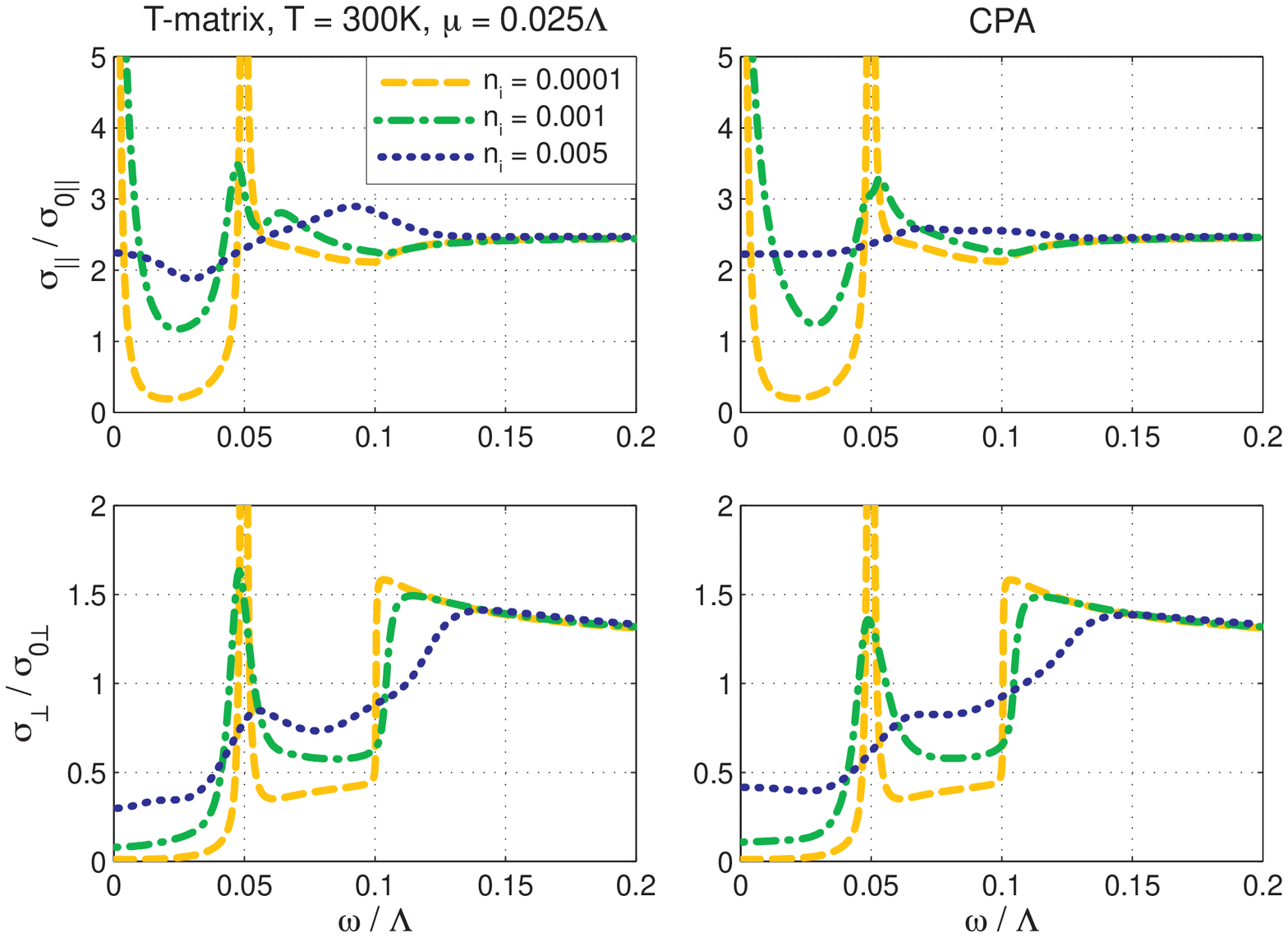}
 \caption
 {[color online] 
  Conductivity as a function of frequency (in units of the cut-off) for the bilayer
  at finite temperature $T = 300 K$ at finite chemical  potential $\mu = 0.025 \Lambda$.
Left: t-matrix; Right: CPA. Top: in plane; Bottom: c-axis.}
 \label{fig_freq_cond_bilayer7}
 \end{figure}

The frequency dependence of the conductivities are pictured in 
Fig.~\ref{fig_freq_cond_bilayer4}-\ref{fig_freq_cond_bilayer7}. 
The figures reveals some interesting
features of the band structure and the semimetallic behavior.
For the case of a finite chemical potential the temperature does not 
make a big difference since it (at $300 \, \text{K}$) 
is still much smaller than the Fermi energy, this is manifested in the
small difference between
Fig.~\ref{fig_freq_cond_bilayer6} and 
Fig.~\ref{fig_freq_cond_bilayer7}.
Near zero chemical potential the effect of the temperature is more
dramatic. The temperature increase the number of carriers and
is responsible for the Drude-like
peaks that appear in the plots for low impurity
concentrations. A well-known feature of semimetals is that
the temperature is an important factor in determining the number of
carriers in the system.
The peak at $\om \sim \tp = .05 \Lambda$ is due to the onset of
interband transitions. 
The frequency-dependence of the absorption of electromagnetic
radiation has also been studied
by Abergel and Falko with similar results,\cite{Abergel_2007}
they also study transitions between Landau levels in a
magnetic field. 


\subsection{Temperature dependence}

\begin{figure}[htb]
\centering
 \includegraphics[scale=.40]{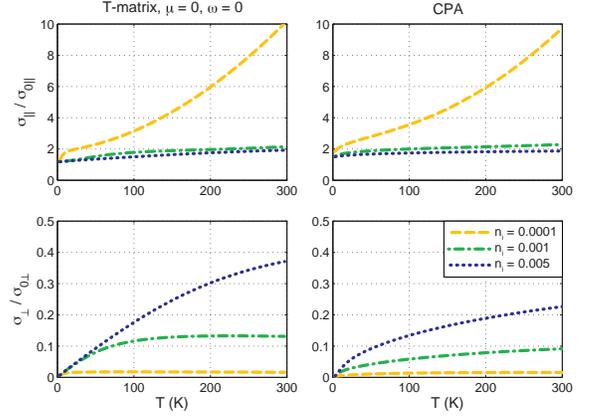}
 \caption
 {[color online] 
   Temperature dependence of the DC conductivities in the bilayer at the Dirac point
   $\mu = 0$. Left: t-matrix; Right: CPA. Top: in plane; Bottom: c-axis.}
 \label{fig_freq_cond_bilayer8}
 \end{figure}

\begin{figure}[htb]
\centering
 \includegraphics[scale=.40]{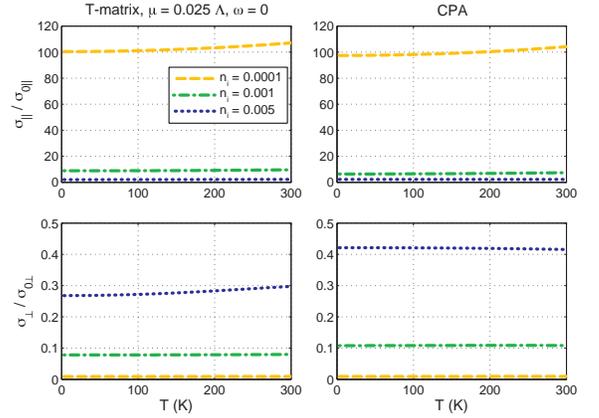}
 \caption{[color online] 
   Temperature dependence of the DC conductivities in the bilayer at
   finite chemical  potential $\mu = 0.025 \Lambda$.}
 \label{fig_freq_cond_bilayer9}
 \end{figure}

The temperature dependence of the DC conductivity can be found in
Fig.~\ref{fig_freq_cond_bilayer8} and Fig.~\ref{fig_freq_cond_bilayer9}.
For the case of a finite chemical potential the in-plane conductivity
curves are flat and proportional to $1/n_i$, as is expected in
a normal Fermi liquid metal.\cite{mahan}
The contribution to the scattering from impurities is
very weakly temperature dependent. Nevertheless, there is a small
temperature dependence for the lowest impurity concentration which is
due to the fact that $T /E_{\text{F}}$ is still not negligible.
Near the Dirac point the characteristics of a semimetal appear
again as the conductivities become temperature dependent.
Note however that we are not considering scattering by phonons
which is important at finite temperatures. 

\section{Results for the conductivities in multilayer graphene}
\label{sec:conductivity_results_g}
Using the same procedure as for the bilayer in the previous
section we have calculated the kernels for arbitrary values of the
self-energies. 
Details of this rather lengthy calculation are provided in
App.~\ref{app:kernels}. 
In this section we present results for the conductivities
using the kernels obtained with t-matrix and CPA self-energies 
discussed in Sec.~\ref{sec:G_multilayer}.

\begin{figure}[htb]
\centering
 \includegraphics[scale=.40]{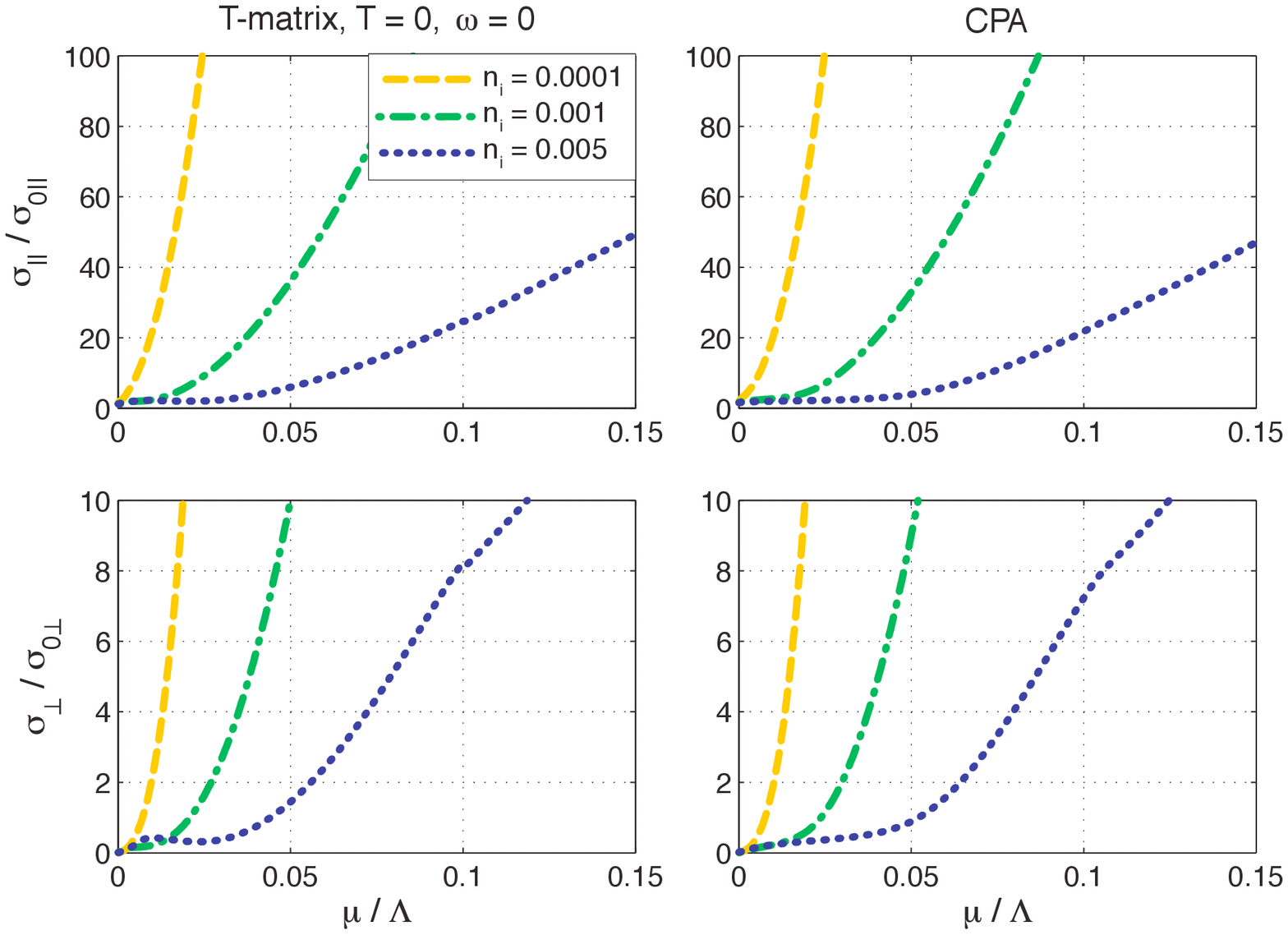}
 \caption{[color online] 
   DC conductivities in the multilayer as a function of the chemical
   potential (in units of the cutoff) at T=0. Left: t-matrix; Right: CPA.
Top: in plane; Bottom: c-axis.}
 \label{fig_DCcond_multilayer1}
 \end{figure}

\begin{figure}[htb]
\centering
 \includegraphics[scale=.40]{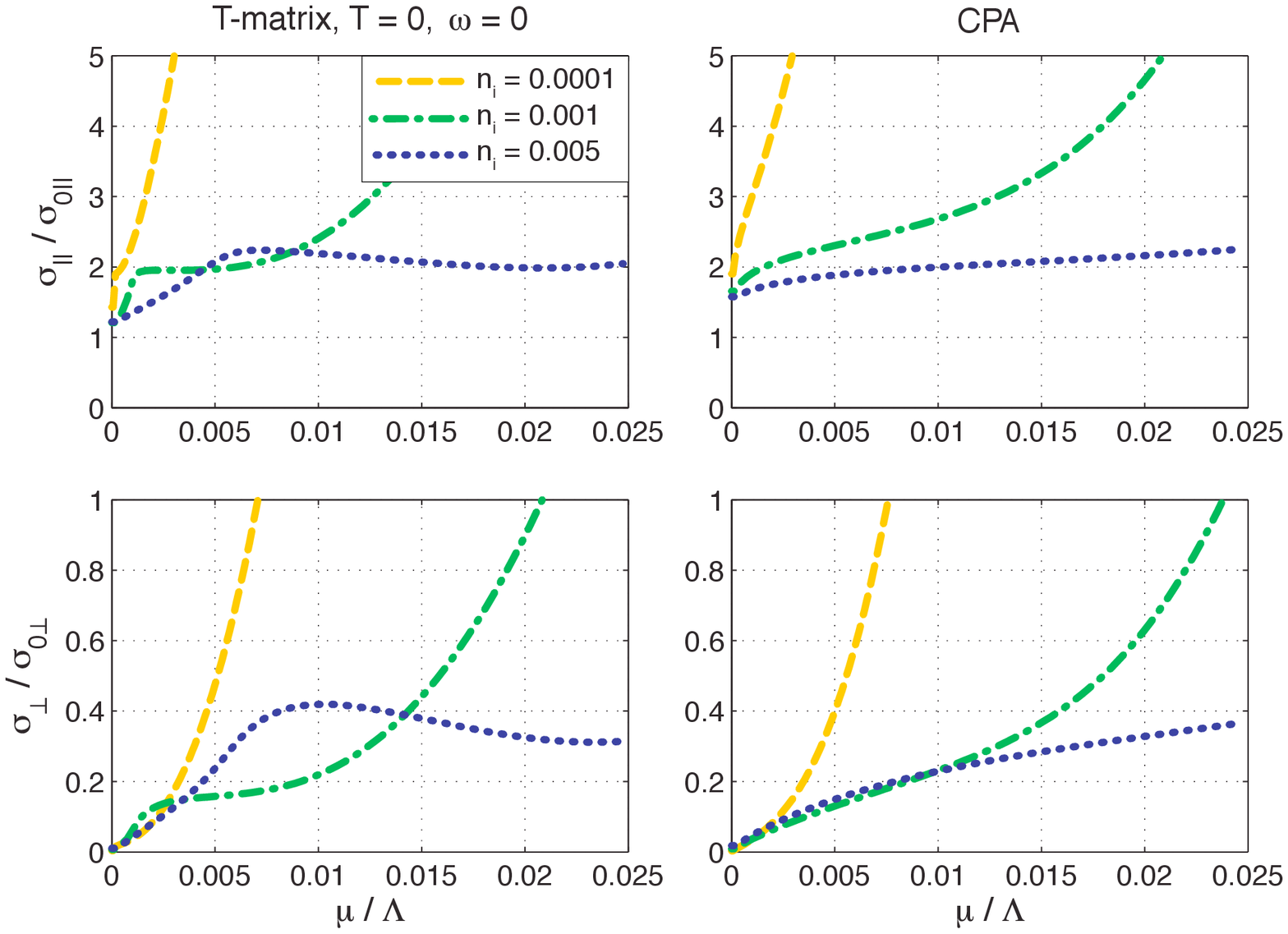}
 \caption{[color online] 
   DC conductivities in the multilayer as a function of the chemical
   potential (in units of the cutoff) at T=0. Left: t-matrix; Right: CPA.
Top: in plane; Bottom: c-axis.}
 \label{fig_DCcond_multilayer2}
 \end{figure}

The DC conductivities in the multilayer as a function of the chemical
potential $\mu$ are pictured in 
Figs.~\ref{fig_DCcond_multilayer1}-\ref{fig_DCcond_multilayer2},
the only difference between the figures are in the scales of the axes.
The property of disorder-enhanced transport in the
perpendicular direction seems to 
survive in this model for the multilayer,
but only for very low values of the chemical potential. 
For larger values of the chemical potential the influence of disorder
becomes more conventional. In this case, because of the finite Fermi
surface, the transport properties are more like in a normal metal.

\subsection{Perpendicular transport near the Dirac point}
\label{sec:perp_transport}
Generalizing Section~\ref{sec:born_self_energy} 
to the multilayer we find again
that $\SA \sim 0$ and $\SB \sim -i \Gamma_{\rB}$ is purely
imaginary in the Born limit at the Dirac point.
Nevertheless, as we shall see it is necessary for the computation
of $\sigma_{\perp}$ that $\SA$ remain finite.
Therefore we take $\SA \sim -i \Gamma_{\rA}$ and assume that
$\Gamma_{\rA} \ll  \Gamma_{\rB}$.
We note that this is also consistent with a self-consistent
version of the Born approximation for weak potentials.
Thus, for $\om \rightarrow 0$ we have
\begin{subequations}
\begin{eqnarray}
  \label{eq:dis_nobias_asymptotic_g3}
  \imag [ \gr_{\text{AA}}^{\text{D}} ] & \sim & 
  \frac{- \Gamma_{\rB} ( k^2 + \Gamma_{\rA} \Gamma_{\rB} )}
  {[ 2 \tp \Gamma_{\rB} \cos(\kp) ]^2 
    + (k^2 + \Gamma_{\rA} \Gamma_{\rB} )^2} , \\
  \imag [ \gr_{\text{AA}}^{\text{ND}} ] & \sim & 0.
\end{eqnarray}
\end{subequations}
Inserting these expressions into 
Eq.~(\ref{eq:dis_nobias_conductivity_perp_graphite}) it is
possible to perform the integrals exactly with the result
\begin{multline}
  \label{eq:xi_multi_asymptotic1}
  \Xi_{\perp,\text{multi}}
  \sim \frac{1}{d}
  \Bigl( \frac{4 d \tp}{3 a t} \Bigr)^2
\\ \times
  \Bigl( \frac{\Gamma_{\rB}}{8 \tp} \Bigr) 
  \log \Bigl[
  \frac{ \sqrt{ 1 + (\Gamma_{\rA}/2 \tp)^2 } + 1 )}
       { \sqrt{ 1 + (\Gamma_{\rA}/2 \tp)^2 } - 1 )}
  \Bigr].
\end{multline}
Thus there is a logarithmic singularity in the limit
$\Gamma_{\rA} \rightarrow 0$ as mentioned above.
Intuitively this singularity comes from ``clean'' chains of atoms
along the A sublattice where transport is unhindered once some
weight has been pushed onto the A sublattice by the impurities
on the B sublattice. 
It is plausible that $\Xi_{\perp,\text{multi}}$
increases with increasing disorder. 
It is so because the first factor grows linearly whereas the second
factor decays only logarithmically with the $\Gamma$ in question.

For the case of vacancies in the CPA a result analogous to the one
in Eq.~(\ref{eq:dis_nobias_S_smallOm}) can be obtained.
In fact the result is the same up to a factor: 
$\SA \rightarrow 2^{1/3} \SA $ and
$\SB \rightarrow 2^{-1/3} \SB $.
Therefore the asymptotic expressions in 
Eq.~(\ref{eq:dis_nobias_asymptotic_g1})
are valid also in the multilayer. In addition one finds
that 
$\gr_{\text{AA}}^{\text{ND}}(\om \rightarrow 0,\vk) \sim 0$.
Thus, asymptotically one finds that 
$\Xi_{\perp,\text{multi}} \sim \om^{2/3}$, which leads to
a temperature dependence of $\sigma_{\perp}$ at the  Dirac point 
that is of the form $T^{2/3}$.
We also note that $\sigma_{\parallel,\text{min}}$ is independent of $\tp$
in the bilayer, thus we conclude that it takes on
the same value in both the bilayer and the multilayer.
Using the fact that $\Xi_{\parallel,\text{multi}} \sim $ 
constant at the Dirac point 
we find that $\sigma_{\parallel} / \sigma_{\perp}$
diverges as $T^{-2/3}$ as $T \rightarrow 0$
as reported previously.\cite{Nilsson2006b}

 \begin{figure}[htb]
\centering
 \includegraphics[scale=.40]{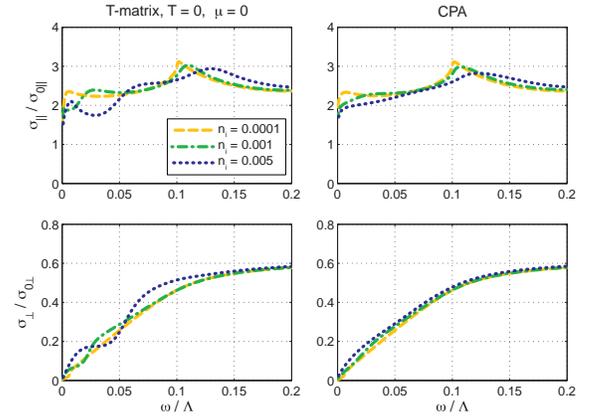}
 \caption{[color online] 
   Conductivity as a function of frequency (in units of the cutoff) at the Dirac point $\mu
   =0$ for $T=0$ in the multilayer. Left: t-matrix; Right: CPA.
Top: in plane; Bottom: c-axis.}
 \label{fig_freq_cond_multilayer1}
 \end{figure}

 \begin{figure}[htb]
\centering
 \includegraphics[scale=.40]{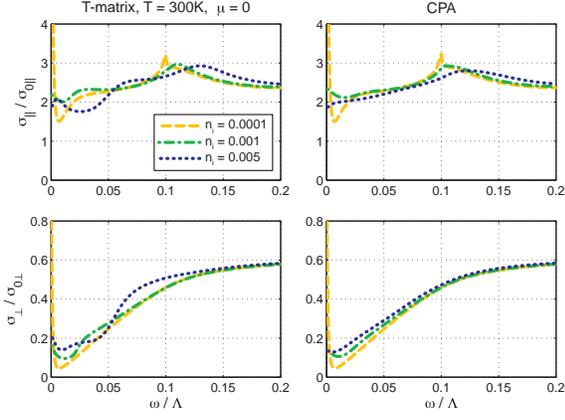}
 \caption{[color online] 
 Conductivity as a function of frequency (in units of the cutoff) at finite temperature
  $T = 300 K$ near the Dirac point $\mu = 0$ in the multilayer. Left: t-matrix; Right: CPA.
Top: in plane; Bottom: c-axis.}
 \label{fig_freq_cond_multilayer2}
 \end{figure}

 \begin{figure}[htb]
\centering
 \includegraphics[scale=.40]{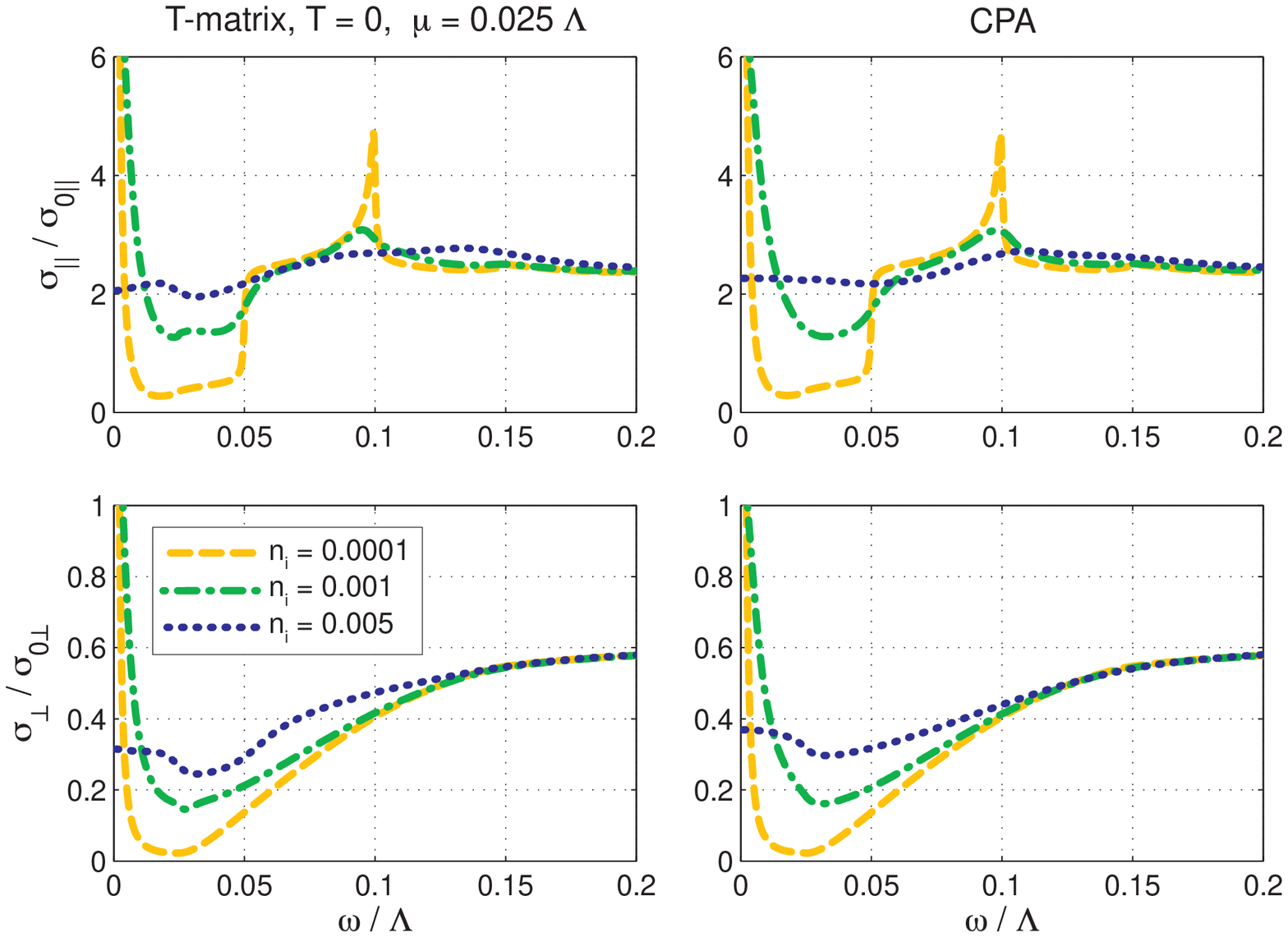}
 \caption{[color online] 
   Conductivity as a function of frequency (in units of the cutoff) at finite chemical
   potential $\mu = 0.025 \Lambda$ at $T=0$ in the multilayer. Left: t-matrix; Right: CPA.
Top: in plane; Bottom: c-axis.}
 \label{fig_freq_cond_multilayer3}
 \end{figure}

 \begin{figure}[htb]
\centering
 \includegraphics[scale=.40]{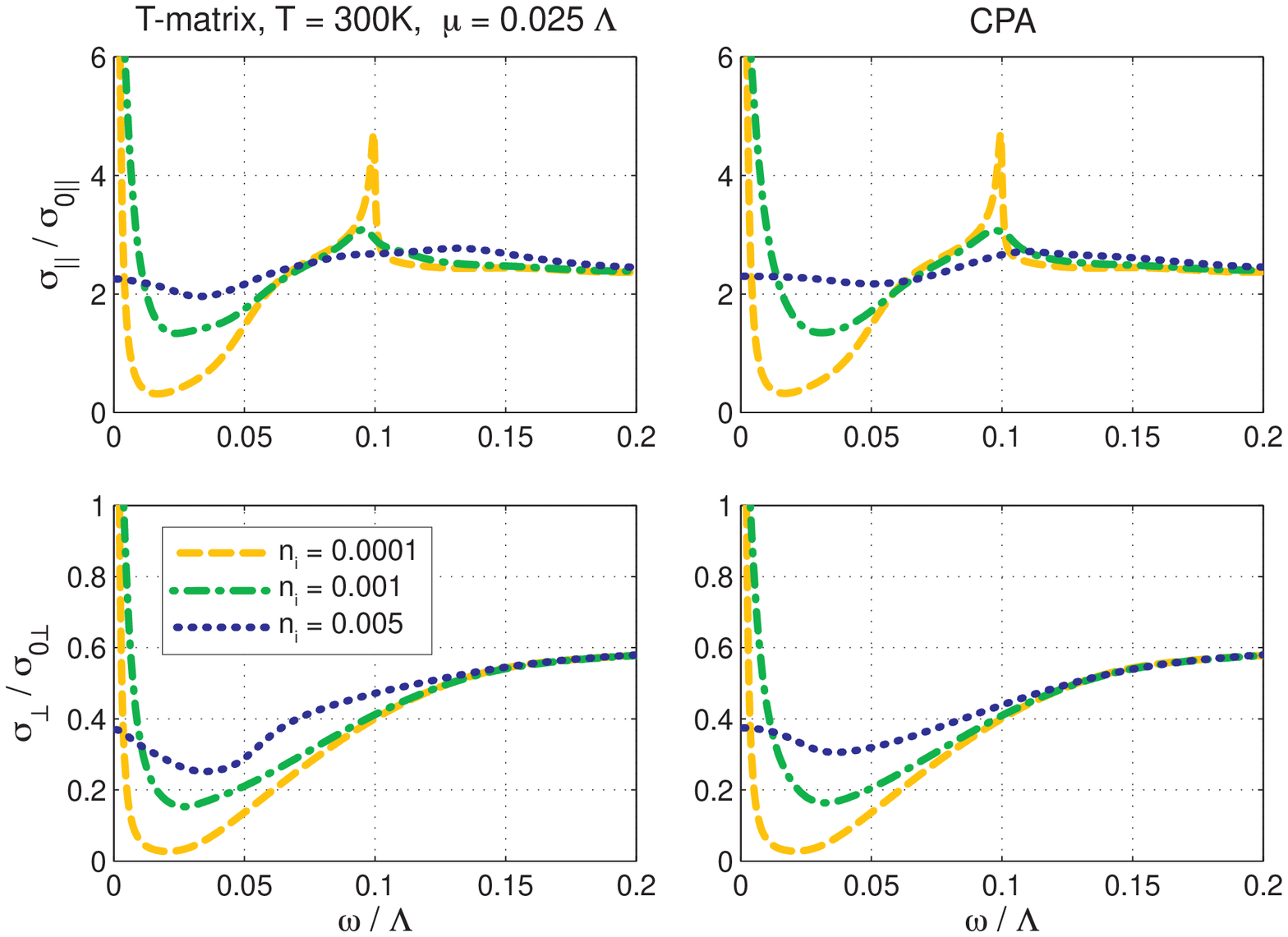}
 \caption
 {[color online] 
   Conductivity as a function of frequency (in units of the cutoff) at finite temperature
   $T = 300 K$ at finite chemical  potential $\mu = 0.025 \Lambda$ in the multilayer. Left: t-matrix; Right: CPA.
Top: in plane; Bottom: c-axis.}
 \label{fig_freq_cond_multilayer4}
 \end{figure}

 \begin{figure}[htb]
\centering
 \includegraphics[scale=.40]{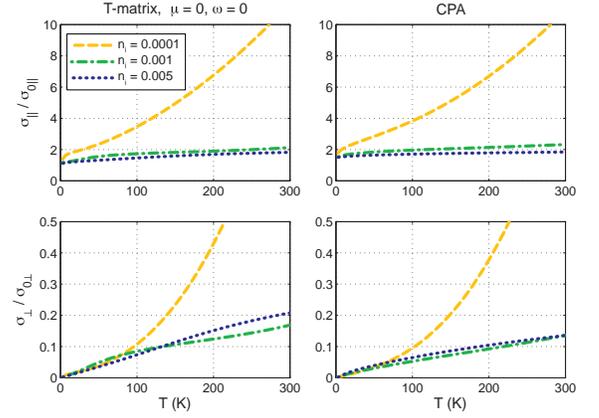}
 \caption{[color online] 
   Temperature dependence of the DC conductivities at the Dirac point
   $\mu = 0$ in the multilayer. Left: t-matrix; Right: CPA.
Top: in plane; Bottom: c-axis.}
 \label{fig_Tdep_cond_multilayer1}
 \end{figure}

 \begin{figure}[htb]
\centering
 \includegraphics[scale=.40]{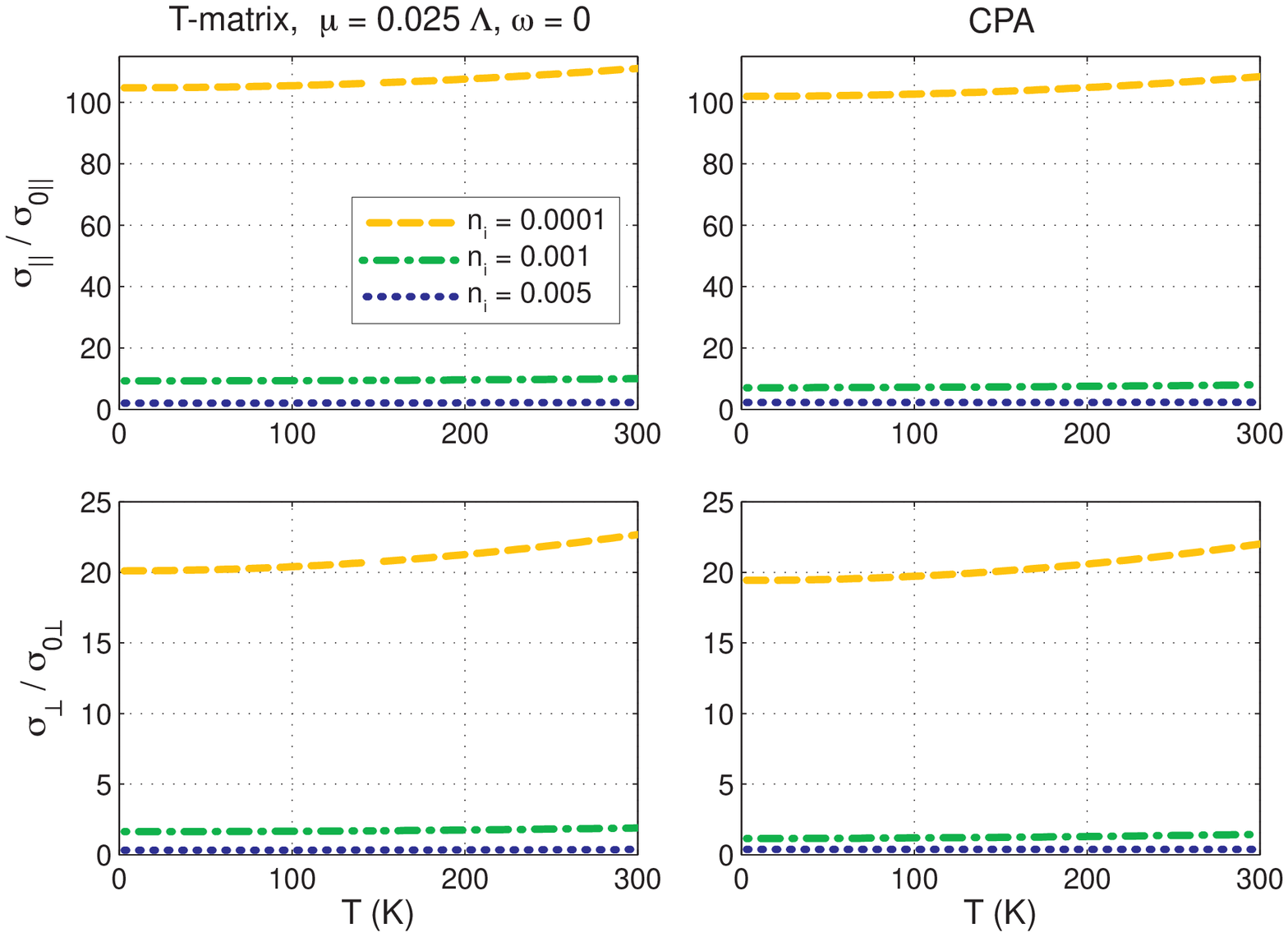}
 \caption
 {[color online] 
   Temperature dependence of the DC conductivities at
   finite chemical  potential $\mu = 0.025 \Lambda$ in the multilayer. Left: t-matrix; Right: CPA.
Top: in plane; Bottom: c-axis.}
 \label{fig_Tdep_cond_multilayer2}
 \end{figure}

\subsection{Frequency and temperature dependence}

The frequency dependence of the conductivities in the multilayer is shown in
Figs.~\ref{fig_freq_cond_multilayer1}-\ref{fig_freq_cond_multilayer4} 
for two different temperatures and both at the Dirac point
and for a finite chemical potential.
For the cleaner systems a Drude-like peak appears at finite temperatures for
both in-plane and perpendicular transport at the Dirac point.
For a finite chemical potential -- because the system is metallic in both
directions -- the system has a Drude peak in the conductivity also at zero temperature. 
Moreover, it can be seen how the suppression of the conductivity in the frequency
range before interband contributions sets in (i.e. at $\om = 2 \mu$) is affected
by both disorder and temperature.

We note that our curves for the frequency-dependent in-plane conductivity
is very similar to the recent results of Ref.~\onlinecite{Kuzmenko_07},
which include both measurements and calculations based on the full SWM model.

Our results for the temperature dependence of the conductivities in the multilayer
are shown in Figs.~\ref{fig_Tdep_cond_multilayer1}-\ref{fig_Tdep_cond_multilayer2}.
At the Dirac point, the in-plane conductivity goes to a finite constant while the
perpendicular conductivity goes to zero as $T \rightarrow 0$. The disorder-enhanced
transport at low temperatures can also be seen in the figure.
For a finite chemical potential, the system behaves like a metal with only a weak
temperature dependence of the conductivities.

%
%

\section{Impurities in the biased graphene bilayer}
\label{sec:BGBimpurities}

In the following sections we study the problem of impurities and
mid-gap states in a biased graphene 
bilayer. We show that the properties of the bound states, such 
as localization lengths and binding energies, can be controlled 
externally by an electric field effect. Moreover, the band gap is
renormalized and impurity bands are created  at finite impurity
concentrations. Using the CPA we
calculate the electronic density of states and its dependence on the
applied bias voltage.
Many of the results we present here were previously reported in a brief
form in Ref.~[\onlinecite{Nilsson2007}].
A recent detailed study of the impurity states in the unitary limit 
in both biased and unbiased
bilayer graphene can be found in Ref.~[\onlinecite{Balatsky_tuning_2007}].

\subsection{Band model}
In this section, we review the properties of the
minimal model introduced in Eq.~(\ref{eq:bias_Hkin0bilayer}).
Throughout this section we use units such that $\vf = \hbar = 1$
unless otherwise specified. 
For numerical estimates we use $\tp=.35 \, \text{eV}$ and
insert the appropriate factors of $\vf = 3 t a /2$ with
$t=3 \, \text{eV}$ and $a=1.42 \, \text{\AA}$.
From Eq.~(\ref{eq:bias_Hkin0bilayer}) 
one finds two pairs of electron-hole symmetric eigenvalues:
\begin{multline}
  \label{eq:BGBi_eigenvalues}
  E_{\pm,s} =
\\ \pm \sqrt{ k^2 + \frac{V^2}{4} + \frac{\tp^2}{2} 
  + s \frac{1}{2}\sqrt{4 ( V^2 +\tp^2) k^2 + \tp^4 }} ,
\end{multline}
where $s=\pm$.
The lowest energy bands (with respect to the ``Dirac point'' at zero energy) 
representing the valence and conduction bands 
are the $E_{\pm,-}$ bands.
The smallest gap takes place at a finite wave vector given by
\begin{equation}
  \label{eq:BGBi_kgap}
  \kg = \frac{V}{2} \sqrt{\frac{V^2 + 2 \tp^2}{V^2 + \tp^2}}, 
\end{equation}
so that the size of the band gap becomes
\begin{equation}
  \label{eq:BGBi_Egap}
  \Eg = \frac{V \tp}{\sqrt{V^2 + \tp^2}}.
\end{equation}
At $k=0$ the distance between the valence and the conduction band is
given by the applied voltage difference $V$.
Note that V should in reality be taken to be not the bare applied
voltage difference but instead the self-consistently determined value
$\VMF$ as discussed in 
Refs.~[\onlinecite{McCann2006a,MacDonald_bilayergap_2006,Nilsson2006_BBB,Castro_PRL_2007}].
Near $\kg$ the energy of the quasi-particles in the conduction band
can be expanded as 
\begin{multline}
  \label{eq:BGBi_mexicanhat_band1}
   E_{+,-} \approx \frac{V \tp}{2 \sqrt{V^2 + \tp^2}} +
  \frac{V (V^2 + 2\tp^2)}{\tp (V^2 + \tp^2)^{3/2}}
  (k-\kg)^2 \\
 \equiv \frac{\Eg}{2} + \frac{(k-\kg)^2}{2 m^{*}},
\end{multline}
and as long as this approximation is valid the density of states 
per unit area is
\begin{equation}
  \label{eq:BGBi_DOS1}
  N(\om) = \frac{k_0}{\pi}\sqrt{\frac{2 m^*}{|\om| - \Eg /2 }},
\end{equation}
for $|\om| \geq \Eg /2$. This includes both the valley and the spin degeneracy.
Notice that the divergence of the density of states (DOS) at the band edge is
similar to what one would get in a truly 1D system.
The fact that a large DOS is accumulated near the band edge has
important consequences for the properties of the bound states as we
shall see in the following.

\subsection{Bare Green's function}

An explicit expression for the bare Green's function, 
which is given by $\GR^{0} = \bigl[ \om - \Hca_{BB}  \bigr]^{-1}$,
is:
\begin{widetext}
\begin{multline}
  \label{eq:BGBi_Green_Glory}
   \GR^0 =  \frac{1}{D} \left(
\begin{array}{ll}
  (\omega - V/2) \bigl[ (\omega + V/2)^2 - k^2 \bigr]
   & \bigl[ (\omega + V/2)^2-k^2 \bigr] k e^{i \phi }
\\
   \bigl[ (\omega +V/2)^2-k^2 \bigr] k e^{-i \phi }
   & (\omega - V/2) \bigl[ (\omega + V/2)^2 - k^2 \bigr]
    - ( \omega + V/2) \tp^2
\\
     \tp ( \omega^2 - V^2/4 ) 
   & \tp (\omega + V/2) k e^{i \phi }
\\
     \tp (\omega - V/2 ) k e^{i \phi } 
   &  \tp k^2 e^{2 i \phi }
\end{array}
\right.
\\
\left.
\begin{array}{ll}
     \tp \left(\omega ^2- V^2/4 \right)
   & \tp (\omega - V/2) k e^{-i \phi }
\\
     \tp (\omega + V/2) k e^{-i \phi } 
   & \tp k^2 e^{-2 i \phi }
\\
     ( \omega + V/2) \bigl[ (\omega - V/2)^2 - k^2 \bigr]
   & \bigl[ (\omega - V/2)^2 - k^2 \bigr] k e^{-i \phi }
\\
     \bigl[ (\omega - V/2)^2 - k^2 \bigr] k e^{i \phi }
   & (\omega + V/2 ) 
   \bigl[ (\omega - V/2 )^2 - k^2 \bigr] - (\omega - V/2)\tp^2
\end{array}
\right),
\end{multline}
\end{widetext}
where
\begin{eqnarray}
\label{eq:BGBi_Det1}
D &=& \bigl[ k^2 - V^2/4 - \om^2 \bigr]^2
+ M^4 ,
\\
\label{eq:BGBi_Mdef}
M^4 &=& \frac{V^2 \tp^2}{4} - \om^2 (V^2 + \tp^2 ).
\end{eqnarray}
So that, for example, the important diagonal components are given by
\begin{subequations}
  \label{eq:BGBi_G0diagonal1}
\begin{eqnarray}
  \GR^{0}_{\rA 1  \rA 1} &=& \frac{(\om - V/2) \bigl[
  (\om +V/2 )^2 -k^2 \bigr]}{D} ,
\\
  \GR^{0}_{\rB 1  \rB 1} &=& 
  \GR^{0}_{\rA 1  \rA 1} 
 - \frac{(\om + V/2)\tp^2}{D} .
\end{eqnarray}
\end{subequations}
The corresponding components for plane 2 are obtained by the
substitution $V \rightarrow -V$. Note that $M > 0$ inside of the gap.

\section{Bound states for Dirac delta potentials}
\label{sec:BGBi_bs_Dirac}
\begin{figure}[htb]
\includegraphics[scale=0.42]{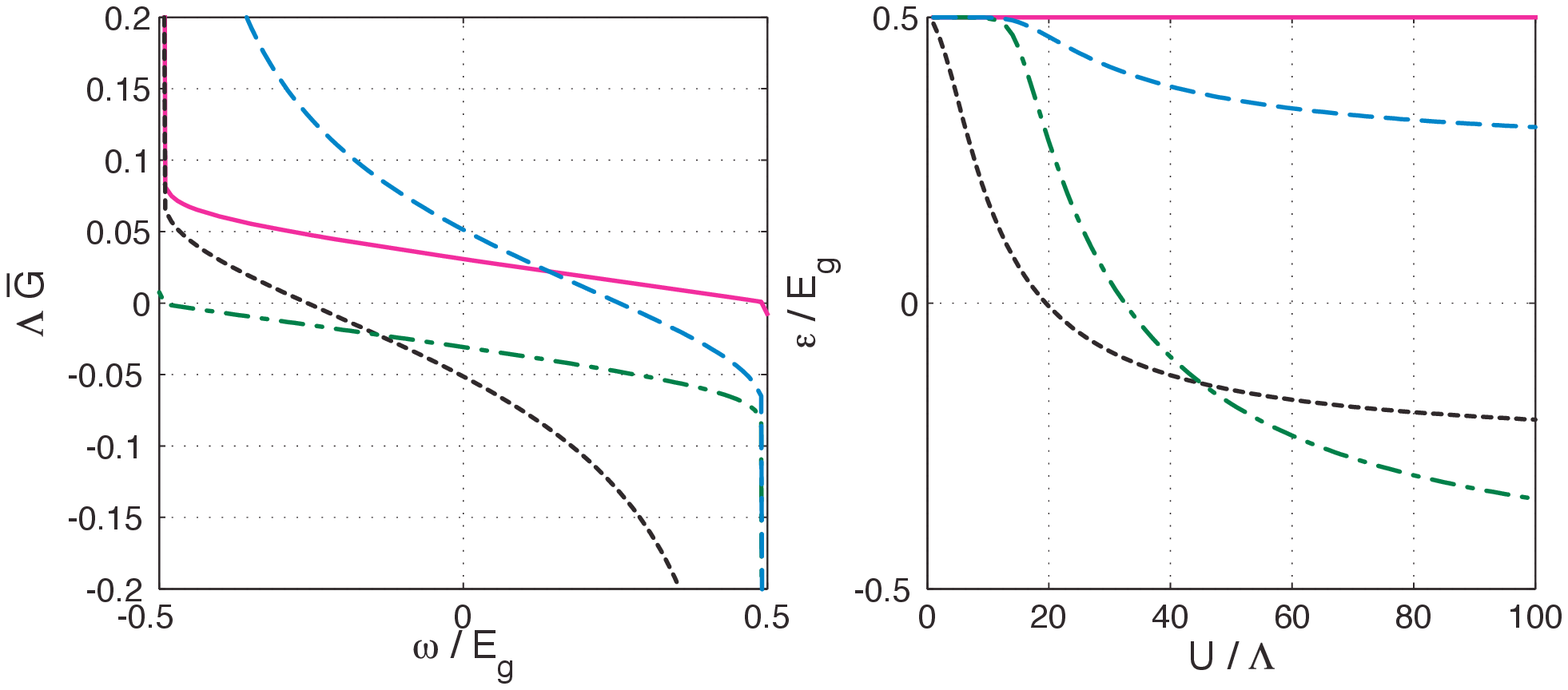}
  \caption
{[color online] 
Left: typical behavior of the $\overline{\GR}^0$'s:
$\overline{\GR}_{\rB 1}^{0}(\om)$ (dotted), 
$\overline{\GR}_{\rB 2}^{0}(\om)$ (dashed), 
$\overline{\GR}_{\rA 1}^{0}(\om)$ (solid), 
$\overline{\GR}_{\rA 2}^{0}(\om)$ (dash-dotted).
Here $V = 50 \, \text{meV}$.
The divergences close to the band edges are clearly visible.  
Right: bound state energies $\e$ for a local potential of strength $-U$,
the labeling of the sublattices is the same as to the left.
  }
  \label{fig_gbars}
\end{figure}

Bound states must be located inside of the gap so that their energies fulfill
$|\e| < \Eg/2 $, otherwise the asymptotic states at infinity are 
not exponentially localized.
If we decode a number (say $N_i$) of local impurities in a matrix of the form
\begin{equation}
  \label{eq:BGBi_12}
  \hat{V} = \text{Diag}[U_1, \, U_2 \,, \ldots , \, U_{N_i}],
\end{equation}
where we let $U_i$ denote the strength of the impurity potential 
that is located at site $\vx_i$.
The total Green's function is then given by
\begin{multline}
\label{eq:BGBi_Green_Tmat_impurity1}
  \GR = \GR^{0} + \GR^{0} \hat{V} \GR
  = \GR^{0} + \GR^{0} \hat{V} \GR^{0} 
  + \GR^{0} \hat{V} \GR^{0} \hat{V} \GR^{0} + \ldots
  \\
  = \GR^{0} + \GR^{0} 
       \bigl[ \hat{V} + \hat{V} \GR^{0} \hat{V}  
       + \hat{V} \GR^{0} \hat{V} \GR^{0} \hat{V} + \ldots \bigr]
  \GR^{0}
\\
  \equiv \GR^{0} + \GR^{0} T \GR^{0}.
\end{multline}
Here $T$ is the t-matrix of the system (see e.g. Ref.~[\onlinecite{JonesMarch2}]).
The interpretation of this expression is most transparent in the real
space picture, where it includes the repeated scattering off of all of the
impurities in every possible way.
Another way of expressing $T$ is 
(decomposing $\hat{V}$ as $\hat{V} = \sqrt{\hat{V}} \sqrt{\hat{V}}$):
\begin{multline}
  \label{eq:BGBi_Tmat_impurityMany}
  T = \sqrt{\hat{V}}( 1 + \sqrt{\hat{V}} \GR^{0} \sqrt{\hat{V}}
  + (\sqrt{\hat{V}} \GR^{0} \sqrt{\hat{V}})^2 + \ldots ) \sqrt{\hat{V}}
  \\ = 
  \sqrt{\hat{V}} 
  \frac{1}{1-\sqrt{\hat{V}} \GR^{0} \sqrt{\hat{V}}} \sqrt{\hat{V}}.
\end{multline}
Bound states generated by the impurities can readily be identified
by the possibly new poles in the full propagator of the system.
Therefore an equation that can be solved to find the energies of the
bound states of the system is given by
\begin{equation}
  \label{eq:BGBi_BSgeneral}
  \text{Det} \Bigl[ \delta_{i,j} - \sqrt{U_i}\GR^{0}_{ij}(\e)\sqrt{U_j}
  \, \Bigr]
  = 0.
\end{equation}
Here $\GR^{0}_{ij}(\e)$ denotes the (real space) 
propagator from site $j$ to site $i$. In principle one can put
in an arbitrary number of impurities in this expression.
However, if two impurities are located too close to 
each other the continuum approximation to the propagators is not
expected to be accurate and one must instead work with the full
tight-binding description 
(see Ref.~[\onlinecite{Wehling2006}] for an illustration of this approach
in monolayer graphene).
If we specialize to one local impurity affecting only one site the calculations
simplify considerably. The Fourier transform of the local potential
is $U / N$ (where $N$ is the number of unit cells in the system)
so that we can write
\begin{equation}
  T = \frac{1}{N} \frac{U}{1-U \overline{\GR}^{0}}.
\end{equation}
As mentioned previously, 
to locate the bound states we must find possible new poles due to
the potential. Explicitly we need $U \overline{\GR}^{0}(\e) = 1$.
Like in Section~\ref{sec:impurities_tmatrix}, $\overline{\GR}^{0}$
is the local propagator at the impurity site that is given by 
the expression in Eq.~(\ref{eq:BGBi_Gbar1}) with ${\GR}^{0}$
taken from Eq.~(\ref{eq:BGBi_Green_Glory}).
Using Eq.~(\ref{eq:BGBi_G0diagonal1}) we can perform the 
integrals exactly in the continuum approximation with the result
\begin{subequations}
\begin{multline}
  \label{eq:BGBi_Gbars0}
  \overline{\GR}_{\rA 1}^{0}  = \frac{ V/2 - \om}{2 \Lambda^2} 
  \Bigl\{
  \log \Bigl( \frac{\Lambda^4}{M^4 + ( V^2 / 4 + \om^2)^2} \Bigr) 
\\ 
-  \frac{ 2 \om V}{M^2}
  \Bigl[ \tan^{-1} \Bigl( \frac{ V^2 / 4 + \om^2}{M^2} \Bigr) +
  \tan^{-1} \Bigl(\frac{\Lambda^2}{M^2} \Bigr) \Bigr]
  \Bigr\},
\end{multline}
%
\begin{multline}
  \overline{\GR}_{\rB 1}^{0}  = \overline{\GR}_{\rA 1}^{0} 
  - \frac{({V/2} + \om) \tp^2}{ M^2 \Lambda^2} 
  \Bigl[ \tan^{-1} \Bigl(\frac{\Lambda^2}{M^2} \Bigr)
\\ +
 \tan^{-1} \Bigl( \frac{ V^2 / 4 + \om^2}{M^2} \Bigr)
   \Bigr],
\end{multline}
\end{subequations}
where $M^2 =  \sqrt{ V^2 \tp^2 / 4 - \om^2 ( V^2 + \tp^2 )}$,
and $\Lambda$ ($\approx 7$ eV) is the high energy cutoff.
The corresponding expressions in plane 2 are obtained by the
substitution $ V \rightarrow - V$.
The typical behavior of $\overline{\GR}^{0}(\om)$ as a function of
the frequency $\om$ is shown in Fig.~\ref{fig_gbars}.

\begin{figure}[htb]
\includegraphics[scale=0.43]{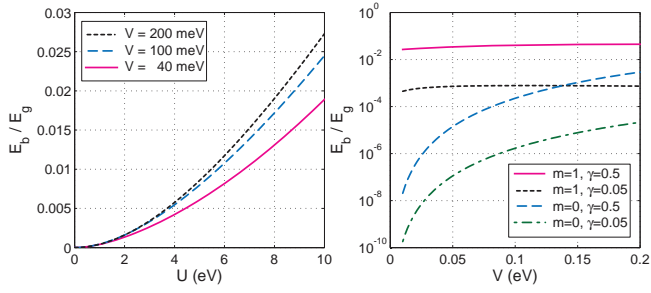}
\caption
{[color online] Left: Bound state binding energies $\Eb$ 
(in units of the gap $\Eg$) for a Dirac delta potential 
of strength $-U$ for different bias $V$. 
Right: Binding energies of a potential well of range $R=10 \, a$ and strength
  $\gamma = \gamma_1 = \gamma_2$ (see the text in Sec.~\ref{sec:potential_well}) 
  for different angular momentum $m$
  channels and external bias potential $V$. 
}
\label{fig:boundstates1}
\end{figure}

From this we conclude that a Dirac delta potential always 
generates a bound state (no matter how weak the potential is)
since $\overline{\GR}^{0}$ diverges as the band edge 
is approached (where $M \rightarrow 0$).
The dependence on the cutoff (except for the overall scale) is
rather weak so that the linear in-plane approximation to the spectrum
should be a good approximation as in the case of
graphene.\cite{Wehling2006}
For a given strength of the potential $U$, there are four different
bound state energies depending on which lattice site it is sitting on.
In Fig.~\ref{fig_gbars} we show the energies of these bound states 
for strong impurities. Even at these scales the bound state coming
from $\overline{\GR}_{\rA 1}^{0}$ is so weakly bound that it is barely
visible in the figure.
In Fig.~\ref{fig:boundstates1} we show the binding energy as a function 
of $U$ and $V$ for the deepest bound state
(coming from $\overline{\GR}_{\rB 1}^{0}$).
In the limit of $U \rightarrow \infty$ the electron-hole symmetry of
the bound state energies is restored.
For illustrative purposes we consider only attractive potentials
in this work, analogous results hold for repulsive potentials
because of the electron-hole symmetry of the model that we
are using.
For smaller values of the potential ($|U| \ll \Lambda$) the binding
energy measured from the band edge $\Eb = \Eg/2 - \e$ 
grows as $U^2$ and the states are weakly bound.
For example, for $ V = 40 \, \text{meV}$ and
$U \lesssim 1 \text{eV}$ one finds
$\Eb \lesssim 4 \times 10^{-4} \Eg$.

\subsection{Angular momentum states}
\label{sec:angularmomentum}
For any potential with cylindrical symmetry it is useful to
classify the eigenstates according to their angular momentum 
$m$.
In the presence of the ``trigonal distortion''
parametrized by $\gamma_3$ the calculations become more involved
because of the broken cylindrical symmetry. We discuss
this issue briefly in Sec.~\ref{sec:BGBi_trigonal}.
The real space continuum version of the Hamiltonian matrix in
Eq.~(\ref{eq:bias_Hkin0bilayer})
that includes a potential, which in general is
allowed to take on different values in the two planes, is 
\begin{equation}
  \label{eq:BGBi_Hkin0bilayer_realspace}
  \Hca_0 =
  \begin{pmatrix}
    V/2 + g_1(r) -i\sigma^{*} \cdot \PD
    & \tp (1+\sigma_z)/2 \\
    \tp(1+\sigma_z)/2 & 
    -V/2 + g_2(r) -i\sigma \cdot \PD
  \end{pmatrix}.
\end{equation}
Here $\sigma_i$ ($i=x,y,z$) are the usual Pauli matrices.
For example, a symmetric Coulomb problem could then be approximated by
taking $g_1(r) = g_2(r) = g/r$.
Going to cylindrical coordinates the derivatives transforms according to
\begin{equation}
  \PD_{x} \pm i \PD_y = e^{\pm i \varphi}(\PD_r \pm \frac{i}{r}\PD_{\varphi}),
\end{equation}
where we use the coordinate convention $x \pm i y = r e^{\pm i \varphi}$.
For the Hamiltonian in Eq.~(\ref{eq:BGBi_Hkin0bilayer_realspace}) one
can now -- in analogy with the usual Dirac equation~\cite{mele} --
construct an angular momentum operator that commutes with the
Hamiltonian.
The angular ($\varphi$) dependence of the angular momentum $m$ eigenstates
are those of the vector:
\begin{equation}
    \label{eq:BGBi_vecm}
     u_{\alpha,m}(\varphi) = e^{i m \varphi}
     \begin{pmatrix}
       1  \\
       e^{-i \varphi} e^{-i \alpha \pi/2} \\
       1  \\
       e^{i \varphi} e^{i \alpha \pi/2}
     \end{pmatrix},
  \end{equation}
where parameter $\alpha$ is introduced for later
convenience, it is used later to obtain more compact expressions.
It is convenient to define the following ``star'' product of two vectors
that results in  another vector with components given by
\begin{equation}
  \label{eq:BGBi_starproduct}
  [\, a \star b \,]_{\alpha j} = a_{\alpha j} \, b_{\alpha j}.
\end{equation}
By writing $\Psi = u_{0,m} \star \psi(r) / \sqrt{r}$
the eigenvalue problem $\Hca_0 \Psi = E \Psi$ is transformed
into a set of coupled ordinary linear differential equations
for the radial wave-function $\psi(r)$:
\begin{widetext}
\begin{equation}
  \label{eq:BGBi_DiffE1}
    \begin{pmatrix}
    g_1(r) + V/2 & -i \partial_r + i \, j /r & \tp & 0 \\
    -i \partial_r -i \, j /r & g_1(r) + V/2  & 0 & 0 \\
    \tp & 0 & g_2(r) - V/2 & -i \partial_r -i \,(j+1)/r \\
    0 & 0 &  -i \partial_r +i \, (j+1)/r & g_2(r) - V/2
  \end{pmatrix}
  \psi(r) = E \psi(r).
\end{equation}
\end{widetext}
Here we have introduced $j=m-1/2$ to render the equations more
symmetric.
If the potential generates bound states inside of the gap these states 
decay exponentially $\sim r^{\gamma} e^{-\kappa r}$ as 
$r \rightarrow \infty$.
Assuming that the potential decays fast enough
the asymptotic behavior of Eq.~(\ref{eq:BGBi_DiffE1}) imply that
the allowed values for $\kappa$ are $\kappa_{\pm}$ satisfying
\begin{subequations}
  \label{eq:BGBi_kappadef}
\begin{eqnarray}
  \kappa_{\pm}  &=& \sqrt{-(\e^2 + V^2/4) \pm i M^2} , \\
  |\kappa|^4 &=& (V^2/4 - \e^2) (V^2/4 - \e^2 + \tp^2) , \\
  \label{eq:BGBi_kappadef2}
  \kappa_{\pm} &=& |\kappa| \exp \Bigl\{ \pm i \bigl[ \frac{\pi}{2} -
  \frac{1}{2} \tan^{-1} \Bigl(\frac{M^2}{\e^2 + V^2/4}\Bigr)  \bigr] \Bigr\}.
\end{eqnarray}
\end{subequations}
So that, for weakly bound states we have
\begin{equation}
  \label{eq:BGBi_kappadef_weak}
  \kappa_{\pm} \approx \frac{M^2}{\sqrt{V^2 + \Eg^2}} 
  \pm i \frac{1}{2}\sqrt{V^2 + \Eg^2},
\end{equation}
leading to a localization length
\begin{equation}
  \label{eq:BGBi_lloc}
  l \sim \frac{2 \kg}{V \tp} \sqrt{\frac{\Eg}{\Eb}},
\end{equation}
that diverges as the band edge is approached and decreases as
the bias voltage increases.

\subsection{Free particle wave functions in the angular momentum basis}
\label{sec:Angular_momentum_eigenstates}
The free particle wave functions in the angular momentum basis can be
conveniently expressed in terms of the following vectors:
\begin{equation}
  \label{eq:BGBi_vvecdef}
  v_{Z,m}(z) =
  \begin{pmatrix}
    Z_{m}(z) \\
    Z_{m-1}(z) \\
    Z_{m}(z) \\
    Z_{m+1}(z)
  \end{pmatrix},
\end{equation}
\begin{equation}
    \label{eq:BGBi_wecdef}
      w(p) =
  \begin{pmatrix}
    \bigl[ (\om + {V/2})^2 - p^2 \bigr] (\om - {V/2}) \\
    \bigl[ (\om + {V/2})^2 - p^2 \bigr] p \\ 
    \tp (\om^2 - V^2 / 4 ) \\ 
    \tp (\om - {V/2}) p
  \end{pmatrix}.
  \end{equation}
The last vector is useful as long as $\om \neq {V/2}$
(cf.\ the discussion of the two eigenvectors in 
Ref.~[\onlinecite{Nilsson2006_BBB}].)
The denominator (actually a determinant) that 
determines the eigenstates is:
\begin{equation}
  \label{eq:BGBi_Detdef}
  D(p,\om) = \bigl[ p^2 - V^2 / 4 - \om^2 \bigr]^2
 + V^2 \tp^2 / 4 - \om^2 ( V^2 + \tp^2 ).
\end{equation}
Then, provided that $D(k,\om) = 0$, ($k>0$) which corresponds to
propagating modes, the eigenfunctions are proportional to:
\begin{equation}
  \label{eq:BGBi_PsiZ}
  \Psi_{Z,m}(\om,k,r) = u_{1,m}  \star v_{Z,m}(k r) \star w(k),
\end{equation}
where $Z_{m}(x) = J_{m}(x)$ or $Y_{m}(x)$ are Bessel functions and the star
product is defined in Eq.~(\ref{eq:BGBi_starproduct}).
If on the other hand $D(i \kappa, \om) = 0$, ($\text{Re}[\kappa] >0$) 
the eigenfunctions are instead:
  \begin{eqnarray}
  \label{eq:BGBi_PsiKI}
  \Psi_{K,m}(\om,\kappa, r) &=& 
  u_{0,m} 
  \star v_{K,m}(\kappa r)
  \star w(i \kappa ) ,
\\
  \Psi_{I,m}(\om,\kappa,r ) &=&
  u_{0,m} 
  \star v_{I,m}(\kappa r)
  \star w(-i \kappa ) ,
\end{eqnarray}
with $I_{m}(x)$ and $K_{m}(x)$ being modified Bessel functions.
That these vectors are indeed free-particle eigenstates can be verified
straightforwardly by applying the Hamiltonian in 
Eq.~(\ref{eq:BGBi_Hkin0bilayer_realspace}) with $g_1 = g_2 = 0$ to them.

\subsection{Local impurity wave functions}
The general expression for the retarded Green's function is
\begin{equation}
  \label{eq:BGBi_Green_exp}
  \GR_{\alpha j , \alpha' j'}(\vx,\vx') = \sum_{n}
  \frac{
  \la \alpha j ,\vx| n \ra  \la n | \alpha' j',\vx' \ra
  }  {\om +i \delta - E_n} 
  ,
\end{equation}
where the sum is over the eigenstates $|n\ra$ 
(with eigenenergy $E_n$) of the system.
Comparing the coefficient of the poles in this expression with
those in Eq.~(\ref{eq:BGBi_Green_Tmat_impurity1}) one can read of the
wave functions of the bound states directly. The result is that the
wave functions are Fourier transforms of the columns of the bare
Green's evaluated with the frequency set to be equal to the energy of
the bound state.
\subsubsection{Impurity on an A1 site}
When the impurity is on an A1 site the expression becomes
[using Eq.~(\ref{eq:BGBi_Green_Glory})]
\begin{equation}
  \label{eq:BGBi_psiA1_1}
  \begin{pmatrix}
    \GR_{\rA 1 \rA 1} \\  \GR_{\rB 1 \rA 1} 
\\  \GR_{\rA 2 \rA 1} \\  \GR_{\rB 2 \rA 1}
  \end{pmatrix}
  =
 \frac{1}{N} \sum_{\vk} \frac{e^{i \vk \cdot \vx}}{D}
 \begin{pmatrix}
   (\om - V/2) \bigl[ (\om + V/2)^2 - k^2 \bigr] \\
   \bigl[ (\om + V/2)^2 - k^2 \bigr] k e^{\mp i \phi}\\
   \tp (\om^2 - V^2/4) \\
   \tp (\om - V/2) k e^{\pm i \phi }.
 \end{pmatrix}.
\end{equation}
Performing the angular average one ends up with Bessel functions:
\begin{equation}
  \label{eq:BGBi_psiA1_2}
  \begin{pmatrix}
    \GR_{\rA 1 \rA 1} \\  \GR_{\rB 1 \rA 1} 
\\  \GR_{\rA 2 \rA 1} \\  \GR_{\rB 2 \rA 1}
  \end{pmatrix}
  =
  \int_0^{\Lambda} \frac{k dk}{\Lambda^2 D}
 \begin{pmatrix}
   (\om - V/2) \bigl[ (\om + V/2)^2 - k^2 \bigr] J_0 (k r) \\
   i \bigl[ (\om + V/2)^2 - k^2 \bigr] k 
   J_1 (k r) e^{-i \varphi} \\
   \tp (\om^2 - V^2/4) J_0 (k r) \\
   i \tp (\om - V/2) k  J_1 (k r) e^{i \varphi}
 \end{pmatrix}.
\end{equation}
There are really two such terms, one for each K-point, 
which corresponds to the different signs of the phases in 
Eq.~(\ref{eq:BGBi_psiA1_1}). 
Note that this state has angular momentum $m=0$ in the language 
of the Section~\ref{sec:angularmomentum}.
The $k$-integral can be performed analytically (taking
$\Lambda \rightarrow \infty$ in the integration limit).
Using $\kappa_{\pm}$ defined in Eq.~(\ref{eq:BGBi_kappadef})
we obtain
\begin{subequations}
\begin{eqnarray}
  \label{eq:BGBi_psiA1_3}
    \GR_{\rA 1 \rA 1} &=& \frac{V/2 - \om}{2} 
    \sum_{\alpha = \pm } 
    \Bigr\{ 1 - i \alpha \frac{V \om}{M^2} \Bigr\} 
    K_0(\kapa r) ,
    \\ 
    \GR_{\rB 1 \rA 1} &=& \frac{-i}{2} 
    \sum_{\alpha = \pm } 
    \Bigr\{ 1 - i \alpha \frac{V \om}{M^2} \Bigr\} 
       \bigl[ \kapa K_1(\kapa r)  \bigr]
       e^{-i \varphi} ,
    \\ 
    \GR_{\rA 2 \rA 1} &=& 
    \frac{-i \tp (V^2/4 - \om^2)}{2 M^2 }
    \sum_{\alpha = \pm } 
    \bigl[ \alpha K_0(\kapa r)  \bigr] ,
    \\ 
    \GR_{\rB 2 \rA 1} &=& 
    \frac{\tp(V/2 - \om)}{2 M^2 }
    \sum_{\alpha = \pm } 
    \bigl[ \alpha \kapa K_1 (\kapa r) \bigr] 
    e^{i \varphi} .
\end{eqnarray}
\end{subequations}
These propagators can also be easily expressed in terms of the
free-particle wave functions:
\begin{equation}
  \label{eq:BGBi_PsiA1}
  \GR_{\alpha j,\rA 1} = \Bigr[
  \frac{ \Psi_{K,0}(\e,\kappa_{+}, r) 
       - \Psi_{K,0}(\e,\kappa_{-}, r) }
       {-i 2 M^2 \Lambda^2}  
  \Bigr]_{\alpha j}.
\end{equation}
This property is not a coincidence since the particles
are essentially free, except for the potential that acts on
the single impurity site at the origin.
\subsubsection{Impurity on a {\rm B1} site}
When there is an impurity on a B1 site one can perform the same
calculation with the result that the wave function becomes
\begin{equation}
  \label{eq:BGBi_PsiB1}
  \GR_{\alpha j,\rB 1} = \Bigl[
  \frac{ \kappa_{+} \Psi_{K,1}(\e,\kappa_{+}, r) 
         - \kappa_{-} \Psi_{K,1}(\e,\kappa_{-}, r) }
       {2 M^2 \Lambda^2 ({V/2}-\e)}
       \Bigr]_{\alpha j}.
\end{equation}
This state has angular momentum $m=1$ in the language of
Sec.~\ref{sec:angularmomentum}.
A similar expression can be obtained when there is an impurity 
on an $\rA 2$ ($\rB 2$) site, where in this case the corresponding
state has angular momentum $m=0$ ($m=-1$).

\subsection{Asymptotic behavior}

The asymptotic behavior of the modified Bessel functions is
$K_{n}(z) \sim \exp(-z)/\sqrt{z}$ as $z \rightarrow \infty$.
Therefore the bound states are exponentially localized 
within a length scale given by [See Eq.~(\ref{eq:BGBi_kappadef})]
\begin{multline}
  \label{eq:BGBi_Lloc}
  l =  \bigl\{ \real ( \kappa_{\pm} ) \bigr\}^{-1}
\\ =
  \biggl( \sqrt[4]{(\e^2+V^2/4)^2+M^4} 
  \sin \Bigr\{ \frac{1}{2} \tan^{-1}
  \Bigl[ \frac{M^2}{(\e^2+V^2/4)^2} \Bigr] \Bigl\} \biggr)^{-1} \\
  \approx \frac{2 \kg}{V \tp} \sqrt{\frac{2}{1 - 2 |\e| / \Eg}},
\end{multline}
where the last line is applicable for weakly bound states close to the
band edge. This is in agreement with the general results above in 
Eq.~(\ref{eq:BGBi_lloc}).
At short distances one may use that $K_{n}(z) \sim 1/z^n$ for $n\geq
0$ and $K_0(z) \sim -\ln z$ to conclude that the 
wave functions are not normalizable in the continuum.
In particular, for an impurity on the $\rA 1$ ($\rB 1$) 
site the wave function on the $\rB 1$ ($\rA 1$) site diverges as $1/r$.
This divergence is however rather weak (i.e., logarithmic) and not real 
since in a proper treatment of the short-distance physics, the divergence 
is cutoff by the lattice spacing $a$ (this is equivalent to cutting of the
$k$-integral in Eq.~(\ref{eq:BGBi_psiA1_2}) at $k=\Lambda$
instead of taking $\Lambda \rightarrow \infty$).

\section{Simple criterion}
\label{sec:simple_estimate_nc}
Using the asymptotic form of the wave functions one can approximate the
wave function as:
\begin{equation}
  \label{eq:BGBi_crit1}
  \psi \sim \frac{A}{\sqrt{2r}}e^{-\kappa r},
\end{equation}
in each plane. Normalization then requires that 
$A = \sqrt{\kappa' /\pi}$, where $\kappa'$ is the real part of
$\kappa_{\pm}$. 
Thus one impurity is interacting with approximately
\begin{equation}
  \label{eq:BGBi_crit2}
  N_{l} = \la \pi r^2 \ra / \frac{3 \sqrt{3}a^2}{4}
\end{equation}
atoms per plane.
For an impurity density of $n_i$, the number of impurities interacting
with a given impurity is given by
\begin{equation}
  \label{eq:BGBi_crit3}
  N_i = \frac{\pi \sqrt{3}}{2} \Bigl( \frac{t}{\kappa'} \Bigr)^2 \nimp.
\end{equation}
A simple estimate of the critical density $n_c$ above which the
interaction between different impurities are important is then given
by $N_i \sim 1$. Writing $\e  = \Eg/2 - \Eb$ one then finds the
following criterion for overlap of impurity wave functions (assuming weakly binding impurities):
\begin{equation}
  \label{eq:BGBi_crit4}
  \nimp \gtrsim n_c = 
  \frac{1}{2 \pi \sqrt{3}} 
  \Bigl( \frac{V \tp}{\kg t} \Bigr)^2 \frac{\Eb}{\Eg}
  \approx 2.5 \times 10^{-3} \frac{\Eb}{\Eg},
\end{equation}
indicating that the critical density increases with the applied gate
voltage. The last step is valid for $V \ll \tp$.
Taking $U \lesssim 1\, \text{eV}$ we found in 
Section~\ref{sec:BGBi_bs_Dirac} that
$\Eb \lesssim 4 \times 10^{-4} \Eg$,
leading to $n_{c} \sim 10^{-6}$.
Hence, even tiny concentrations of impurities lead to wave function overlap.
This result shows that even a small amount of impurities can have strong
effects in the electronic properties of the BGB.

\section{Variational calculations}
\label{sec:variational}
For general potentials it is not possible to solve for the
bound states in closed form. Nevertheless, for estimates and
to gain intuition about the problem 
it is fruitful to study the problem with
variational techniques. In this section we consider two
different variational approaches.

\subsection{Variational calculation I}
\label{sec:variational_I}
Using Eq.~(\ref{eq:BGBi_DiffE1}) one can show the existence of bound states
variationally.
For simplicity we consider only the case $m=0$ ($j=-1/2$)
and a symmetric potential $g_1 = g_2 = g(r)$.
We use the simple trial wave function $\psi_2 = A \exp(-k r/2)$.
The following integrals are useful in the process:
\begin{subequations}
\begin{eqnarray}
  \label{eq:BGBi_varintegrals1}
  \int_{0}^{\infty} |\psi_2|^2  dr &=& 1 ,
\\
  \int_{a}^{\infty} \frac{|\psi_2|^2}{r}  dr &=& k E_1(ka) ,
\\
  \int_{0}^{\infty} \Theta(R-r) |\psi_2|^2  dr &=& 1-\exp(-k R).
\end{eqnarray}
\end{subequations}
Here $a$ is a cutoff on the order of the lattice spacing needed to
regularize the integral. $E_1(x) = \int_{x}^{\infty}dr \exp(-r)/r$ 
is an exponential integral (see e.g. Ref.~[\onlinecite{Gradshteyn}]).
The eigenvalues ($\e_k$) of the kinetic term is given by the equation
\begin{equation}
  \label{eq:BGBi_Variational_kin111}
  0 = \Bigl\{  {\e}^2 + \frac{V^2}{4}  
  -\frac{k^2}{4} \bigl[E_1(ka)^2 -1 \bigr] \Bigr\}^2
  + \frac{V^2 \tp^2}{4} - {\e}^2 (V^2 + \tp^2 ),
\end{equation}
which is the same as the equation for the 
bare bands [cf.\ Eq.~(\ref{eq:BGBi_Det1})]
 with the substitution
$k^2 \rightarrow k^2 [ E_1(ka)^2 -1]/4$.
Provided that $ka \lesssim 0.26$ ($k \lesssim 1.2 \, \text{eV}$)
the new ``momentum'' is real. 
For an attractive potential we may then construct a wave packet
corresponding to the $E_{+,-}$ band leading to a positive 
contribution from the kinetic term.

\begin{figure}[htb]
  \includegraphics[scale=0.42]{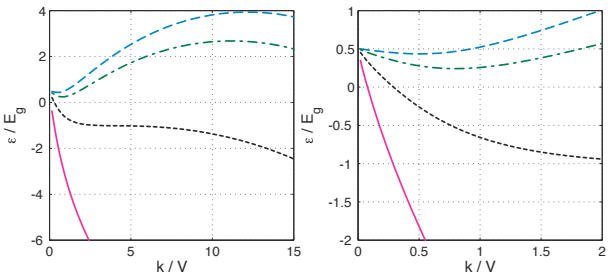}
  \caption{[color online]
    Variational energy for a Coulomb potential
    as a function of the variational parameter $k$
    for $V = 50 \, \text{meV}$. 
    From top to bottom: $\alpha = .033$, $.1$, $.33$, and  $1$.
    Left: large view; Right: zoom in for small $k$.
}
\label{fig:Variational1}
\end{figure}

\begin{figure}[htb]
  \includegraphics[scale=0.42]{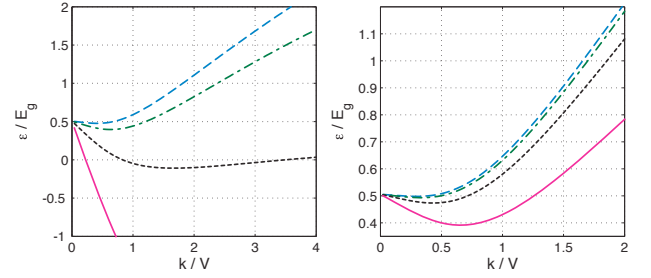}
  \caption{[color online] 
 Variational energy for a short-range potential of range $R$ 
as a function of the variational parameter $k$ for $V = 50 \, \text{meV}$. 
From top to bottom: $U = .033$, $.1$, $.33$, and  $1$ eV.
Left: $R = 10 \, a$; Right: $R = 1 \,a$.}
\label{fig:Variational2}
\end{figure}

We consider two types of potentials: 
one of the Coulomb type, $g_C = -\alpha/r$, 
characterized by the dimensionless strength $\alpha$; 
and a local potential, $g_L = -U \Theta(R-r) $, 
characterized by the strength $U$ and the range $R$.
The total variational energies for the two types of potentials are:
\begin{subequations}
  \begin{eqnarray}
    \label{eq:BGBi_Variational_E1}
    E_{\text{var},C} &=& E_{+,-}\Bigl(\frac{k}{2} \sqrt{E_1(ka)^2 -1}\Bigr) 
    - \alpha k E_1(ka) , \\
    E_{\text{var},L} &=& E_{+,-}\Bigl(\frac{k}{2} \sqrt{E_1(ka)^2 -1}\Bigr) 
    - U [ 1-e^{-k R} ].
  \end{eqnarray}
\end{subequations}
Some typical results obtained from these expression are shown in 
Figs.~\ref{fig:Variational1} and \ref{fig:Variational2}.
For a sufficiently strong potential it is favorable for the state to
become very localized close to the impurity, and the  
assumed ``bound state'' is located inside of the continuum
of the valence band.
This is problematic as
it leads to a breakdown of the picture of a bound
state coming only from the states in the $E_{+,-}$ band.
The state can no longer be considered a true bound state
since it is allowed to hybridize with the states in the valence band
and hence leak away into infinity. This state can therefore only be 
regarded as a resonance.
%
Nevertheless, for weak Coulomb potentials there are indeed bound
states inside of the gap, and for short-range potentials the
variational treatment give results that are consistent with the more rigorous
study coming up in Section~\ref{sec:potential_well}.


\subsection{Variational calculation II}
\label{sec:variational_II}
Another simple variational approach is to construct a
wave packet with angular momentum $m$ and a momentum close 
to $k_g$ from the $E_{+-}$ band using the free particle
solutions in Eq.~(\ref{eq:BGBi_PsiZ}) according to:
\begin{equation}
  \label{eq:BGBi_Variational_state_II}
  \Psi_{\text{var}}(\xi) = \int_{k_g-\xi}^{k_g+\xi} dp \,
  \Psi_{J,m}(E_{+-}(p),p,r) \, \sqrt{\frac{p}{4 \pi \xi}},
\end{equation}
where we assume that $\xi \ll \kg$. The factor $\sqrt{p/\xi}$ is
included to generate a properly normalized variational state.
Since this state is built up of eigenstates of the kinetic term
the contribution from the kinetic energy to the variational energy
becomes:
\begin{equation}
  \label{eq:BGBi_Variational_E_II_k}
  E_{\text{var,kin}} \approx \frac{1}{\xi} \int_{k_g-\xi}^{k_g+\xi} dp
  \frac{p^2}{2 m^*} = \frac{\xi^2}{3 m^{*}},
\end{equation}
using Eq.~(\ref{eq:BGBi_mexicanhat_band1}). The leading contribution
to the interaction energy for small $\xi$ becomes:
\begin{multline}
  \label{eq:BGBi_Variational_E_II_i}
  E_{\text{var,int}} = \int d^2\vx 
  \int_{k_g-\xi}^{k_g+\xi} dp
  \int_{k_g-\xi}^{k_g+\xi} dp' 
\\ \times
  \frac{\sqrt{p \, p'}}{{4 \pi \xi}}
  \Psi^{\dag}_{J,m}(E_{+-}(p),p,r) g(r) \Psi_{J,m}(E_{+-}(p'),p',r)
\\
  \approx 
  2 \xi \kg \int dr r 
  \Psi^{\dag}_{J,m}(E,\kg,r) g(r)
  \Psi_{J,m}(E,\kg,r)\Bigr|_{E = E_{+-}(\kg)}
\\
  \equiv - 2 \xi \kg \frac{U}{\widetilde{R}^2}.
\end{multline}
Therefore, the variational calculation shows that 
{\it for any} $m$, a weak attractive potential of strength 
$\propto U$ leads to a weakly bound state with binding energy
$\Eb \propto U^2$.
This can be understood by noting that for each value of $m$, due to
$\kg$ being nonzero, the problem maps into a 1D system with an
effective local potential. It is well known (see e.g. Ref.~[\onlinecite{LL_quantum}])
that in 1D a weak
attractive potential ($\propto U$) always leads to a bound state with
binding energy $\Eb \propto U^2$.
Thus the result is a direct
consequence of the peculiar topology of the BGB band edge --
see however Sec.~\ref{sec:BGBi_trigonal}.

\section{Potential well}
\label{sec:potential_well}

For the case of the simple local ``potential well'' defined by
the potentials
$g_{Lj} = -U_j \Theta(R-r) \equiv -\gamma_j \Theta(R-r) /R$ 
it is possible to
make analytic progress with the continuum problem.
In Sec.~\ref{sec:Angular_momentum_eigenstates} we gave the
explicit form of the eigenstates for a constant potential in the
angular momentum basis.
Bound states are possible when the two solutions for $r<R$ and the two
solutions for $r>R$ are not linearly independent at $r=R$.
This can be tested by evaluating the $4 \times 4$ 
determinant of the matrix built up by the four eigenstate spinors.
Given $U_1$, $U_2$ and $R$ the resulting determinant is a function of
the energy $\om$.
Zeros of the determinant inside of the band gap corresponds to
the bound states that we are searching for.
Inside of the potential region the effective frequency and bias
are given by:
\begin{subequations}
\begin{eqnarray}
  \widetilde{\om} &=& \om + (U_1+U_2)/2 , \\
  \widetilde{V} &=& V + ( U_2 - U_1 ) .
\end{eqnarray}
\end{subequations}
The determinant is given by one of the following expressions:
\begin{widetext}
\begin{eqnarray*}
  \label{eq:BGBi_finterange_det}
  D_0(\om) &=& \text{Det} \bigl[
  \Psi_{K,m}(\om,\kappa_{+},R), \,
  \Psi_{K,m}(\om,\kappa_{-},R), \,
  \Psi_{I,m}(\widetilde{\om},\widetilde{\kappa}_{+},R), \,
  \Psi_{I,m}(\widetilde{\om},\widetilde{\kappa}_{-},R) \bigr],
\\
  D_1(\om) &=& \text{Det} \bigl[
  \Psi_{K,m}(\om,\kappa_{+},R), \,
  \Psi_{K,m}(\om,\kappa_{-},R), \,
  \Psi_{J,m}(\widetilde{\om},p_{+},R), \,
  \Psi_{I,m}(\widetilde{\om},\widetilde{p}_{-},R) \bigr],
\\
  D_2(\om) &=& \text{Det} \bigl[
  \Psi_{K,m}(\om,\kappa_{+},R), \,
  \Psi_{K,m}(\om,\kappa_{-},R), \,
  \Psi_{J,m}(\widetilde{\om},p_{+},R), \,
  \Psi_{J,m}(\widetilde{\om},p_{-},R) \bigr],
\end{eqnarray*}
\end{widetext}
depending on whether there are zero, one or two propagating
modes at the chosen energy inside of the potential region.
Here, 
\begin{eqnarray}
  \label{eq:BGBi_QW_1}
  \kappa_{\pm} &=& \sqrt{-(\om^2 + V^2/4) \pm i M^2} ,
\\  \label{eq:BGBi_QW_2}
  p_{\pm} &=& 
  \sqrt{ ( \widetilde{\om}^2 + \widetilde{V}^2 /4 ) 
    \pm \sqrt{-\widetilde{M}^4}} ,
\\  \label{eq:BGBi_QW_3}
  \tilde{p}_{-} &=& \sqrt{
    \sqrt{-\widetilde{M}^4} - ( \widetilde{\om}^2 + \widetilde{V}^2 /4)
    } ,
\end{eqnarray}
where $\widetilde{M}$ is given by Eq.~(\ref{eq:BGBi_Mdef})
with the substitutions
$V \rightarrow \widetilde{V}$ and
$\om \rightarrow \widetilde{\om}$.

By monitoring the zeros of $D_{n}$ as a function of the radius $R$
and the strengths $\gamma_j$ we have studied the 
binding energies and find that the deepest
bound states are in one of the angular momentum channels $m=0,\pm 1$
for a substantial parameter range.
Since these types of states are also present for the Dirac delta potential
we argue that the physics of short-range potentials can be 
approximated (except for the short-distance physics)
by Dirac delta potentials with a strength tuned to give the 
correct binding energy.
A typical result for the binding energies is shown in
Fig.~\ref{fig:boundstates1}.

\begin{figure}[htb]
  \includegraphics[scale=0.42]{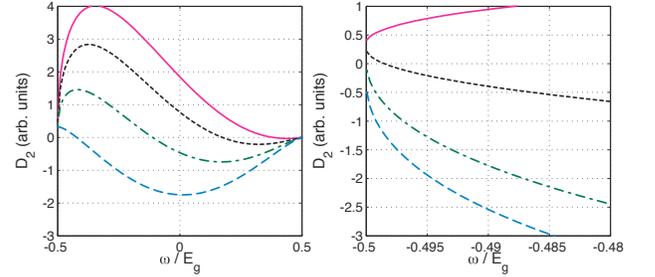}
  \caption{[color online] Plot of $\imag[D_2(\om)]$ as a function of the frequency $\om$
  inside of the band gap for $m=0$, $r=10\, a$, $V = 50 \, \text{meV}$ 
  and $\gamma_1 = \gamma_2 = \gamma$.
  Left: from top to bottom $\gamma = 1.9, \, 2.0, \, 2.1, \, 2.2$;
  Right: zoom in near the lower band edge,
  from top to bottom $\gamma = 2.1, \, 2.3, \, 2.5, \, 2.7$.
}
\label{fig:moving_bs}
\end{figure}

A feature of potentials with a finite range is that upon increasing
the potential strength the binding energies can be made to
increase until the state merges with the continuum of the
lower band and becomes a resonance.
This is illustrated in Fig.~\ref{fig:moving_bs} where we have plotted
$D_2(\om)$ for different values of the strength of the potential;
and it can be seen how the zeros of $D_2$ moves across the gap
and ultimately disappears into the valence band.
Notice that this is consistent with the interpretation of the variational
calculation of Sec.~\ref{sec:variational_I}.
We expect a similar behavior to occur for a strong Coulomb potential,
but this interesting case is beyond the scope of this study.
Another related example of this phenomenon (without a hard gap) is
the problem of a strong Coulomb impurity in monolayer graphene
that has acquired much interest 
recently.\cite{Levitov_2007a,Pereira2007,Levitov_2007b,Fogler_hypercritical_2007}

The important case of a screened Coulomb potential generally
requires a different approach.
Nevertheless, we do not anticipate any qualitative discrepancies
between a potential well and a screened Coulomb potential.
We expect the screening wave vector to be roughly
proportional to the density of states at the Fermi energy;
and once the range and the strength of the potential have
been estimated a potential well can be used
to approximate the binding energies. 
We also note that the asymptotic behavior in 
Eq.~(\ref{eq:BGBi_kappadef}) is quite general for a decaying potential.

\subsection{Polarization function}
\label{subsec:polarization}
Since we were just discussing the issue of screening it
fits well to briefly discuss the issue of the dielectric
function in the biased bilayer (see also Ref.~\onlinecite{Stauber_ring_2007}).
With the introducing of the symmetric ($+$) and antisymmetric
($-$) densities (see e.g. Ref.~[\onlinecite{Nilsson2006a}])
\begin{equation}
  \label{eq:BGBi_pol1}
  \rho_{\pm}(\vq) = \sum_{\vk}\psi^{\dag}(\vk+\vq)
  \begin{pmatrix}
    1 & 0 & 0 & 0 \\
    0 & 1 & 0 & 0 \\
    0 & 0 & \pm 1 & 0 \\
    0 & 0 & 0 & \pm 1
  \end{pmatrix}
  \psi(\vk).
\end{equation}
The usual manipulations then gives the retarded response in the
symmetric channel as\cite{GG_2005}
\begin{multline}
  \label{eq:BGBi_pol3}
  \chi_{++}(-\vq, \om)
  =2 \sum_{l,l'}\int \frac{d\vk}{(2\pi)^2}
| v^{\dag}_{l}(\vk) v_{l'}(\vk+\vq) |^2
\\ \times
  \frac{n_F \bigl[E_l(\vk)\bigr] - n_F\bigl[ E_{l'}(\vk+\vq) \bigr]}
  {E_l(\vk) - E_{l'}(\vk+\vq) + \om + i\delta }
   .
\end{multline}
Here $ v_{l}(\vk)$ is the spinor wave function of band $l$ at momentum
$\vk$. 
At half filling this expression only have contributions from $l\neq l'$
and since the wave function 
overlap at $\vq = 0$ between different bands for a fixed value of $\vk$
is zero we conclude that $\chi_{++}(\vq,0) \propto q^2$ in the
limit of $q \rightarrow 0$.
The expression is expected to be dominated by the transitions between the $E_{-,-}$
and $E_{+,-}$-bands leading to $\chi_{++}(\vq,0) \sim - 2 q^2 / V$.
The Random Phase Approximation (RPA) dielectric function is then given by 
$\e(\vq) \approx 1- (2 \pi e^2/q)\chi_{++}(\vq,0)$ which imply that 
$\e(\vq) \sim 1$ as $q\rightarrow 0$. From this we can conclude
that the BGB is unable to contribute to the screening of
the long-range part of the Coulomb interaction.
Note that the dimensionality of the system is crucial for this argument.
In three dimensions, where the Coulomb interaction goes as $1/q^2$,
the same argument as in the above
usually gives a large contribution to $\e$
for a semiconductor.\cite{Ziman1972}
For the unbiased bilayer at $\mu=0$ in the low-energy approximation of 
Eq.~(\ref{eq:nobias_Hkin0bilayer}) one finds (using RPA) 
a screening wave vector  that is proportional to $\tp$.\cite{Nilsson2006a}
This is in agreement with what one expects for an electron gas
in 2D where the screening wave vector is proportional to the the
effective mass. For a more detailed discussion of the unbiased graphene
bilayer dielectric function including the trigonal warping
see Ref.~\onlinecite{Chakraborty_2007}.

\section{Coherent potential approximation}
\label{sec:CPA_biased}
As discussed above in Sec.~\ref{sec:simple_estimate_nc}, 
for a finite density $\nimp$ of impurities, 
the bound states can interact with each other leading to the possibility
of band gap renormalization and the formation of impurity bands.
A simple, but crude, theory of these effects is the 
CPA.\cite{Soven67,Velicky68}
In this approximation, the disorder is treated as a self-consistent medium
with recovered translational invariance.
The medium is described by a set of four local self-energies which are
allowed to take on different values on all of the inequivalent lattice
sites in the problem.
In fact this section is a straightforward extension to the biased case
of the methods
applied in Sec.~\ref{sec:impurities_tmatrix} for the unbiased case.
The self-energies are chosen so that there is no scattering on average 
in the effective medium. It has been argued that the CPA is the best 
single-site approximation to the full solution of the 
problem.\cite{Velicky68} 

In the following we often suppress the frequency dependence of
the self-energies for brevity. 
The expression for the diagonal elements of the Green's function
$\GR$ is given in App.~\ref{app:Gdetail_biased}.
We follow the standard approach to derive the 
CPA (see for example Refs.~[\onlinecite{Velicky68,JonesMarch2}]), 
and we obtain the self-consistent equations:
\begin{equation}
  \label{eq:BGBi_finiteU}
  \Sigma_{\alpha j} = \frac{n_i U} {1-(U-\Sigma_{\alpha j})
  \overline{\GR}_{\alpha j}}.
\end{equation}
%
%
The limit of site dilution (or vacancies)
used in Sec.~\ref{sec:impurities_tmatrix}
is obtained in the limit $U \rightarrow \infty$
leading to the self-consistent equations:
\begin{equation}
  \label{eq:BGBi_GbarAA1}
  \Sigma_{\alpha j}  = - \frac{n_i}{\overline{\rm G}_{\alpha j} }.
\end{equation}
An explicit expression for the local propagators 
$\overline{\GR}_{\alpha j}$ is given in App.~\ref{app:Gdetail_biased}.
Using the expressions obtained there
the self-consistent equations for $U \rightarrow \infty$ becomes:
\begin{subequations}
\begin{eqnarray}
\frac{n_i}{\Sigma_{\rA 1}} = -\overline{\GR}_{\rA 1}  &=& \beta_1 
(\xi_1 - \alpha_2 \beta_2 \xi_0)  ,\\
\frac{n_i}{\Sigma_{\rB 1}} = -\overline{\GR}_{\rB 1}  &=& 
\alpha_1 (\xi_1 - \alpha_2 \beta_2 \xi_0) +\tp^2 \beta_2 \xi_0 .
\end{eqnarray}
\end{subequations}
From these equations it is straightforward to obtain the density of
states (DOS) on the different sublattices $\alpha j$
from $\rho_{\alpha j}(\om) = -\imag \overline{\text{G}}_{\alpha j}(\om
+ i \delta) 
/ \pi$. In the clean case, one finds:
\begin{subequations}
\label{eq:BGBi_DOS_BGB1}
\begin{eqnarray}
\rho_{\rA 1}^0 &=& \biggl|
   \frac{\om - {V/2}}{2 \Lambda^2} \biggl[ \chi
 - \frac{ \om V (2-\chi) }{\sqrt{( V^2 + \tp^2) \om^2  - V^2
   \tp^2 /4 }} \biggr] \biggr|,
\\
\rho_{\rB 1}^0 &=& \biggl| \rho_{\rA 1}^0 + 
\frac{\tp^2 ( \om + {V/2})  (2 - \chi) }{2 \Lambda^2 \sqrt{( V^2 +
    \tp^2) \om^2 - V^2 \tp^2 / 4}} \biggr|,
\end{eqnarray}
\end{subequations}
for $|\om| \geq \Eg/2$.
Here $\chi = (0,1,2)$ for ($|\om|  \leq {V/2} $,  
${V/2} \leq |\om| \leq \sqrt{\tp^2 + V^2 / 4}$,  $\sqrt{\tp^2 +
  V^2 / 4} \leq |\om| $). 
The corresponding quantities in plane 2 are obtained by the
substitution $ V \rightarrow - V$.
In the limit of $ V \rightarrow 0$ we recover the
known unbiased result of
Eq.~(\ref{eq:dis_nobias_DOS0bilayer}).
Notice that the square-root singularity starts to appear 
already above $V/2$ on the B1 sublattice. 
There is also a divergence on the A1 sublattice but the coefficient in
front is usually much smaller.
The DOS on the A1 sublattice vanishes at $\om = V/2$
while the DOS on the B1 sublattice is finite there.

\begin{figure}[htb]
  \includegraphics[scale=0.42]{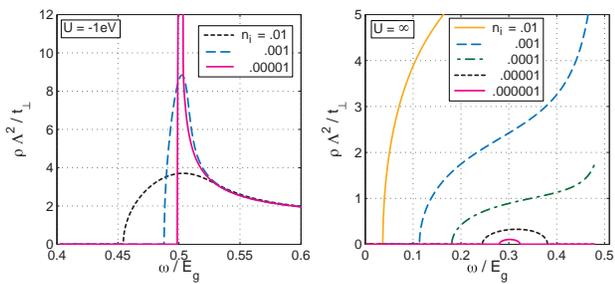}
  \caption
{[color online] Left: DOS as a function of the energy 
(in units of $E_g$) close to the conduction band edge 
for different impurity concentrations (see inset),  
$U=-1$ eV. Right: Details of the DOS 
inside of the gap for different impurity 
concentrations for $U \to \infty$.
In both cases $V = 40 \,\text{meV}$.}
\label{fig:dos1}
\end{figure}

\begin{figure}[htb]
\includegraphics[scale=0.42]{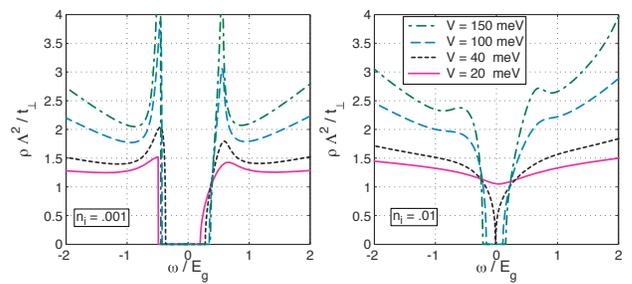}
\caption
{[color online] DOS as a function of energy (in units of $E_g$)
for different values of the applied bias $V$ (see inset) and $U= -10$ eV.
Left: $n_i = 10^{-3}$; Right: $n_i = 10^{-2}$.}
\label{fig:dos2}
\end{figure}

The numerically calculated density of states for $U \to \infty$
is shown in Fig.~\ref{fig:dos1}.
The impurity band evolves from the single-impurity B$2$ bound
state which for the parameters
involved is located at $\e \approx 0.3 \Eg$.
Further evidence for this interpretation is that the total integrated
DOS inside the split-off bands for the two lowest
impurity concentrations is equal to $n_i$.
It is worth mentioning that the width of the impurity band in the
CPA is likely to be overestimated. The reason for this 
is that that the use of effective atoms, all of which have some
impurity character, increases the interaction between the 
impurities.\cite{Velicky68}
For smaller values of the impurity strength the single-impurity bound
states are all weakly bound (cf.\ Fig.~\ref{fig:boundstates1})
and the ``impurity bands'' merge with the bulk bands
as shown in Figs.~\ref{fig:dos1}
and~\ref{fig:dos2}.
The bands have been shifted rigidly by the amount $n_i U $ for a more
transparent comparison between the different cases. 
The smoothening of the singularity and the band gap
renormalization is apparent. Observe also that the band edge moves further into
the gap at the side where the bound states are located.
It is likely that the CPA gives a better approximation for these states
since by Eq.~(\ref{eq:BGBi_kappadef_weak}) they are weakly damped 
almost propagating modes. Notice that the gap and the whole structure of
the DOS in the region of the gap is changing with $V$,
and in particular the possibility that the actual gap closes before $V=0$ 
because of impurity-induced states inside of the gap.
Finally we note that this observation is consistent with the results of 
numerically exact calculations using the recursive Green's function method 
for strong disorder.\cite{Castro_gaped_2007}

\section{Effects of trigonal distortions}
\label{sec:BGBi_trigonal}

Before we conclude this paper we would like to briefly comment on the effect
of the $\gamma_3$-term on our findings in the previous sections.
The effects of $\gamma_3$ on the spectrum of the BGB was discussed in 
Sec.~\ref{sec:simplebands},
where it was shown that
this term breaks the cylindrical symmetry and 
leads to the ``trigonal distortion'' of the bands.
In the BGB the result is three copies of a more conventional elliptic
dispersion at the lowest energies near the band edge.
Using the same method as in Section~\ref{sec:BGBi_bs_Dirac} we
find that, also for an elliptical band edge, a Dirac delta potential always
generates a bound state in 2D.
The divergence of $\overline{\GR}^{0}$ is generally only logarithmic 
as the band edge is approached however, 
whereas the divergence is an inverse square root without $\gamma_3$.
More confined bound states with larger binding energies sample a
larger area of the BZ.
Therefore we do not expect that the small details at the band edge to
significantly modify the results that we obtain with the minimal
band model for these states.

Another observation is that when there is a finite density of
impurities in the BGB the self-energies can become quite large as we
have seen in the previous section. 
Consider the case $V=50 \, \text{meV}$, 
for which $(V-\Eg)/\Eg \approx 0.01$.
Therefore, by looking at Fig.~\ref{fig_bilayer_bands2},
we see that $\gamma_3$ smooth out the square-root singularity
on this scale.
Comparing with Fig.~\ref{fig:dos1} we see that for an impurity
of strength $U = -1 \, \text{eV}$ the trigonal distortion would
correspond to a density of impurities of around $\nimp \sim .001$.
In the case that the gap is filled up with
impurity-induced states (see Fig.~\ref{fig:dos2}), the disorder-induced
energy scale is much larger than that generated by $\gamma_3$.
Therefore we argue that the possibility that the gap closes before
$V =0$ is robust to the presence of a $\gamma_3$,
even if it is as large as the values quoted in the graphite literature.

\section{Conclusion and outlook}
\label{sec:conclusions}

Graphene research is one of the fastest growing fields in condensed matter 
research since the isolation by Novoselov {\it et al.} of the first
graphene flake in 2004.\cite{Novoselov2004} Three years after that, and
after hundreds of theoretical papers on the subject,\cite{ahcn_review} 
the physics of single layer graphene is relatively well understood and 
very few controversies remain. Meanwhile, the study of multilayer graphene,
and particularly bilayer graphene, continues to be, experimentally and
theoretically, an open field of research. Partially, this can be assigned
to the natural attraction of researchers to the ``one atom thin'' material
and its unusual electronic spectra. Nevertheless, graphene bilayer is equally
thin (two atoms thick, indeed) and has also 2D Dirac spectrum (albeit massive)
with unusual properties. Graphene bilayer is also more prone to show strong
electron-electron correlation effects, such as magnetism \cite{Nilsson2006a} 
and charge density wave \cite{DWBZB07}
phases, because of its finite density of states at the Dirac point, unlike its
single layer counterpart where interactions are at most marginal in renormalization
group sense.\cite{Peres2005_ferro} Interestingly enough, the study of 
the effect of electron-electron interactions in graphene is a field in its
infancy.\cite{KNCN07,MacDonald_pseudospin_2007}

Furthermore, graphene bilayer is equally easy (or hard)
to find or produce epitaxially. Besides its intriguing electronic properties, the graphene
bilayer is a promising candidate to bulk electronic devices with properties
that are insensitive to surface (edge) defects such as graphene nanoribbons and 
quantum dots. Perhaps even more interesting is the fact that graphene bilayer is
the only known material that has an electronic gap between conduction and valence
bands that can be fully controlled by the application of a transverse electric
field (a tunable gap semiconductor), as has been demonstrated 
experimentally.\cite{Castro_PRL_2007,OHLMV07}
This property opens up an enormous number of possible ways to use bilayer graphene,
from transistors to lasers working in the terahertz regime. 

Nevertheless, in order to be able to use graphene bilayer (and multilayers) in device
applications, one has to understand how material issues, such as disorder, affect
its electronic properties. This was the main aim of this work, namely, to understand
how disorder affects the electronic properties of bilayer (and multilayer) graphene
in its most basic model. We have shown that the electronic self-energies can be 
calculated analytically within the T-matrix and CPA presenting some unusual features that can
be measured either by transport or spectroscopy (ARPES and STM). We have calculated
a series of important physical properties such as spectral functions, and frequency
dependent conductivities. We have also studied the problem of bound states in 
the biased bilayer graphene and their effect in the electronic structure and have
shown that the properties of these bound states can be equally well controlled
by applied transverse fields. We also point that we have left open issues associated
with trigonal distortions. At this point in time, it is not clear that such effects,
associated with $\gamma_3$, are going to be the same as observed in 3D graphite
and more experimental studies are needed in order to investigate the issue. 
We hope that our results will stimulate more experimental studies of these amazing new materials.

\begin{acknowledgments}
We are always very thankful to Prof. A. Geim for his level headed comments and criticisms.
J.N. acknowledges funding from the Dutch Science Foundation FOM 
during the final stages of this work.
N.M.R.P. thanks ESF Science Programme INSTANS 2005-2010
and FCT under the grant POCTI/FIS/58133/2004. F.G. acknowledges
funding from MEC (Spain) through grant FIS2005-05478-C02-01,
and the European Union contract 12881 (NEST).
\end{acknowledgments}

\appendix

%
%
\section{Minimal conductivity of the graphene bilayer including trigonal warping}
\label{app:minimal}
In this appendix we provide an alternative derivation of the value
of the minimal conductivity of the graphene bilayer in the
presence of trigonal warping.
The conductance of a wide strip 
of graphene at the Dirac 
point is mediated
by evanescent modes that connect the leads. 
We define the hamiltonian as
\begin{equation}
{\cal H} \equiv \left( \begin{array}{cc} 0 &v_F ( k_x + i k_y ) \\ v_F ( k_x -
    i k_y ) &0 \end{array} \right) ,
\end{equation}
and use the Landauer formalism described in
Refs.~\onlinecite{Katsnelson06a,Beenakker06}.
We take the width $W$ of the sample to be much larger
than its length $L$. 
If we assume that the leads on the 
right and on the left are heavily doped clean graphene, the incoming, reflected,
and outgoing waves can be approximated as:
\begin{subequations}
\label{pwavefunctions}
\begin{eqnarray}
\Psi_{in} &\equiv &\frac{1}{\sqrt{2}} \left( \begin{array}{c} 1 \\ 1
  \end{array} \right) e^{i k_y y}  , \\
\Psi_{ref} &\equiv &\frac{r ( k_y )}{\sqrt{2}} \left( \begin{array}{c} 1 \\ - 1
  \end{array} \right) e^{i k_y y}  , \\
\Psi_{trans} &\equiv &\frac{t ( k_y )}{\sqrt{2}} \left( \begin{array}{c} 1 \\ 1
  \end{array} \right) e^{i k_y y}  ,
\end{eqnarray}
\end{subequations}
where $t(k_y )$ [$r(k_y )$] is the transmission [reflection] amplitude
of the mode with perpendicular momentum $k_y$.
The wavefunction in the central region, $0 \le x \le L$,  at zero
energy, can be written as
\begin{equation}
\Psi \equiv A \left(  \begin{array}{c}  e^{- k_y x} \\ 0 \end{array} \right)
  e^{i 
  k_y y} +
B  \left( \begin{array}{c} 0 \\ e^{ k_y x} \end{array} \right) e^{i k_y y} .
\label{pwavefunctions_2}
\end{equation}
The matching conditions at the edges at $x = 0$ and $x=L$ are:
\begin{subequations}
\begin{eqnarray}
1 + r ( k_y ) &= &\sqrt{2} A  , \\
1 - r ( k_y ) &= &\sqrt{2} B   , \\
t ( k_y ) &= &\sqrt{2} A e^{- k_y L }  , \\
t ( k_y ) &= &\sqrt{2} B e^{ k_y L },
\end{eqnarray}
\end{subequations}
resulting in the transmission probability
\begin{equation}
 T ( k_y )  \equiv  | t ( k_y ) |^2 = \frac{1}{\cosh^2 ( k_y L )} .
\label{pmatching}
\end{equation}
The conductance per channel is thus given by
\begin{equation}
G = \frac{e^2}{h} \frac{W}{2 \pi} \int_{- \infty}^{\infty} d k_y T ( k_y )  =
\Bigl( \frac{e^2}{\pi h} \Bigr) \frac{W}{ L}  .
\end{equation}

We will now extend this result to the anisotropic Dirac equation. 
The Hamiltonian is
\begin{equation}
{\cal H} \equiv \left( \begin{array}{cc} 0 & v_x k_x + i v_y k_y \\ v_x k_k -
    i v_y k_y &0 \end{array} \right),
\end{equation}
where the Fermi velocities ($v_x$ and $v_y$)
are allowed to be different in the $x$ and $y$ directions.
We use the incoming and outgoing wavefunctions wavefunctions in
Eq.~(\ref{pwavefunctions}), and generalize the wavefunctions in the graphene
junction, Eq.~(\ref{pwavefunctions_2}) to
\begin{equation}
\Psi \equiv A \left( \begin{array}{c}  e^{- \kappa x} \\ 0 \end{array} \right)
  e^{i 
  k_y y} +
B \left( \begin{array}{c} 0 \\  e^{ \kappa x} \end{array} \right) e^{i k_y y} ,
\label{pwavefunctions_3}
\end{equation}
where the Dirac equation implies that $\kappa = v_y k_ y / v_x$. 
Matching the wave functions at the contacts with the leads, we find that 
the generalization of Eq.~(\ref{pmatching}) is
\begin{equation}
 T ( k_y )  = \frac{1}{\cosh^2 \left( \frac{v_y k_y L}{v_x} \right)} ,
\end{equation}
so that
\begin{equation}
G = \Bigl( \frac{e^2}{\pi h} \Bigr)  \Bigl( \frac{v_x}{v_y} \Bigr) \frac{W}{  L} .
\end{equation}
If the junction is rotated by an angle $\theta$ with respect to the main axes
of the anisotropic Dirac equation, the Hamiltonian becomes:
\begin{widetext}
\begin{equation}
{\cal H} \equiv \left( \begin{array}{cc} 0 &v_a [ k_x \cos ( \theta ) - k_y
    \sin ( \theta ) ] + i v_b [ k_x  \sin ( \theta ) + k_y \cos ( \theta ) ]
    \\ v_a [ k_x \cos ( \theta ) - k_y
    \sin ( \theta ) ] - i v_b [ k_x  \sin ( \theta ) + k_y \cos ( \theta ) ]
    &0 \end{array} \right),
\end{equation}
\end{widetext}
where $v_a$ and $v_b$ are the Fermi velocities along the two principal axes. 
The wave function in the central region is now
\begin{equation}
\Psi \equiv A \left( \begin{array}{c}  e^{- \kappa x} e^{i k' x} \\ 0
  \end{array} \right) 
  e^{i 
  k_y y} +
B \left( \begin{array}{c} 0 \\  e^{ \kappa x}  e^{i k' x} \end{array}
  \right) e^{i k_y y} ,
\label{pwavefunctions_4}
\end{equation}
where
\begin{subequations}
\begin{eqnarray}
\kappa &= &\frac{v_a v_b}{v_a^2 \cos^2 ( \theta ) + v_b^2 \sin^2 ( \theta )} k_y
, \\
k' &= &\frac{\sin ( \theta ) \cos ( \theta ) ( v_a^2 - v_b ^2 )}{v_a^2 \cos^2
  ( \theta ) + v_b^2 \sin^2 ( \theta )}  k_y .
\end{eqnarray}
\end{subequations}
Using this we also obtain
\begin{equation}
 T ( k_y )  = \frac{1}{\cosh^2 \left( \frac{v_a v_b k_y L}{v_a \cos^2 (
  \theta ) + v_b^2 \sin^2 ( \theta )} \right)} ,
\end{equation}
leading to
\begin{equation}
G =  \Bigl( \frac{e^2}{\pi h} \Bigr)
	\Bigl( \frac{v_a^2 \cos^2 ( \theta ) + v_b^2 \sin^2 ( \theta )}{v_a v_b} \Bigr)
  	\frac{W}{ L} .
\end{equation}

In a graphene bilayer, including trigonal warping but ignoring terms that
couple sites in the same sublattice, we have four Dirac points. One of them
is isotropic, with $v_a = v_b$, and the three others are
 anisotropic, with $v_b = 3 v_a$. The principal axes at these three Dirac
 points form angles with respect to a barrier which can be parametrized as
 $\theta_0 , \theta_0 + 2 \pi / 3$ and $\theta_0 + 4 \pi / 3$, where
 $\theta_0$ depends on the orientation of the barrier. 
 The conductance is therefore given by
\begin{equation}
G =  \Bigl( \frac{e^2}{\pi h} \Bigr)  \frac{W}{ L} \left[ 1 + \frac{3}{2} \left( \frac{v_a}{v_b} +
    \frac{v_b}{v_a} \right) \right] .
\end{equation}
This expression is independent of the angle $\theta_0$. For $v_b / v_a = 3$,
we find $G = 6 \times [e^2/ (\pi h)] \times (W /  L ) $ per channel, in agreement
with Refs.~\onlinecite{Ando06a,Cserti2007b}.

\section{Density of states in multilayer graphene}
\label{app:DOSgraphite}
In this appendix we derive explicit expressions for the 
DOS in graphene multilayers. The expressions are used in
Section~\ref{sec:G_multilayer}.
To calculate the DOS in graphite we must perform two integrals to
get $\overline{\GR}$. One integral we need is 
\begin{equation}
  \label{eq:dis_nobias_integralI1}
  I_{1} = \int_0^{\Lambda^2}  d (p^2) \int_{-\pi/2}^{\pi/2} 
  \frac{dk_{\perp}}{\pi} 
  \Bigl[ \frac{1}{2 D_{-}} + \frac{1}{2 D_{+}} \Bigr].
\end{equation}
First we perform the perpendicular integral using complex variables to
rewrite the integral as a contour integral around the unit circle and
then picking up the pole inside:
\begin{multline*}
  \int_{-\pi/2}^{\pi/2}dk_{\perp} \Bigl[ \frac{1}{D_{-}} + \frac{1}{D_{+}}
  \Bigr]
  \\
  = \frac{1}{\om - \Sigma_B}
  \oint \frac{dz}{i}
  \frac{1}{\tp(z^2 + 1) - A z} 
\\
  = 
  - \frac{2 \pi}{\om -\Sigma_B} 
  \frac{\sign[\real(A)]}{\sqrt{A^2 - 4 \tp^2}},
\end{multline*}
where $A \equiv  p^2/(\om - \Sigma_B) - (\om - \Sigma_A)$.
Note that the function $\sign[\real(A)]$ 
changes sign just where the branch
of the square root does.
Moreover the square root is purely imaginary
there. Therefore the function is actually continuous across the
point where $\real(A) = 0$.
From now on in this appendix, as well as in
Appendix~\ref{subsec:multi_kernel},
we choose
the branch of the square root such that the sign of the 
real part is included, with this convention $A$ and
$\sqrt{A^2 - 4 \tp^2}$ always lies in the same quadrant of
the complex plane.
Because of the form of $A$ we can use the integral formula
\begin{multline}
  \label{eq:dis_nobias_xi1}
  \xi^{(1)}(p^2) = 
  \int  d (p^2) \frac{1}{\sqrt{A^2 - 4 \tp^2}}
\\
  =
  (\om-\Sigma_B)  
  \log \Bigl[A + \sqrt{A^2 - 4 \tp^2} \Bigr],
\end{multline}
directly,
since the argument of the log does not cross any branch cut.
Thus 
the result of the integral is
\begin{equation}
  \label{eq:dis_nobias_I1result1}
  I_{1}  = - \log \Bigl[
  \frac{A(\Lambda^2) + \sqrt{A^2(\Lambda^2) - 4 \tp^2}}
  {A(0) +  \sqrt{A^2(0) - 4 \tp^2}}
  \Bigr].
\end{equation}
Finally to leading order in $\Lambda$ we get
\begin{equation}
  \label{eq:dis_nobias_I1result2}
  I_{1}  = - \log \Bigl[
  \frac{2 \Lambda^2 }
  {-(\om - \SB)(\om - \SA + \sqrt{(\om - \SA)^2 - 4 \tp^2}}
  \Bigr].
\end{equation}
%
%
Similarly the integral
\begin{equation}
  \label{eq:dis_nobias_I2def}
  I_{2} = \int_0^{\Lambda^2}  d (p^2) \int_{-\pi/2}^{\pi/2} \frac{dk}{\pi} 
  \Bigl[ \frac{-2\tp \cos(k)}{2 D_{-}} + \frac{2\tp \cos(k)}{2 D_{+}} \Bigr], 
\end{equation}
can be written as
\begin{equation}
  I_{2}
  =
  \frac{1}{\om - \Sigma_B}
  \Bigl[ \Lambda^2 - \int_0^{\Lambda^2}d(p^2)
  \frac{ A}{\sqrt{A^2 - 4 \tp^2}} \Bigr].
\end{equation}
Now we may use the integral formula
\begin{equation}
  \label{eq:dis_nobias_integral1}
  \int  d (p^2) \frac{ A }{\sqrt{A^2 - 4 \tp^2}}
  =
  (\om - \Sigma_B)  
  \sqrt{A^2 - 4 \tp^2} \bigr),
\end{equation}
and one can once again convince oneself 
that there are no contribution from
the possible crossing of the branch at $\real(A) = 0$. 
With the help of this the resulting expression becomes
\begin{multline}
  \label{eq:dis_nobias_I2result1}
  I_{2} = \frac{1}{\om - \Sigma_B} \Bigl\{
  \Lambda^2 - (\om - \Sigma_B) \Bigl[
  \sqrt{A^2(\Lambda^2) - 4 \tp^2} 
\\
  +\sqrt{A^2(0) - 4 \tp^2} \Bigl] \Bigr\}.
\end{multline}
Finally, keeping only the leading term 
in the expansion in $\Lambda$ we get
\begin{equation}
  \label{eq:dis_nobias_I2result2}
  I_{2} = (\om - \Sigma_A) -
  \sqrt{(\om - \Sigma_A)^2 - 4 \tp^2}.
\end{equation}
%
%


%
%
%
\section{Conductivity kernels}
\label{app:kernels}
In this appendix we derive formulas for the
conductivity kernels that we use in 
Sec.~\ref{sec:conductivity_results_b} and Sec.~\ref{sec:conductivity_results_g}.
First we rewrite the kernels with the identity
\begin{multline}
  \label{eq:dis_nobias_kernelidentity}
  \imag \bigl[ \GR_{1}(\e ) \bigr]
  \imag \bigl[ \GR_{2}(\e + \om ) \bigr]
\\
  = \frac{1}{2} \real \bigl[ \GR_{1}^{A}(\e) \GR_{2}^{R}(\e +\om )
  - \GR_{1}^{R}(\e) \GR_{2}^{R}(\e + \om) \bigr]
\\
  = \frac{1}{2} \real \Bigl[ \sum_{\gamma = \pm} 
  \gamma \GR_{1}(E_{1} -i \gamma \Gamma_{1})
  \GR_{2}(E_{2}+i \Gamma_{2}) \Bigr],
\end{multline}
and introduce the notations
\begin{subequations}
\begin{eqnarray}
  \label{eq:dis_nobias_kernel_notation}
 E_{A} &=&   \e -\SA(\e), \\
 E_{B}  &=& \e -\SB(\e) , \\
 \tE_{A} &=&  \e + \om - \SA(\e +\om) , \\
  \tE_{B} &=& \e + \om - \SB(\e +\om) .
\end{eqnarray}
\end{subequations}

\begin{widetext}
\subsection{Bilayer graphene}

The integrals we need for the bilayer are easy to obtain since there
is never any problems with branches of the logarithms. 
The kernels can be expressed in terms of the following integrals:
\begin{equation}
  \label{eq:dis_nobias_kernelintegral1}
  \int_0^{\Lambda^2} d(p^2) 
  {\rm g_{AA}^{D}}(\e)
  {\rm g_{BB}^{D}}(\e+\om) 
  =
  \frac{1}{4}\sum_{\alpha,\beta} 
  \frac{E_B (\tE_A + \beta \tp)}
  {E_B (E_A + \alpha \tp) -\tE_B (\tE_A +\beta \tp)}
  \log \Bigl[
  \frac{-\tE_B (\tE_A +\beta \tp)}{-E_B (E_A + \alpha \tp)}
  \Bigl],
\end{equation}
\begin{equation}
  \label{eq:dis_nobias_kernelintegral2}
  \int_0^{\Lambda^2} d(p^2) 
  {\rm g_{BB}^{D}}(\e)
  {\rm g_{AA}^{D}}(\e+\om)
 =
  \frac{1}{4}\sum_{\alpha,\beta} 
  \frac{\tE_B (E_A + \alpha \tp)}
  {\tE_B (\tE_A + \beta \tp) -E_B (E_A +\alpha \tp)}
  \log \Bigl[
  \frac{-E_B (E_A +\alpha \tp)}{-\tE_B (\tE_A + \beta \tp)}
  \Bigl],
\end{equation}
\begin{multline}
  \label{eq:dis_nobias_kernelintegral3}
  \int_0^{\Lambda^2} d(p^2) 
  {\rm g_{AB}^{ND}}(\e)
  {\rm g_{AB}^{ND}}(\e+\om)
  =
  \frac{1}{4}\sum_{\alpha,\beta} 
  \frac{\alpha \beta}
  {E_B (E_A + \alpha \tp) -\tE_B (\tE_A +\beta \tp)}
\times
\\
  \Bigl\{
  E_B (E_A + \alpha \tp) \log \Bigl[
  \frac{\Lambda^2}{-E_B (E_A + \alpha \tp)} \Bigr]
  -
  \tE_B (\tE_A + \beta \tp) \log \Bigl[
  \frac{\Lambda^2}{-\tE_B (\tE_A + \beta \tp)} \Bigr]
  \Bigl\},
\end{multline}
and
\begin{multline}
  \label{eq:dis_nobias_kernelintegral4}
  \int_0^{\Lambda^2} d(p^2) 
  \bigl[ {\rm g_{AA}^{D}}(\e) {\rm g_{AA}^{D}}(\e+\om)
  - {\rm g_{AA}^{ND}}(\e) {\rm g_{AA}^{ND}}(\e+\om) \bigr]
\\
  =
  \frac{1}{2} \sum_{\alpha} 
  \frac{E_B \tE_B}
  {E_B (E_A + \alpha \tp) -\tE_B (\tE_A -\alpha \tp)}
  \log \Bigl[
  \frac{-\tE_B (\tE_A -\alpha \tp)}{-E_B (E_A + \alpha \tp)}
  \Bigl].
\end{multline}
In fact, although the cutoff $\Lambda$ enter the expression in
Eq.~(\ref{eq:dis_nobias_kernelintegral3}), the final result
is actually independent of $\Lambda$.
For the DC conductivity it is convenient to work out that for
two retarded propagators we have the relation
\begin{equation}
  \label{eq:dis_nobias_retret_kernel}
  \int_0^{\Lambda^2} d(p^2) 
  \bigl[ {\rm g_{AA}^{D}}(\e) {\rm g_{BB}^{D}}(\e)
  + {\rm g_{AB}^{ND}}(\e) {\rm g_{AB}^{ND}}(\e) \bigr]
  = - 1.
\end{equation}

\subsection{Multilayer graphene}
\label{subsec:multi_kernel}

For multilayer graphene we have to perform two integrals, they are
\begin{equation}
  \label{eq:dis_nobias_J1kerneldef}
  J_1 = \int_0^{\Lambda^2}d(p^2)
  \int_{-\pi/2}^{\pi/2} \frac{dk}{\pi} 
  \bigl[ {\rm g_{AA}^{D}} \widetilde{\rm g_{BB}^{D}}
  +
  {\rm g_{BB}^{D}} \widetilde{\rm g_{AA}^{D}}
  +
  2 {\rm g_{AB}^{ND}} \widetilde{\rm g_{AB}^{ND}}
  \bigr],
\end{equation}
and
\begin{equation}
  \label{eq:dis_nobias_J2kerneldef}
  J_2 = \int_0^{\Lambda^2}d(p^2)
  \int_{-\pi/2}^{\pi/2} \frac{dk}{\pi} 
  \bigl[ {\rm g_{AA}^{D}} \widetilde{\rm g_{AA}^{D}}
  +
  {\rm g_{AA}^{ND}} \widetilde{\rm g_{AA}^{ND}}
  \bigr] \sin^2(k).
\end{equation}
Exactly like in the case above we find $J_1 = -2$ when $\om =0$ and
both the propagators are retarded.
First we perform the perpendicular
integral using a contour integral as we did when we computed the DOS 
in Appendix~\ref{app:DOSgraphite}. In the following the branch of the
square root includes the sign of the real part 
$\sqrt{A^2-4\tp^2} \equiv \sign[\real(A)] \sqrt{A^2-4\tp^2}$. 
This implies that $A$
and $\sqrt{A^2-4\tp^2}$ always lies in the same quadrant.
The results we need are
\begin{multline}
  \label{eq:dis_nobias_integralg1}
  \int_{-\pi/2}^{\pi/2} \frac{dk}{\pi} 
  {\rm g_{AA}^{D}}(\e)
  {\rm g_{BB}^{D}}(\e+\om)
  =
  -\frac{1}{2 \tE_B}
  \biggl\{
  \frac{2 }{\sqrt{A^2- 4 \tp^2}}
\\
+
  \frac{p^2}{\tE_B} 
  \Bigl[ \frac{1}{A - \tA} - \frac{1}{A + \tA} \Bigl]
  \frac{1}{\sqrt{A^2- 4 \tp^2}}
+
  \frac{p^2}{\tE_B}
  \Bigl[ \frac{1}{\tA - A} - \frac{1}{A + \tA} \Bigl]
  \frac{1}{\sqrt{\tA^2- 4 \tp^2}}
  \biggr\},
\end{multline}
\begin{equation}
  \label{eq:dis_nobias_integralg2}
  \int_{-\pi/2}^{\pi/2} \frac{dk}{\pi} 
  {\rm g_{AB}^{ND}}(\e)
  {\rm g_{AB}^{ND}}(\e+\om)
  =
  -\frac{p^2}{2 E_B \tE_B}
  \Bigl\{
  \Bigl[\frac{1}{A - \tA} + \frac{1}{A + \tA}\Bigr]
  \frac{1}{\sqrt{A^2- 4 \tp^2}}
  +
  \Bigl[\frac{1}{\tA - A} + \frac{1}{A +\tA}\Bigr]
  \frac{1}{\sqrt{\tA^2- 4 \tp^2}}
  \Bigr\},
\end{equation}
and 
\begin{equation}
  \label{eq:dis_nobias_integralg3}
  \int_{-\pi/2}^{\pi/2} \frac{dk}{\pi} 
  \bigl[ {\rm g_{AA}^{D}}(\e) {\rm g_{AA}^{D}}(\e+\om)
  + {\rm g_{AA}^{ND}}(\e) {\rm g_{AA}^{ND}}(\e+\om) \bigr] \sin^2(k)
\\
  =
  \frac{1}{4 \tp^2}
  \Bigl\{
  -1 + \frac{1}{A- \tA}
  \Bigr[
   \sqrt{A^2- 4 \tp^2}
  -  \sqrt{\tA^2- 4 \tp^2}
  \Bigr].
\end{equation}
Where
\begin{eqnarray}
  \label{eq:dis_nobias_integralg4}
  A &=& p^2/E_B - E_A , \\
  \tA &=& p^2/ \tE_B - \tE_A.
\end{eqnarray}
Adding up all the contributions for the $J_1$ we get after
some rearrangements
\begin{multline}
  \label{eq:dis_nobias_J1working1}
  J_1 = \int_0^{\Lambda^2} d(p^2)
  \biggl\{
  -\Bigl[ \frac{1}{\tE_B} \frac{1}{\sqrt{A^2 -4 \tp^2}}
  +
  \frac{1}{E_B} \frac{1}{\sqrt{\tA^2 -4 \tp^2}} \Bigr]
  +\frac{d_{-}^2}{2 d_{+}} \Bigl[1+\frac{c_{+}}{p^2-c_{+}}\Bigr]
  \times \Bigl[ \frac{1}{\sqrt{A^2 -4 \tp^2}}
  + \frac{1}{\sqrt{\tA^2 -4 \tp^2}} \Bigr]
\\
  - \frac{d_{+}^2}{2 d_{-}} \Bigl[1+\frac{c_{-}}{p^2-c_{-}}\Bigr]
  \times \Bigl[ \frac{1}{\sqrt{A^2 - 4 \tp^2}}
  - \frac{1}{\sqrt{\tA^2 -4 \tp^2}} \Bigr]
\biggr\} ,
\end{multline}
where
\begin{eqnarray}
  \label{eq:dis_nobias_kernel_cdefs}
  c_{\pm} &=& \frac{E_A \pm \tE_A}{d_{\pm}}
  = E_B \tE_B \frac{E_A \pm \tE_A}{\tE_B \pm E_B}  
  \\
  d_{\pm} &=& \frac{1}{E_B} \pm \frac{1}{\tE_B}
  = \frac{\tE_B \pm E_B}{E_B \tE_B}.
\end{eqnarray}
It is convenient to define the integral
\begin{multline}
  \label{eq:dis_nobias_kernel_xi2def}
  \xi_{\pm}^{(2)}(p^2) = \int d(p^2) \frac{1}{p^2 - c_{\pm}}
  \frac{1}{\sqrt{A^2 - 4 \tp^2}}
\\
=
\frac{1}{ \sqrt{B_{\pm}^2 - 4 \tp^2}}
\Bigl\{
\log(p^2 -c_{\pm})
- \log \bigl[ A B_{\pm} +
\sqrt{A^2-4\tp^2}
 \sqrt{B_{\pm}^2-4\tp^2} -4\tp^2 \bigl]
\Bigl\},
\end{multline}
in which we have introduced
\begin{eqnarray}
  \label{eq:dis_nobias_kernel_Bdef}
  B_{\pm} = c_{\pm} / E_B - E_A,
\end{eqnarray}
and the square roots are again chosen such that
$B_{\pm}$ and $ \sqrt{B_{\pm}^2-4\tp^2}$ are in
the same quadrant in the complex plane.
Using this we may write
\begin{equation}
  \label{eq:dis_nobias_integralg5}
  \int d(p^2) \frac{1}{p^2 - c_{\pm}}
   \sqrt{A^2 - 4 \tp^2}
= 
\sqrt{A^2 - 4 \tp^2}
+
B_{\pm} \log \bigl[A +  \sqrt{A^2 - 4 \tp^2} \bigl]
+
\bigl[ B_{\pm}^2 - 4 \tp^2 \bigr] \xi_{\pm}^{(2)}(p^2) ,
\end{equation}
If we use this formulas blindly and just plug in the endpoints the
resulting expressions are
\begin{multline}
  \label{eq:dis_nobias_J1result1}
  J_1 = \biggl\{
  \Bigl[ \frac{d_{-}^2}{2 d_{+}} - \frac{d_{+}^2}{2 d_{-}} -
  \frac{1}{\tE_B} \Bigl]
  \xi^{(1)}(z)
+
  \Bigl[ \frac{d_{-}^2}{2 d_{+}} + \frac{d_{+}^2}{2 d_{-}} -
  \frac{1}{E_B} \Bigl]
  \widetilde{\xi}^{(1)}(z)
\\
+ 
\frac{d_{-}^2 c_{+}}{2 d_{+}}
\Bigl[ \xi_{+}^{(2)}(z) + \widetilde{\xi}_{+}^{(2)}(z) \Bigr]
- \frac{d_{+}^2 c_{-}}{2 d_{-}}
\Bigl[ \xi_{-}^{(2)}(z) - \widetilde{\xi}_{-}^{(2)}(z) \Bigr]
\biggl\}_{0}^{\Lambda^2},
\end{multline}
and
\begin{multline}
  \label{eq:dis_nobias_J2result2}
  J_2 = \frac{1}{4 \tp^2} \Bigl\{
  - z + \frac{1}{d_{-}} \Bigl[
  \sqrt{A^2 - 4\tp^2} 
  - \sqrt{\tA^2 - 4\tp^2}
  + \frac{B_{-}}{E_B}\xi^{(1)}(z) 
  - \frac{\widetilde{B}_{-}}{\tE_B}\widetilde{\xi}^{(1)}(z)
\\
  + \bigl[ B_{-}^2 - 4 \tp^2 \bigr] \xi_{-}^{(2)}(z)
  - \bigl[ \widetilde{B}_{-}^2 - 4 \tp^2 \bigr] \widetilde{\xi}_{-}^{(2)}(z)
  \Bigl]
  \biggl\}_{z=0}^{\Lambda^2}.
\end{multline}
\end{widetext}
One must be careful with the imaginary part of $\xi^{(2)}$ however.
First we note that $\widetilde{B}_{+} = - B_{+}$ 
and $\widetilde{B}_{-} = B_{-}$, which implies that
the $\log(p^2 - c_{\pm})$ term in $\xi^{(2)}$ does not contribute.
Secondly, we write 
$A = 2 \tp \cosh(\alpha)$ and $B = 2\tp \cosh(\beta)$ and use
hyperbolic trigonometric identities to write:
\begin{multline}
\log \Bigl[ A B + \sqrt{A^2-4\tp^2} \sqrt{B^2-4\tp^2} -4\tp^2 \Bigr]
\\
= \log \{ 4 \tp^2 [\cosh(\alpha+\beta) - 1] \}
\\
= \log(8 \tp^2)
+ 2 \log \Bigl[ \sinh \Bigl(\frac{\alpha+\beta}{2}\Bigr) \Bigr].
\end{multline}
By convention $\real(\alpha)>0$ and $-\pi \leq \imag(\alpha) < \pi$ 
(and the same goes for $\beta$). Therefore
$-\pi < \imag(\alpha+\beta)/2 <\pi$ and the argument of 
the log never crosses the branch cut along the negative real
axis. Therefore the representation of $A$ and $B$ in terms of
hyperbolic functions automatically takes care of the phase
information of the argument.
Finally, the case when $B$ is purely real and negative
(this is relevant for the calculation of the DC conductivity) 
requires that one should choose $\imag(\beta) < 0$ so that 
$\sign[\real(B)]\sqrt{B^2 - 4\tp^2} = 2 \tp \sinh(\beta)$.

%
%
\section{Local propagator in the biased graphene bilayer}
\label{app:Gdetail_biased}
In this appendix we provide some details on the calculation
of the local propagator in the BGB that we use in 
Sec.~\ref{sec:CPA_biased}.
Introducing the notations:
\begin{subequations}
\label{eq:appBGBi_green3}
\begin{eqnarray}
\alpha_1 &=& \om - V/2 - \Sigma_{\rA 1} , \\
\beta_1 &=&  \om - V/2 - \Sigma_{\rB 1} , \\
\alpha_2 &=& \om + V/2 - \Sigma_{\rA 2} , \\
\beta_2 &=&  \om + V/2 - \Sigma_{\rB 2} ,
\end{eqnarray}
\end{subequations}
one can write
\begin{subequations}
\label{eq:appBGBi_green4}
\begin{eqnarray}
\GR_{\rA 1  \rA 1} &=& \frac{\beta_1 (\alpha_2 \beta_2 - k^2)}{D} , \\
\GR_{\rB 1  \rB 1} &=& \frac{\alpha_1 (\alpha_2 \beta_2 - k^2) - \tp^2 \beta_2}{D}.
\end{eqnarray}
\end{subequations}
The equations for the corresponding quantities
 in layer 2 are obtained by exchanging
the indeces $1\leftrightarrow2$ everywhere.
The denominator can be written as
\begin{equation}
D = (\alpha_1 \beta_1 -k^2) (\alpha_2 \beta_2 - k^2) - \tp^2 \beta_1 \beta_2
\equiv (k^2 - z_{-} )(k^2 - z_{+}),
\end{equation}
where we have defined
\begin{equation}
z_{\pm} = \frac{\alpha_1 \beta_1 +\alpha_2 \beta_2}{2} \pm \frac{1}{2}
\sqrt{(\alpha_1 \beta_1 - \alpha_2 \beta_2)^2 + 4 \tp^2 \beta_1 \beta_2}.
\end{equation}
Introducing the integrals
\begin{multline}
\Lambda^2 \xi_0 = \int_0^{\Lambda^2}d(k^2)\frac{1}{D} 
\\ =
        \frac{1}{z_+ - z_-} \Bigl [\ln \Bigl( \frac{\Lambda^2-z_+}{-z_+} \Bigr)
        - \ln \Bigl( \frac{\Lambda^2-z_-}{-z_-} \Bigr)  \Bigr]
        \\
        \approx 
        \frac{1}{z_+ - z_-} \Bigl [\ln \Bigl( \frac{\Lambda^2}{-z_+} \Bigr)
        - \ln \Bigl( \frac{\Lambda^2}{-z_-} \Bigr)  \Bigr],
\end{multline}
and
\begin{multline}
  \Lambda^2 \xi_1 = \int_0^{\Lambda^2}d(k^2)\frac{k^2}{D} 
\\ =
  \frac{1}{z_+ - z_-} \Bigl [ z_+ \ln \Bigl( \frac{\Lambda^2-z_+}{-z_+} \Bigr)
      -z_- \ln \Bigl( \frac{\Lambda^2-z_-}{-z_-} \Bigr)  \Bigr]
        \\
        \approx 
        \frac{1}{z_+ - z_-} \Bigl [z_+ \ln \Bigl( \frac{\Lambda^2}{-z_+} \Bigr)
        - z_- \ln \Bigl( \frac{\Lambda^2}{-z_-} \Bigr)  \Bigr],
\end{multline}
we can easily compute $\overline{\text{G}}$.
Using the explicit form of the propagators in
Eq.~(\ref{eq:appBGBi_green4}) and  the same continuum
approximation as in Eq.~(\ref{eq:BGBi_Gbar1}) we obtain
\begin{subequations}
\begin{eqnarray}
-\overline{\GR}_{\rA 1}  &=& \beta_1 (\xi_1 - \alpha_2 \beta_2 \xi_0) ,\\
-\overline{\GR}_{\rB 1}  &=& \alpha_1 (\xi_1 - \alpha_2 \beta_2 \xi_0) 
+\tp^2 \beta_2 \xi_0 .
\end{eqnarray}
\end{subequations}
Again the corresponding quantities in plane 2 are obtained by
exchanging the indeces $1\leftrightarrow2$ everywhere.

\bibliography{graphite}

\begin{thebibliography}{84}
\bibitem[{not()}]{note_localization}
\bibinfo{note}{This approximation gives a good account of the electronic
  properties as long as the localization length is larger than the sample size.
  This condition is not constraining in 2D due to the exponetially large
  localization length.}
\end{thebibliography}

\end{document}